\documentclass[a4paper,11pt,fleqn]{article}
\pdfoutput=1 
\usepackage[utf8]{inputenc}
\usepackage{jheppub} 

\usepackage{amssymb,amsmath,amsfonts}
\usepackage{mathtools}
\usepackage{graphicx}
\usepackage{multirow}
\usepackage{verbatim}
\usepackage{appendix}
\usepackage{fancybox}
\usepackage{array}
\usepackage{url}
\usepackage{float}
\usepackage[mathscr]{euscript}
\usepackage{bbm} 
\usepackage{slashed}
\usepackage{tikz-cd}

\newcommand{\bea}{\begin{eqnarray}}
\newcommand{\eea}{\end{eqnarray}}
\newcommand{\be}{\begin{equation}}
\newcommand{\ee}{\end{equation}}

\newcommand{\Z}{{\mathbb Z}}
\newcommand{\R}{{\mathbb R}}
\newcommand{\C}{{\mathbb C}}
\newcommand{\Q}{{\mathbb Q}}

\def\frak{\mathfrak}

\def\tilde{\widetilde}
\renewcommand{\bar}{\overline}
\renewcommand{\hat}{\widehat}
\def\^{{\wedge}}
\def\*{{\star}}

\def\CA{{\mathcal A}}

\def\CC{{\mathcal C}}
\def\CD{{\mathcal D}}

\def\CH{{\mathcal H}}

\def\CM{{\mathcal M}}
\def\CN{{\mathcal N}}
\def\CO{{\mathcal O}}

\def\CR{{\mathcal R}}
\def\CS{{\mathcal S}}
\def\CT{{\mathcal T}}

\def\ra{\rightarrow}
\def\da{\downarrow}

\newcommand{\Tr}{{\rm Tr}}
\def\Beta{{\mathrm{B}}}
\def\Tor{{\mathrm{Tor}\,}}
\def\Ker{{\mathrm{ker}\,}}
\def\Spin{\mathrm{Spin}}
\def\String{\mathrm{String}}
\def\TMF{\mathrm{TMF}}
\def\tmf{\mathrm{tmf}}
\def\Tmf{\mathrm{Tmf}}
\def\KO{\mathrm{KO}}
\def\dim{{\mathrm{dim}\,}}

\title{4-manifolds and topological modular forms}

\author[1,2]{Sergei Gukov}
\author[1,3]{Du Pei}
\author[4,5]{Pavel Putrov}
\author[6]{Cumrun Vafa}

\affiliation[1]{Walter Burke Institute for Theoretical Physics, California Institute of Technology, Pasadena, CA 91125, USA}
\affiliation[2]{Max-Planck-Institut f\"{u}r Mathematik, Vivatsgasse 7, D-53111 Bonn, Germany}
\affiliation[3]{Center for Quantum Geometry of Moduli Spaces, Department of Mathematics, University
of Aarhus, DK-8000, Denmark}
\affiliation[4]{ICTP, Strada Costiera 11, Trieste 34014, Italy}
\affiliation[5]{School of Natural Sciences, Institute for Advanced Study, Princeton, NJ 08540, USA}
\affiliation[6]{Jefferson Physical Laboratory, Harvard University, Cambridge, MA 02138, USA}

\emailAdd{gukov@theory.caltech.edu, pei@caltech.edu, putrov@ictp.it, vafa@physics.harvard.edu}

\preprint{CALT-TH-2018-034}

\abstract{We build a connection between topology of smooth 4-manifolds and the theory of topological modular forms by considering topologically twisted compactification of 6d $(1,0)$ theories on 4-manifolds with flavor symmetry backgrounds. The effective 2d theory has $(0,1)$ supersymmetry and, possibly, a residual flavor symmetry. The equivariant topological Witten genus of this 2d theory then produces a new invariant of the 4-manifold equipped with a principle bundle, valued in the ring of equivariant weakly holomorphic (topological) modular forms. We describe basic properties of this map and present a few simple examples. As a byproduct, we obtain some new results on 't Hooft anomalies of 6d $(1,0)$ theories and a better understanding of the relation between 2d $(0,1)$ theories and TMF spectra.

}

\begin{document}
\maketitle

\section{Introduction}

The existence of non-trivial superconformal theories in six dimensions has been one of the major discoveries of the past few decades in string theory.  These come in two classes depending on the number of supersymmetries:  $(2,0)$ or $(1,0)$.
The case of $(2,0)$ has been the most studied one and it comes in ADE-types \cite{Witten:1995zh}.  The A-type is realized by parallel M5 branes in M-theory \cite{Strominger:1995ac}.  The $(1,0)$ is far more extensive in variety and a recent classification of them has been proposed in \cite{Heckman:2013pva,Bhardwaj:2015xxa,Heckman:2015bfa}
.  They are related to singularities in elliptic Calabi-Yau threefolds.\footnote{When the elliptic fibration is trivial this gives the ADE type singularities leading to the special case with enhanced $(2,0)$ supersymmetry.}  These theories are interesting in their own right in 6 dimensions as novel quantum systems which are decoupled from gravity.  Moreover there has been tremendous activity by studying their compactifications down to 4 and 3 dimensions to get novel theories even in lower dimensions.

On the other hand $(2,0)$ theories have been used to obtain invariants for manifolds in 4 and 3 dimensions \cite{Vafa:1994tf, Kapustin:2006pk,  Witten:2011zz, Gadde:2013sca, Dedushenko:2017tdw, Dimofte:2011ju, Dimofte:2011py, Gukov:2016gkn, Gukov:2017kmk}.  Namely one considers topologically twisted theories by embedding an $SU(2)_+$ part of spin connection $SU(2)_+\times SU(2)_-$ for 4-manifolds or the $SU(2)$ holonomy for the three manifold, with an $SU(2)$ subgroup of R-symmetry group which is $SO(5)$, leading to supersymmetric theories in lower dimensions.  In particular for the generic 4-manifolds this leads to is $(0,2)$ in 2 dimensions and for the case of 3-manifolds leading to $\CN =2$ theories in 3d.    However there are far more 6 dimensional $(1,0)$ theories and it is natural to ask what kinds of invariants do they lead to when compactifying them to lower dimensions.  Unlike the $(2,0)$ case, the R-symmetry for these theories is exactly $SU(2)$ so it is the most economical one to allow defining topologically twisted theories for 4- and 3-manifolds!  Twisted compactifications of them on 4-manifolds lead to $(0,1)$ supersymmetric theories in 2d\footnote{If it is a K\"ahler manifold, the supersymmetry of the 2d theory will be enhanced to $(0,2)$ while for hyper-K\"ahler manifolds it leads to $(0,4)$ theories.} and ${\CN= 1}$ supersymmetric theories in 3d. Topological aspects of the resulting $2d$ and $3d$ theories can be viewed as invariants associated to 4- and 3-manifolds.
In other words we associate to each three-manifold a 3d $\CN=1$ supersymmetric quantum field theory
\be
M_3 \leadsto T[M_3],
\ee
where $T$ labels a particular 6d (1,0) Theory.  Similarly to each 4-manifold it associates a 2d $(0,1)$ theory,
\be
M_4 \leadsto T[M_4].
\ee

Evaluating an elliptic genus \cite{Witten:1986bf} of $T[M_4]$ thus should produce an invariant of $M_4$ valued in the ring of modular forms MF. The ordinary elliptic genus is believed to be extendable to the so-called topological Witten genus $\sigma$, valued in the ring $\pi_*\text{TMF}$ of (stable) homotopy groups of the spectrum of topological modular forms TMF.\footnote{See Section \ref{sec:TMF2D} and Appendix \ref{app:TMF} for details.} Thus, naively, we have the following diagram:
\begin{equation}
\begin{tikzcd}
{\frac{\left\{\begin{array}{c} \text{Spin} \\ \text{4-manifolds} \end{array}\right\}}{\sim\;\text{diffeomorphism}}}  \ar[r,"T"] &
\frac{\left\{\begin{array}{c} \text{(0,1) theories} \end{array}\right\}}{\sim \;\text{SUSY deformations}} \ar[r,"\sigma"] \ar[dr,"\text{elliptic genus}" description]& \pi_*\text{TMF} \ar[d]
\\
	& & \text{MF}\subset \Z[[q]]
\end{tikzcd}
\label{map-to-TMF-naive}
\end{equation}
However, the 6d $(1,0)$ theories are typically richer than their $(2,0)$ counterparts:  They typically come equipped with additional global symmetries.  This allows one to turn on background gauge fields in the flavor group $G$.  In the 4d case this amounts to choosing an instanton background and in 3d case to a flat bundle in $G$.\footnote{We can use abelian subfactors in $G$ and turn on constant fluxes in addition to the above choices.}
Due to various technicalities, discussed in detail in the main part of the paper, one needs to refine the statement as follows. Let the flavor symmetry of the 6d theory be $G$, and the maximal order  (modulo --- in the multiplicative sense --- perfect squares) of the elements of the defect group \cite{DelZotto:2015isa} be $N_0$.  This is related to the analog notion of ``spin structure'' for the 6d $(1,0)$ theory and is needed to define the partition function of $(1,0)$ theories on 6-manifolds.  Fix a subgroup $G'\subset G$ for which we have turned on the background field on the 4-manifold and let
\begin{equation}
	G_\text{2d} \; := \; \text{Centralizer}_{G}(G')
\end{equation}
be its centralizer subgroup.  This corresponds to the group which can be viewed as the surviving portion of the flavor symmetry group in 2d.  Then there is a following diagram of maps:
\begin{equation}
\begin{tikzcd}
\frac{\left\{\begin{array}{c} \text{Spin} \\ \text{4-manifolds} \\ \text{w/ maps to $BG'$} \end{array}\right\}}{\sim\;\text{diffeomorphism, homotopy}}  \ar[r,"T"] &
\frac{\left\{\begin{array}{c} \text{relative} \\ \text{(0,1) theories} \\ \text{w/ symmetry $G_{\text{2d}}$} \end{array}\right\}}{\sim \;\text{SUSY deformations}} \ar[r,"\sigma"] \ar[dr,"\text{equivariant elliptic genus}" description]& \pi_*\text{TMF}_{G_{\text{2d}}}(\Gamma_0(N_0)) \ar[d] \\
& & \text{MF}_{G_{\text{2d}}}(\Gamma_0(N_0))\subset R(G_{\text{2d}})[[q]]
\end{tikzcd}
\label{map-to-TMF-complicated}
\end{equation}
where $\text{MF}_{G_{\text{2d}}}(\Gamma_0(N_0))$ is the ring of weakly holomorphic $G_{\text{2d}}$-equivariant modular forms with respect to congruence subgroup $\Gamma_0(N_0)$ and $R(G_{\text{2d}})$ is the representation ring of $G_{\text{2d}}$.
In general, we do not expect the map $\sigma$, the equivariant topological Witten genus, to be defined for all 2d theories obtained in this way, but only for those theories whose space of bosonic zero-modes is $G_{\text{2d}}$-compact ({\it i.e.} fixed loci under the $G_{\text{2d}}$-action are all compact). However, as we will see later, the composition $\sigma \circ T$ is still expected to be defined for (almost) all 4-manifolds if a generic map to $BG'$ is chosen.  This is because twisting the 2d partition function with flavor symmetry $G_{\text{2d}}$ will get rid of zero modes and renders the path-integral finite.

To get invariants for the three manifolds in this way, the easiest thing to consider is $M_3\times S^1$ and turn on instanton flavor background leading again to $(0,1)$ theory in $2d$ for each choice of instanton background.\footnote{If $G$ is trivial, there are other ways, discussed in the paper, where we can obtain invariants for $M_3$.}

Finally the main question is how do we actually compute the corresponding modular forms for arbitrary $M_4$ with some $G$-bundle on it.  It turns out that this is not easy, unfortunately.  Nevertheless for some special cases of the 6d theory and for some special $M_4$ such as product of a pair of Riemann surfaces, we have managed to compute it.  The idea is to use the knowledge of the 4d $\CN =1$ theory obtained for these theories when we compactify from 6d to 4d on a Riemann surface \cite{Bah:2012dg, Razamat:2016dpl, Kim:2017toz}, and use this to compute the partition function on $T^2$ times another Riemann surface.
Part of the difficulty in computing the partition functions is that the 6d $(1,0)$ theories do not typically have a tangible field theoretic description.  However, compactifying them on a circle, and going to 5d, they do seem to have convenient gauge theoretic descriptions.  Viewing this circle as one of the two circles of elliptic genus torus $T^2$, we formulate the necessary computation to obtain Witten genus from this 5d perspective.  Even though we have not used this picture to do explicit computations, we believe this may hold the key to more general approach to such computations in the future.

From a mathematical perspective, since 6d $(1,0)$ theories are classified by singularities of elliptic Calabi--Yau threefolds,  we can summarize the maps by saying that each singular elliptic Calabi--Yau $ECY$ gives rise to a map 
from 4-manifolds (which of course includes $M_3\times S^1)$ 
to topological modular forms:
$$ECY:\ \ \ \ \{\text{4-manifolds}\}\rightarrow \pi_*(\TMF).$$
If resulting 6d $(1,0)$ theory is relative or has flavor symmetries, the target should of course be $\pi_*\TMF_{G_{2d}}(\Gamma_0(N_0))$. As we will explain later, this map can be upgraded into a functor between categories. 
Although this does not necessarily provide stronger invariants, it incorporates 3-manifolds into the story in an interesting way.

In a sense the new invariants we associate to four manifolds are extensions of Donaldson's invariants:  If we reverse the order of compactification and first compactify on $T^2$ and then on the four-manifold, the theory has $\CN=2$ supersymmetry in four dimensions.  It differs from the usual twist studied by Witten \cite{Witten:1988ze}, in that it includes extra degrees of freedom coming from six dimensions.  These extra fields would lead to a modular partition function instead of what one has in the case of Donaldson theory. In that sense they have a feature more similar to $\CN=4$ Yang--Mills theory which leads to modular partition functions \cite{Vafa:1994tf}.  On the other hand they categorify the four manifold invariants, in the sense that if we compactify the 6d $(1,0)$ theory on a four manifold the Hilbert space of supersymmetric states of the $(0,1)$ theory in 2d together with all the residual flavor symmetries which act on them.  Thus we end up with a rich class of invariants for four manifolds which from the physics perspective is rather interesting and one expects them to lead to new mathematical insights in understanding invariants for four manifolds.
\bigskip

The paper is organized as follows.
The maps $T$ and $\sigma$ are carefully introduced in sections~\ref{sec:setup} and \ref{sec:TMF2D}, respectively.
In particular, many parts of section~\ref{sec:setup} are devoted to various aspects of compactification on
4-manifolds, from six dimensions down to two dimensions. The latter, then, becomes the main subject of section~\ref{sec:TMF2D},
where connections to TMF are introduced and, in particular, what one might call ``equivariant TMF'' is proposed based
on physics of 2d $(0,1)$ theories.
The discussion in sections~\ref{sec:setup} and \ref{sec:TMF2D} applies to general $(0,1)$ theories in six and two dimensions.
Concrete examples, which illustrate this general discussion, are worked out in section~\ref{sec:examples}.
The ``mathematical content'' of 2d $(0,1)$ theories $T[M_4]$ and 3d $\mathcal{N}=1$ theories $T[M_3]$
involves delicate invariants of 4-manifolds and 3-manifolds, which sometimes can be formulated using standard
gauge theory techniques. This is explained in section~\ref{sec:BPS-equations}. In Appendix~\ref{app:TMF} we review some relevant facts about the spectrum of the topological modular forms.  In Appendix~\ref{app:BF} we give a brief review of Bauer-Furuta invariants. Finally, in Appendix~\ref{sec:geography}, we look back at the ``big picture'' and ask how far, beyond the existent
set of invariants, can 6d $(1,0)$ theories take us in exploring the wild world of smooth 4-manifolds.

\section{From six to two dimensions, via 4-manifolds}
\label{sec:setup}

\subsection{Topological twist}
\label{sec:twist}

Given an arbitrary 6d (1,0) superconformal theory, one can consider it on a 6-manifold of the form
\begin{equation}
	M_4\times T^2_\tau
\end{equation}
where $M_4$ is an oriented Spin 4-manifold and $T^2_\tau$ is a flat 2-torus with complex structure parameter $\tau$. We work in the Euclidean signature, so that the Riemannian holonomy group is contained in an $SO(4)_E\times U(1)_E$ subgroup of $SO(6)_E$. Along the 4-manifold $M_4$, we perform a partial topological twist to preserve at least one supercharge. The supersymmetry algebra of the 6d theory contains an $SU(2)_R$ R-symmetry. The topological twist is then realized by identifying the $SU(2)_R$ principle bundle with the $SU(2)_+$ factor of $\text{Spin}(4)_E=SU(2)_+\times SU(2)_-$ bundle, the lift of the $SO(4)_E$ orthonormal tangent frame bundle of $M_4$. Later in the paper we will further elaborate on the requirement of $M_4$ being Spin and when this condition can be relaxed. After the twist, the supercharges transform as
\begin{equation}
	\begin{array}{ccccc}
		\mathbf{(4,2)} & \rightarrow &
		\mathbf{((2,1)_{+\frac{1}{2}}\oplus (1,2)_{-\frac{1}{2}},2)} & \rightarrow
		& \mathbf{(1,1)_{+\frac{1}{2}}\oplus(3,1)_{+\frac{1}{2}}\oplus (2,2)_{-\frac{1}{2}}} \\
\\
		\text{Spin}(6)_E\times SU(2)_R & \supset & \text{Spin}(4)_E\times U(1)_E\times SU(2)_R  & \supset &  {SU}(2)_\text{diag}\times SU(2)_- \times U(1)_E\,.
	\end{array}
\label{SUSY-twist}
\end{equation}
The supercharge in the $\mathbf{(1,1)_{+\frac{1}{2}}}$ representation becomes a scalar on the $M_4$, and therefore defines a globally constant supersymmetry transformation along $M_4$ of general holonomy. By taking the size of $M_4$ to be small compared to the size of $T^2_\tau$, one obtains an effective 2d theory which, by analogy with~\cite{Gadde:2013sca}, we shall denote as $T[M_4]$ without making explicit the dependence on the choice of the parent 6d theory. From the viewpoint of this 2d theory, the supercharge in the $\mathbf{(1,1)_{+\frac{1}{2}}}$ representation will be the supercharge of the two-dimensional $(0,1)$ supersymmetry algebra.

While in general the physical theory $T[M_4]$ may depend on the conformal class of the metric on $M_4$, in this paper we study supersymmetry-protected quantities that are invariant under diffeomorphism of $M_4$ and, therefore, are independent of the choice of the metric.
One such invariant that will play a central role in this paper is the partition function of the 6d theory on $M_4\times T^2_\tau$ with an odd Spin structure on $T^2_\tau$. As it is protected by supersymmetry, one indeed expects it to depend only on the diffeomorphism class of $M_4$, which is completely determined by the topology and smooth structure.

In particular, this partition function should not depend on the relative size of $M_4$ and $T^2_\tau$. This allows different interpretations of the 6d partition function:
\begin{equation}
	Z_\text{6d}[M_4\times T^2_\tau] \; = \; Z_{T[M_4]}[T^2_\tau] \; = \; Z_\text{5d}[M_4\times S^1](q) = \; Z_\text{4d}[M_4](\tau) \,.
	\label{Z6d}
\end{equation}
The second quantity in (\ref{Z6d}) is the elliptic genus of $T[M_4]$,
\begin{equation}
	Z_{T[M_4]}[T^2_\tau] \; := \; \text{Tr}_\text{R}(-1)^F q^{L_0},
	\label{EllipticGenus}
\end{equation}
given by a trace over the Hilbert space of $T[M_4]$ on a circle in the Ramond sector \cite{Witten:1986bf}. Here, as usual, $q=\exp(2\pi i \tau)$. The third quantity in (\ref{Z6d}) is the partition function of the effective 5d theory obtained by compactification of the 6d theory on a circle with fixed value $q$ of the holonomy of $U(1)$ Kaluza-Klein (KK) graviphoton symmetry along $S^1$. Finally, $Z_\text{4d}[M_4](\tau)$ is the partition function of the effective 4d theory obtained by further compactifying the 5d theory on a circle, \textit{with all KK modes included}. The 4d theory has $\CN=2$ supersymmetry and is topologically twisted on $M_4$. The 4d R-symmetry is $SU(2)_R$ where $SU(2)_R$ is inherited from the parent 6d theory. Therefore, from the 4d point of view the topological twist is the usual Donaldson--Witten twist performed by identifying the $SU(2)_R$ with the $SU(2)_+$ factor of $\text{Spin}(4)_E=SU(2)_+\times SU(2)_-$ space-time symmetry. The same conclusion, of course, follows from the fact that 4d $\CN=2$ theories (unlike $\CN=4$) have an essentially unique topological twist on manifolds of generic holonomy\footnote{Note, twisting the $SU(2)_-$ subgroup of $\text{Spin}(4)_E$ is related to this twist by an orientation reversal.}.

Generically the 4d theory is not superconformal and does not have extra $U(1)_r$ R-symmetry. Due to the KK modes it has hypers with mass of order $\tau$ which explicitly brake it. The parameter $\tau$ can also appear in 4d theory as a holomorphic gauge coupling constant. In certain special cases one can expect an effective 4d description in terms of a superconformal 4d theory where $\tau$ in (\ref{Z6d}) plays the role of a holomorphic exactly marginal coupling. We will elaborate on the 4d and 5d interpretations on the partition function in Section \ref{sec:BPS-equations}.

Note that, apart from the usual elliptic genus (\ref{EllipticGenus}), it is believed that there exist other quantities invariant under SUSY-preserving deformations of 2d $(0,1)$ theories valued in finite cyclic subgroups of the commutative ring $\pi_*\text{TMF}$ \cite{ST1,ST2}. Here, TMF is the spectrum of a generalized cohomology theory known as ``topological modular forms'' and $\pi_*$ denotes stable homotopy groups (see {\it e.g.}~\cite{hopkins1995topological}). Roughly speaking, the commutative ring $\pi_*\text{TMF}$ can be understood as an extension of the subring of weakly holomorphic modular forms, where the ordinary elliptic genus takes values, by the ideal of $\pi_*\text{TMF}$ generated by all torsion elements. For details see section~\ref{sec:TMF2D} and Appendix~\ref{app:TMF}.

Specifically, for each 2d $(0,1)$ theory with the gravitational anomaly
\be
c_\text{R}-c_\text{L} \; = \; \frac{d}{2}
\ee
there should be an invariant, which we call the \textit{topological Witten genus},\footnote{In the mathematical literature, ``topological Witten genus'' usually refers to the map 
$$\Omega^\String_* \cong \pi_*(\mathrm{MString})\ra\pi_*(\mathrm{tmf})$$
 induced by the String-orientation of $\mathrm{tmf}$: $\mathrm{MString}\ra\mathrm{tmf}$. Our version of the topological Witten genus map reduces to this definition when we take the $(0,1)$ theory to be the sigma-model with a string manifold as the target.} valued in the abelian group $\pi_d \text{TMF}$. The value of the invariant in the free part of $\pi_d \text{TMF}$ coincides with the usual elliptic genus and can be non-zero only when $d\equiv 0\bmod 4$, but there are in addition torsion-valued invariants. The simplest example is the mod-2 index first mentioned in \cite{Witten:1986bf}, while the more general torsion-valued invariants are currently understood only in terms of 2d $(0,1)$ sigma-models with compact target space. We expect that such invariants, much like the usual elliptic genus discussed earlier, all have appropriate counterparts on the 4-manifold side. Since they are still associated to an elliptic curve, we expect that on the 5d/4d side one would need to consider certain observables of the same 5d/4d theory or its mild modification (such as {\it e.g.}~orbifolding). Such observables then should produce torsion-valued invariants of smooth 4-manifolds.

Although in this paper we do not provide an explicit realization of such torsion-valued invariants directly in terms of 4-manifolds, it is natural to expect that, at least in some cases, such invariants are closely related to (equivariant) Bauer-Furuta invariant \cite{bauer2004stable,bauer2004stable2} (also see Appendix \ref{app:BF} for a brief review) and its possible generalizations. Bauer-Furuta invariant and its generalizations are valued in the (equivariant) stable homotopy groups of spheres. The value of an invariant is given by the homotopy class of a certain ``monopole map'', determined by the action of the supercharge on the space of fields in an $\CN=2$ 4d topologically twisted gauge theory on $M_4$. The possibility of the relation to the invariants constructed via compactifications of 6d (1,0) theories is supported by the fact that the torsion part of the ring $\pi_*\text{tmf}$ in low degrees is very close to $\pi_*\mathbb{S}$, the ring of stable homotopy groups of spheres. Moreover, there is a canonical map $\mathbb{S}\rightarrow\text{tmf} $ between the corresponding spectra. It would be interesting to investigate if one can use a similar approach to define invariants valued in the torsion subgroups of $\pi_*\text{TMF}$ for general $(0,1)$ 2d theories.

\subsection{Flux compactifications of 6d $(1,0)$ theories}

For a generic 6d theory and a 4-manifold $M_4$, the most basic partition function (\ref{Z6d}) is often divergent due to presence of non-compact bosonic zero-modes. As we illustrate explicitly in section~\ref{sec:KK-reduction}, these non-compact zero modes originate from 6d bosonic zero modes and then persist in 2d and 5d/4d descriptions as well. Thus, from the 5d/4d perspective, they make the partition function $Z_\text{5d}[M_4\times S^1](q)=Z_\text{4d}[M_4](\tau)$ ill-defined because the moduli spaces of the solutions of the corresponding BPS equations are non-compact. (And, as will be discussed in section~\ref{sec:anomaly-enhancement}, their virtual fundamental class is ill-defined.) This problem, of course, also appears at the level of the 2d partition function, the elliptic genus (\ref{EllipticGenus}), because such non-compact zero-modes are present in the 2d theory $T[M_4]$ as well. The presence of non-compact bosonic zero-modes also makes the torsion-valued invariants mentioned earlier ill-defined. And, even when there are no such bosonic zero-modes, there might be fermionic zero-modes which can force that partition function to vanish. However, all such issues can be fixed by turning on non-trivial background fields for global symmetries.

An important feature that distinguishes 6d $(1,0)$ theories from 6d $(2,0)$ theories is that the former generically have non-trivial flavor symmetries. Using these symmetries, one can define more general compactifications on $M_4\times T^2_\tau$ by turning on background vector fields for the global symmetry $G$, while still preserving at least one real supercharge. Similar general backgrounds were considered for compactifications of 6d $(1,0)$ theories on Riemann surfaces in \cite{Razamat:2016dpl,Kim:2017toz,Kim:2018lfo}. On a Riemann surface one can turn on fluxes for an abelian subgroup of the global symmetry $G$ as well as the holonomies consistent with the chosen fluxes. On 4-manifolds there is a new feature: namely, apart from turning on fluxes and holonomies, one can also consider configurations with non-trivial instanton number for a non-abelian subgroup of $G$. In order to preserve supersymmetry, the background field should be self-dual. This is the same condition as imposed on dynamical gauge fields of 4d $\CN=2$ theories by supersymmetric localization, except in the present context they are fixed background fields and are not integrated over.

\subsection{4-manifolds with special holonomy}
\label{sec:special-holonomy}

In special cases when $M_4$ admits a metric of reduced holonomy, the supersymmetry of the 2d theory $T[M_4]$ is enhanced. In particular, for a generic K\"ahler $M_4$ the holonomy group is $U(2)\subset SO(4)$ and the supercharges in (\ref{SUSY-twist}) are further decomposed as follows,
\begin{equation}
	\begin{array}{rcl}
\mathbf{(1,1)_{+\frac{1}{2}}\oplus(3,1)_{+\frac{1}{2}}\oplus (2,2)_{-\frac{1}{2}}} & \rightarrow &
\mathbf{ 1_{0,+\frac{1}{2}}\oplus 1_{0,+\frac{1}{2}}\oplus 1_{-1,+\frac{1}{2}}\oplus 1_{+1,+\frac{1}{2}}\oplus 2_{-\frac{1}{2},-\frac{1}{2}}\oplus 2_{+\frac{1}{2},-\frac{1}{2}} }  \\
\\
		{SU}(2)_\text{diag}\times SU(2)_- \times U(1) & \supset &  SU(2)_- \times U(1)_\text{diag} \times U(1)
	\end{array}.
\end{equation}
Both of the two supercharges in $\mathbf{ 1_{0,+\frac{1}{2}} }$ representation can be made constant along the $M_4$ and the effective 2d theory $T[M_4]$ will have $(0,2)$ supersymmetry.

When $M_4$ is hyper-K\"ahler, the holonomy is further reduced to $SU(2)\subset U(2)$ and there will be four constant supercharges forming $(0,4)$ super-Poincar\'e algebra in 2d. Furthermore, in the hyper-K\"ahler case the twisted theory is equivalent to the untwisted one, since the $SU(2)_+$ bundle used for twisting becomes trivial.

Another special case is when $M_4 = M_3\times S^1$. The holonomy is then $SO(3)_\text{3d}\subset SO(4)$ with the corresponding Spin lift $SU(2)_\text{3d}\, {\subset}\, \text{Spin}(4)=SU(2)_+\times SU(2)_-$, where $SU(2)_\text{3d}$ is embedded as the diagonal subgroup. The supercharges then transform in the following representations:
\begin{equation}
	\begin{array}{rcl}
\mathbf{(1,1)_{+\frac{1}{2}}\oplus(3,1)_{+\frac{1}{2}}\oplus (2,2)_{-\frac{1}{2}}} & \quad \rightarrow \quad &
\mathbf{ 1_{+\frac{1}{2}}\oplus 3_{+\frac{1}{2}}\oplus 1_{-\frac{1}{2}}\oplus 3_{-\frac{1}{2}} } \\
\\
		{SU}(2)_\text{diag}\times SU(2)_- \times U(1) & \quad \supset \quad &  SU(2)_\text{3d} \times U(1)
	\end{array}.
\end{equation}
The two supercharges in representations $\mathbf{ 1_{\pm\frac{1}{2}} }$ are constant along $M_3$ and form the $(1,1)$ superalgebra of the 2d theory $T[M_4]$. Alternatively, one can compactify the 6d theory directly on $M_3$ and obtain an effective 3d $\CN=1$ theory which, by analogy with \cite{Dimofte:2010tz}, we denote $T[M_3]$.

Note that the case $M_4=M_3\times S^1$ is somewhat degenerate, because when the supersymmetry is enhanced to $(1,1)$ the elliptic genus (\ref{EllipticGenus}) of $T[M_4]$ becomes $q$-independent. One way to avoid this is to consider instead
\begin{equation}
	Z'_{T[M_4]}[T^2_\tau] \; := \; \Tr_\text{R}(-1)^{F_R}q^{L_0}.
	\label{EllipticGenusAlt}
\end{equation}
From the 2d point of view, this means choosing different Spin structure on $T^2_\tau$ for left- and right-moving fermions. Such background, however, generically does not have a natural lift to 6d theory on $M_4\times T^2_\tau$ because {\it e.g.} a single 6d spinor field can produce 2d fermions of different chirality, as will be explicitly demonstrated in section~\ref{sec:KK-reduction}. Another issue with (\ref{EllipticGenusAlt}) is that it cannot be lifted to a partition function of the 3d theory $T[M_3]$ on $T^3=S^1\times T^2_\tau$ because $F_R$, the chiral fermion number symmetry, emerges only in the strict 2d limit. A more clever way to produce invariants of 3-manifolds in this framework is to construct a certain 4-manifold for each $M_3$ in a ``canonical way,'' other than just $M_3\times S^1$. We shall pursue this approach in section~\ref{sec:3-manifolds}.

\subsection{Anomaly polynomial reduction}
\label{sec:anomaly-reduction}

Some basic information about the effective 2d theory can be obtained from its anomaly polynomial $I_4$. It can be obtained by integrating the degree-8 anomaly polynomial $I_8$ of the 6d theory over $M_4$. The anomaly polynomials of general 6d theories were studied in \cite{Ohmori:2014kda,Cordova:2015fha}. Various explicit examples can be found in \cite{Ohmori:2015pua}.

The anomaly polynomial of a generic 6d (1,0) theory with global symmetry $G$ has the following form,
\begin{equation}
	I_8=\alpha c_2(R)^2+\beta c_2(R)p_1(T)+\gamma p_1(T)^2 +\delta p_2(T)+I_8^{(\text{flavor})},
	\label{anomaly-6d}
\end{equation}
where $c_2(R)$ is the second Chern class for the R-symmetry bundle, $p_1(T)$ and $p_2(T)$ are Pontryagin classes of the tangent bundle, and $\alpha,\beta,\gamma,\delta$ are real coefficients. In the above formula, we explicitly separated the contribution coming from the 't Hooft anomalies for the global symmetry $G$, possibly mixed with other symmetries,
\begin{equation}
	I_8^{(\text{flavor})}=\omega_4^{(1)}(G)+\omega_2^{(1)}(G)c_2(R)+\omega_2^{(2)}(G)p_1(T).
\end{equation}
Here, $\omega_n^{(i)}(G)$ can be understood as elements of $H^{2n}(BG,\mathbb{Q})$,\footnote{When $G=SU(N)$ or $U(N)$ they coincide with Chern classes up to factors, $\omega_n^{(i)}\propto c_n^{(i)}$.} so that the whole anomaly polynomial is $I_8\in H^8(BSO\times BG\times BSU(2)_R,\mathbb{Q})$ and defines a Chern--Simons-like invertible TQFT in 7 dimensions. The question of proper quantization of the coefficients will be addressed in section~\ref{sec:anomaly-quantization}.

The anomaly polynomial of the effective 2d theory $T[M_4]$ can be obtained by integrating the above mentioned 8-form over a 4-manifold. When no background for the flavor symmetry $G$ is turned on, and when the 4-manifold $M_4$ is of generic holonomy, the anomaly polynomial of $T[M_4]$ reads
\begin{equation}
	I_4 = \frac{c_\text{R}-c_\text{L}}{24}p_1(T)+\omega_2^{(2d)}(G)
\label{I4-polynomial}
\end{equation}
with
\begin{equation}
	c_\text{R}-c_\text{L}=18 \cdot  (\beta -8 \gamma -4\delta )\sigma+12 \beta  \chi
\end{equation}
 and 
\begin{equation}
	\omega^{(2d)}_2(G)=-\frac{2\chi+3\sigma}{4}\,\omega_2^{(1)}(G)+3\sigma\,\omega_2^{(2)}(G)\,\in H^4(BG,\mathbb{Q})\,,
\end{equation}
where $\chi$ and $\sigma$ denote Euler characteristic and signature of $M_4$.

\subsection{Central charges and K\"ahler 4-manifolds}
\label{sec:Kahler-central}

If the 4-manifold $M_4$ is K\"{a}hler, the effective 2d theory has $(0,2)$ supersymmetry and one can make a naive prediction for individual values of the central charges $c_\text{L}$ and $c_\text{R}$. This is because for K\"{a}hler $M_4$, the $U(1)_R$ subgroup of the six-dimensional $SU(2)_R$ symmetry remains unbroken after the topological twist. This $U(1)_R$ can be identified with an R-symmetry in the 2d theory. Assuming that $U(1)_{R}$ is exactly the R-symmetry of the IR SCFT ({\it i.e.}~it does not mix under the RG flow with any $U(1)$ global symmetries either originating from 6d, or emerging in the infrared), we have
\begin{equation}
I_4 = \frac{c_\text{R}-c_\text{L}}{24}p_1(T)+\frac{c_\text{R}}{6}\,c_1(R^\text{2d})^2+\omega_2^{(2d)}(G),
\end{equation}
where
\begin{equation}
c_\text{R}=18\cdot(3 \alpha \sigma-2 \beta \sigma+2 \alpha \chi).
\label{cR-Kahler}
\end{equation}

When the 4-manifold is hyper-K\"ahler (in the closed case this means it is $K3$ or $T^4$) we have $2\chi+3\sigma=0$ and therefore 
\begin{equation}
c_\text{R}=-36 \beta \sigma
\label{cR-hyper-Kahler}
\end{equation}
which, taking into account the fact that $\sigma/16 \in \Z$ and the quantization condition (\ref{6d-anomaly-coef-quantization}) on $\beta$, is always a multiple of $6$, which is consistent with the small $\CN=4$ 2d superconformal algebra in the right-moving sector of the 2d theory $T[M_4]$.

When the R-symmetry is $SU(2)_R$, or if the 6d theory has no $U(1)$ factors in the flavor symmetry $G$ (and no accidental symmetries appear in the IR) then there should be no mixing between 2d R-symmetry and flavor symmetries. However, even in these cases it may happen that the formulas (\ref{cR-Kahler})-(\ref{cR-hyper-Kahler}) do not give the right value of the right-moving central charge of the 2d (1,0) SCFT $T[M_4]$ in the IR. It is possible that the UV $U(1)_R$ ($SU(2)_R$ in the hyper-K\"ahler case) R-symmetry ``splits'' into a $U(1)$ (resp. $SU(2)$)
left-moving global symmetry times $U(1)$) (resp. $SU(2)$) right-moving R-symmetry. Then the anomaly
of $U(1)$ ($SU(2)$) UV symmetry gives just the difference between the levels of those chiral IR $U(1)$'s ($SU(2)$'s), not just the
level of the IR $U(1)$ ($SU(2)$) right-moving R-symmetry. A similar scenario was also proposed of for effective (0,4) 2d theories obtained by compactification of 6d (2,0) theories on $S^2 \times \text{Riemann surface}$ \cite{Putrov:2015jpa}. 

This phenomenon can be already observed in the case of theory of a single free (1,0) 6d hypermultiplet (cf. Section \ref{sec:KK-reduction}). The formulas above give $c_\text{R}=0$ for any K\"ahler 4-manifold. When $\sigma>0$ this indeed agrees with the fact 2d theory consists of $\sigma/8$ (0,2) Fermi multiplets with R-charge zero. However, when $\sigma<0$ the 2d theory is $-\sigma/8$ (0,2) chiral multiplets with R-charge 1 and the actual $c_\text{R}=-3\sigma/16$. This is explained by observing that $U(1)_R$, the maximal torus of the six-dimensional $SU(2)_R$, acts on both left- and right-moving scalars of the 2d chiral multiplets with the same charge. Its anomaly is zero and not the same as the anomaly of the $U(1)$ R-symmetry acting on just the right-moving scalar.

In Appendix \ref{sec:geography} we discuss the relation between conditions $c_\text{R}>0$, $c_\text{L}>0$ and the geography of complex surfaces.

\subsection{Turning on flavor symmetry background}
\label{sec:background-flavor}

In this section we consider the effect of turning on a non-trivial bundle for a subgroup of the flavor symmetry $G'\subset G$. In particular, as we explain below, in general it modifies the anomaly polynomial of the effective 2d theory. For calculation of anomalies, only the isomorphism class of a $G'$-gauge bundle over $M_4$ matters. Equivalently, the relevant information is given by the homotopy class of the map to the classifying space of $BG'$:
\begin{equation}
	\mu: M_4 \longrightarrow BG'.
\end{equation}
Note that for a $U(1)_i$ subgroup of $G'$ the choice of the homotopy class of the map to $BU(1)=B^2\Z$ is equivalent to the choice of flux $c_1\left(U(1)_i\right)\in H^2(M_4,\Z)$. For any simple Lie group $G_j$ inside $G'$ the choice of the map to $BG_j$ in particular involves the choice of the instanton number in $H^4(M_4,\Z)$. It is given by the pullback of the free generator of $H^4(BG_j,\Z)\cong\Z$. Note, that supersymmetry requires a positivity condition on the instanton numbers (defined with the proper sign), and unless it is satisfied, the supersymmetry protected quantities, such as the partition function, will vanish.

As in the introduction, let
\begin{equation}
	G_{\text{2d}} \; := \; \text{Centralizer}_{G'}(G)
\end{equation}
be the centralizer subgroup of $G'$ in $G$. It has a meaning of the flavor symmetry that remains unbroken in the effective 2d theory after turning on a generic background for the $G'$-bundle on $M_4$. Sometimes, we will denote the effective 2d theory as $T[(M_4,\mu)]$ in order to emphasize the dependence on the topological type of a background flavor symmetry bundle. Thus, the compactification of the 6d theory on 4-manifolds defines the following map,
\begin{equation}
\begin{array}{crcl}
 T: &  \{\text{4-manifolds with $G'$ bundles} \} & \longrightarrow &
\{ \text{2d $(0,1)$ theories with symmetry $G_{\text{2d}}$}\} \\
{} & (M_4,\mu) & \longmapsto & T[(M_4,\mu)]
\end{array}.
\label{flavor-background-map}
\end{equation}

In what follows we determine explicitly the anomaly polynomial of $T[(M_4,\mu)]$. Since $G'$ and $G_{\text{2d}}:=\text{Centralizer}_G(G')$ are two commuting subgroups of $G$, the multiplication map
\begin{equation}
 G_{\text{2d}} \times G' \hookrightarrow G \times G \longrightarrow G
\end{equation}
is a homomorphism and therefore induces a continuous map
\begin{equation}
	\phi:\; BG_{\text{2d}} \times BG'	\rightarrow BG.
\end{equation}

The maps $\mu$, and $\phi$ can be used to construct the following map reducing cohomological grading by 4:
\begin{equation}
	\Phi\,=\, \left(\text{id}_{H^*(BG_{\text{2d}},\Q)}\otimes \left(\int_{M_4} \circ \;\mu^* \right)\right)\circ \phi^*\;:\; H^*(BG,\Q)\;\rightarrow H^{*-4}(BG_{\text{2d}},\Q)
	\label{Phi-operation}
\end{equation}
where
\begin{equation}
	\mu^*:\,H^*(BG',\Q) \longrightarrow  H^*(M_4,\Q)
\end{equation}
and
\begin{equation}
	\int_{M_4}: H^*(M_4,\Q)\rightarrow \Q
\end{equation}
is the pairing with the fundamental class of $M_4$, supported in degree 4. We will also need the following map preserving cohomological grading:
\begin{equation}
	\Psi\,=\, \left(\text{id}_{H^*(BG_{\text{2d}},\Q)}\otimes \epsilon^*\right)\circ \phi^*\;:\; H^*(BG,\Q)\;\rightarrow H^{*}(BG_{\text{2d}},\Q)
\end{equation}
where
\begin{equation}
	\epsilon: \text{pt} \longrightarrow BG'
\end{equation}
with the pullback being the projection on the unit in the cohomology ring,
\begin{equation}
	\epsilon^*:\,H^*(BG',\Q) \longrightarrow  H^*(\text{pt},\Q)\cong \Q.
\end{equation}

The anomaly polynomial of the effective 2d theory on $M_4$ of generic holonomy is then modified to
\begin{equation}
	I_4 = \frac{c_\text{R}-c_\text{L}}{24}p_1(T)+\omega_2^{(2d)}(G_{{\text{2d}}})
\end{equation}
with
\begin{equation}
	c_\text{R}-c_\text{L}=18 \cdot  (\beta -8 \gamma -4\delta )\sigma+12 \,\beta  \chi
+24\,\Phi(\omega_2^{(2)}(G))
\end{equation}
and
\begin{equation}
	\omega^{(2d)}_2(G_{\text{2d}})=-\frac{2\chi+3\sigma}{4}\,\Psi(\omega_2^{(1)}(G))+3\sigma\,\Psi(\omega_2^{(2)}(G))+\Phi(\omega_4^{(1)}(G))\,\in H^4(BG_{\text{2d}},\mathbb{Q})\,.
\end{equation}

In a more formal way, the relation between $I_8$ and $I_4$ can be described as follows. Assuming the coefficients of the anomaly polynomials are rational numbers (the question of the proper quantization condition on the coefficients will be discussed in section~\ref{sec:anomaly-quantization}), the anomaly polynomial of the 6d theory can be understood as a $\Q$-valued bordism invariant of Spin manifolds with $SU(2)_R$ and $G$-bundles,
\begin{equation}
 I_8 \in \text{Hom}\left(\Omega_8^\text{Spin}\left(BSU(2)_R\times BG\right),\Q\right),
\label{anomaly-polynomial-6d-bordism}
\end{equation}
where, as usual, $\Omega_d^{\xi}(X)$ denotes $d$-dimensional bordism group of manifolds equipped with a $\xi$-structure  and a map to $X$. The elements of the group are represented by pairs $(M_d,\alpha)$ where $M_d$ is a $d$-manifold equipped with a $\xi$-structure and $\alpha: M_d\rightarrow X$.
Similarly,
\begin{equation}
 I_4 \in \text{Hom}\left(\Omega_4^\text{Spin}\left(BG_{\text{2d}}\right),\Q\right).
\end{equation}
Then, consider the map
\begin{equation}
 \begin{array}{rrcl}
  \Theta_{(M_4,\mu)}:\; &
\Omega_*^\text{Spin}(BG_{\text{2d}}) &
\rightarrow &
\Omega_{*+4}^\text{Spin}(BSU(2)_R\times BG)
\\
& (M_d,\alpha) &
\mapsto &
(M_d \times M_4, \lambda \times \phi(\alpha \times \mu))
 \end{array}
\label{bordism-map-simple}
\end{equation}
where $\lambda:M_4\rightarrow SU(2)_R$ is the map determined by the topological twisting procedure, that is the projection onto the first component of the Spin structure map $M_4\rightarrow  B\mathrm{Spin}(4) = BSU(2)_+\times BSU(2)_-$. Let $\Theta_{(M_4,\mu)}^*$ denote the induced map
\be
\Theta_{(M_4,\mu)}^*:\; \text{Hom}\left(\Omega_{*+4}^\text{Spin}\left(BSU(2)_R\times BG\right),\Q\right)\ra\text{Hom}\left(\Omega_{*}^\text{Spin}\left(BG_{\text{2d}}\right),\Q\right).
\ee
Then, the relation between the two anomaly polynomials can be concisely written as
\begin{equation}
 I_4=\Theta_{(M_4,\mu)}^*(I_8).
\end{equation}

The anomaly polynomials describe only perturbative 't Hooft anomalies of the respective theories.
The explicit relation between non-perturbative anomalies in 6d and 2d can be obtained in a similar way. We will further elaborate on this in section~\ref{sec:anomaly-quantization}.

One of the nice bonus features of the flavor symmetry background, already mentioned earlier, is that it helps to regularize the partition function \eqref{Z6d} which otherwise might be ill-defined due to bosonic zero-modes. When this happens, the extension of the elliptic genus to the value of the topological Witten genus in $\pi_*\text{TMF}$ may also be ill-defined. There is, however, a simple toy example of the map from 4-manifolds to $\pi_* \text{TMF}$ that avoids this problem and is completely well-defined, even without flavor symmetry backgrounds.

\subsection{A toy model}
\label{sec:toy-model}

There is a simple map
\begin{equation}
	\mathfrak{T}: \{\text{4-manifolds}\} \longrightarrow \pi_* \text{TMF}
\end{equation}
where the right-hand side is actually the ordinary, familiar version of the TMF ({\it i.e.}~non-equivariant and level-1).
 This map is quite simple and depends only on the topology of $M_4$.
It vanishes on many 4-manifolds, but at the same time the image contains some
non-trivial torsion elements and has simple behavior under the connected sum.

Specifically, the map is given by post-composing the map
\begin{equation}
	M_4 \longmapsto \text{ 2d $(0,1)$ lattice CFT with} \;\;\Gamma:= H^2(M_4,Z)/\text{Tor}\, H^2(M_4,Z)
	\label{toy-map-to-TMF}
\end{equation}
with the topological Witten genus map. The $(0,1)$ lattice SCFT above contains $b_2^-$ left-moving real compact bosons, $b_2^+$ right-moving real compact bosons and their super-partners, which are $b_2^+$ right-moving real fermions. The compact bosons are valued in the $H^2(M_4,\mathbb{R})/\Gamma$ torus with chirality determined by the $\pm 1$ eigenvalue of the Hodge star acting on $H^2(M_4,\mathbb{R})$, viewed as the space of harmonic 2-forms. This is the direct $(0,1)$ analogue of the $(0,2)$ lattice CFT considered in \cite{Dedushenko:2017tdw}. Note that, while the 2d theory depends on the conformal class of the metric on $M_4$, theories associated to homeomorphic 4-manifolds can be continuously connected while preserving $(0,1)$ supersymmetry and, therefore, have the same TMF class. The composition with the topological Witten genus gives a topological modular form of degree determined by $b_2^\pm$ of the manifold,
\begin{equation}
	M_4\longmapsto \mathfrak{T}[M_4]\in\pi_d \text{TMF}, \qquad d=3b_2^+ - 2b_2^-.
\end{equation}

Although the map (\ref{toy-map-to-TMF}) does not arise from any physical 6d theory, it exhibits many qualitative features of the full-fledged map in (\ref{map-to-TMF-complicated}). This is because the 2d lattice SCFT can be understood as a subsector of the 2d theory produced by compactification of a single 6d
$(1,0)$ tensor multiplet, which will be discussed in detail in section \ref{sec:KK-reduction}. More precisely, this is the sector that arises from the reduction of the self-dual 2-form field, arguably the most non-trivial ingredient of 6d theories! At the same time, unlike 2d theories $T[M_4]$ that arise from compactification of a full 6d SCFTs, the lattice SCFT above has no non-compact bosonic zero-modes and is always an absolute Spin-theory. The absoluteness follows from the fact that the lattice $H^2(M_4,\Z)$ is self-dual for closed 4-manifolds. So, there is a unique partition function for a given Spin structure on $T^2_\tau$. Another simplification compared to the map $M_4\mapsto T[M_4]$ is that the definition of the map (\ref{toy-map-to-TMF}) requires neither Spin structure on $M_4$, nor smooth structure.

In certain cases the lattice SCFT can be given a nice compact supersymmetric sigma-model description. For example, when $M_4=S^2\times S^2$, it can be described as a $(0,1)$ sigma-model with target $S^1$ with an odd Spin structure, {\it cf.}~\cite{Dedushenko:2017tdw}. The size of $S^1$ is given by the ratio of the sizes of $S^2$'s in $M_4$. The topological Witten genus in this case is given by the value of a mod-2 index of the Dirac operator on the target space. The value in this case is non-trivial and is given by the generator
\begin{equation}
 \eta \in \pi_1 \text{tmf} \simeq \Z_2 \subset \pi_1 \text{TMF} \simeq \pi_1 \text{tmf}\left[\Delta^{-24}\right].
\end{equation}

Another example worth mentioning is $M_4=\mathbb{CP}^2$. The $(0,1)$ lattice SCFT in this case can be described in terms of the following free fields: a compact real chiral (right-moving or, equivalently, holomorphic) boson $\phi$ and a real right-moving fermion $\psi_3$. In terms of these fields, the supercharge can be written as $Q= \bar{\partial} \phi \psi_3$. According to the well-known bosonisation, a theory of a free compact real chiral boson is equivalent to a theory of one free complex right-moving fermion $\psi \equiv \psi_1 + i\psi_2$. Moreover, under the bosonisation map, the fields are related in such a way that $\bar{\partial}\phi = \psi_1\psi_2$. Therefore, we arrive at a theory of three right-moving fermions with the supercharge $Q=\psi_1\psi_2\psi_3$. The value of the topological Witten genus for the latter is believed to be, {\it cf.} \cite{HenriquesST},
\begin{equation}
	\nu \in \pi_3 \text{tmf} \cong \Z_{24},
\end{equation}
where $\nu$ is the generator of the $\Z_{24}$ group.\footnote{We thank E.~Witten for pointing this out to us. }
This fact can be understood as follows. The theory of three free real fermions with the supercharge $Q=\psi_1\psi_2\psi_3$ can be interpreted as the $(0,1)$ supersymmetric $SU(2)$ WZW \cite{DIVECCHIA1985701,Kazama:1988qp} with zero bosonic level (the total level of the affine $SU(2)$ symmetry differs by $+2$ from the bosonic one). In the UV such theory can be described as an $S^3$ sigma-model with a String-structure given by the generator $[S^3]\ \in H^3(S^3,\Z) \cong \Z$ or, in the string terminology, one unit of NS-NS flux (see {\it e.g.}~\cite{Braun:2004qg}). Here we use the fact that on oriented $S^3$ the space of String-structures can be canonically identified with $H^3(S^3,\Z)$.

Note that the map $\mathfrak{T}$, unlike the full-fledged map $T$ given by twisted
compactification of a 6d SCFT, is multiplicative in the $\pi_*\text{TMF}$ ring
under the connected sum operation. This is because connected sum operation corresponds to
stacking 2d lattice SCFTs up to continuous deformation, {\it i.e.},
\begin{equation}
	\mathfrak{T}( M_4 \# M_4' ) = \mathfrak{T}(M_4)\cdot\mathfrak{T}(M_4').
\end{equation}

\begin{table}[t!]
\begin{center}
    \begin{tabular}{| c | c| c|}
    \hline
	$M_4$ & $\mathfrak{T}[M_4]\in \pi_d \text{TMF}$ &  $d=3b_2^+-2b_2^-$
\\ \hline \hline
$S^2\times S^2$ & $\eta$ & 1
\\ \hline
Enriques surface (``$\tfrac{1}{2}$K3'') & ${\eta}\cdot [{E_4}/{\Delta}]$ & $-15$
\\ \hline
 $\mathbb{CP}^2$ & $\nu$ & 3
\\ \hline
K3 & 0 & $-29$
\\ \hline
$\bar{\mathbb{CP}}^2$ & 0 & $-2$
\\ \hline
    \end{tabular}
    \end{center}
\caption{Some examples of values of the map $\mathfrak{T}: \{\text{4-manifolds}\} \rightarrow \pi_*\text{TMF}$. As in the rest of the paper $\eta$ and $\nu$ denote the generators of $\pi_1\text{tmf}\cong \Z_2$ and $\pi_3\text{tmf}\cong \Z_{24}$ respectively, and $[{E_4}/{\Delta}]$ is a generator of $\pi_{-16}\TMF$ that will be explained later. The graded ring $\pi_* \text{tmf}$ is understood as a subring of $\pi_*\text{TMF} := \pi_*\text{tmf}[\Delta^{-24}]$). }
\end{table}
Note that we have the relations $\eta^3=12\nu$, $\eta^4=0$, so in particular taking connected sum with $S^2
\times S^2$ is a nilpotent operation of order-4. The same is true for the $\mathbb{CP}^2\#$ operation, since
$\nu^4=0$.

\subsection{$T[M_4]$ for 6d $(1,0)$ hyper, vector and tensor multiplets}
\label{sec:KK-reduction}

In this section we consider the KK reduction of three basic 6d $(1,0)$ multiplets --- tensor, hyper, and vector --- on a 4-manifold. For the sake of technical simplicity we assume that the homology of $M_4$ has no torsion. Then, compactification of free
6d $(1,0)$ multiplets on $M_4$ with topological twist described earlier gives the following 2d (0,1) content:

\vspace{2ex}
\underline{6d $(1,0)$ tensor multiplet}:

\begin{itemize}
 \item $(0,1)$ $\Gamma= H^2(M_4,\Z)$ lattice CFT described in detail in section~\ref{sec:toy-model}. Note that in case when $H^2(M_4,\Z)$ has torsion, the lattice SCFT will be stacked with a 2d TQFT, which is a $\text{Tor}\, H^2(M_4,Z)$-finite-group gauge theory.

\item $b_1$ $(0,1)$ vector multiplets (equivalent to Fermi multiplets on-shell).

\item $b_0$ $(0,1)$ chiral multiplet ($b_0=1$ for connected $M_4$).

\end{itemize}

As one can see, the presence of $b_0$ non-compact chiral multiplets makes the elliptic genus, and, more generally, topological Witten genus ill-defined. On the other hand, if one reduces the tensor multiplet on $T^2_\tau$, it will produce a 4d $\CN=2$ vector multiplet.

For completeness, let us also write the formula for the resulting gravitational anomaly in 2d, which can be obtained by combining the contributions from the fields above,
\begin{equation}
 \Delta(c_\text{R}-c_\text{L})_{\text{(tensor)}}=\frac{3}{2}b_2^+-b_2^-+\frac{1}{2}b_0-\frac{1}{2}b_1 =
\frac{\chi+5\sigma}{4}.
\end{equation}
Equivalently it can be obtained by integration of the 6d anomaly polynomial
\begin{equation} I_8^\text{(tensor)}=\frac{c_2(R)^2}{24}+\frac{c_2(R)p_1(T)}{48}+\frac{23p_1(T)^2-116p_2(T)}{5760}
	\label{I-tensor}
\end{equation}
over $M_4$, as described in section \ref{sec:anomaly-reduction}.

\vspace{2ex}
\underline{6d $(1,0)$ hyper multiplet}:

\begin{itemize}
\item $\sigma > 0 \Rightarrow \sigma/4$ $(0,1)$ Fermi multiplets,

 \item  $\sigma < 0 \Rightarrow |\sigma|/4$ $(0,1)$ chiral multiplets.
\end{itemize}
The field content depends on the sign of the signature of $M_4$, and when $\sigma=0$, the 2d theory is trivial.

The results above are given by counting harmonic spinors on a 4-manifold. Note that, naively, we get $h^\pm$ copies 2d (0,1) Fermi/chiral multiplet, where $h^\pm$ denote the number of chiral/anti-chiral harmonic spinors. However, since each pair of Fermi and chiral multiplets can be given a mass, only the difference $h^+-h^-$ matters and its value is determined by the index theorem. Note, in this simple example one can see explicitly that, in order to define a map from 4-manifolds to 2d $(0,1)$ theories one needs 4-manifolds to be Spin; otherwise, the signature is not divisible by 4 in general. For smooth Spin manifolds, $\sigma \in 16 \, \Z$ by Rokhlin theorem, and the formulas above make sense.

Again, for $\sigma <0$ and without flavor symmetry backgrounds, the presence of $|\sigma|/4$ non-compact chiral multiplets will make the topological Witten genus ill-defined.

Its gravitational anomaly in 2d is
\begin{equation}
 \Delta(c_\text{R}-c_\text{L})_{\text{(hyper)}}= -
\frac{\sigma}{8}
\end{equation}
and can be equivalently derived by integrating the 6d anomaly polynomial
\begin{equation}
 I_8^\text{(hyper)}=\frac{7p_1(T)^2-4p_2(T)}{5760}
\end{equation}
over $M_4$.

\vspace{2ex}
\underline{6d $(1,0)$ vector multiplet}:

\begin{itemize}
 \item $b_2^-$ $(0,1)$ Fermi multiplets.

\item $b_1$ $(0,1)$ compact chiral multiplets. The compactness follows from the fact that the scalar fields are given by holonomies of the vector field on the 4-manifold. After taking into account large gauge transformations, they are effectively valued in $H^1(M_4,\R)/H^1(M_4,\Z)\cong T^{b_1}$.

\item $b_0$ $(0,1)$ vector multiplets (equivalent to Fermi multiplets on shell).

\end{itemize}

Unlike the cases of tensor and hyper-multiplets, here one finds no non-compact bosonic zero-modes for any $M_4$. If one reverses the order of compactification, the 6d vector multiplet on $T^2_\tau$ gives a 4d $\CN=2$ vector multiplet, same as in the case of 6d tensor multiplet.

The corresponding gravitational anomaly is
\begin{equation}
 \Delta(c_\text{R}-c_\text{L})_{\text{(vector)}}=-\frac{1}{2}b_2^+ + \frac{1}{2}b_1-\frac{1}{2}b_0 =
-\frac{\chi+\sigma}{4}.
\end{equation}
Equivalently, it can be obtained by integrating the 6d anomaly polynomial
\begin{equation}
 I_8^\text{(vector)}=-\frac{c_2(R)^2}{24}-\frac{c_2(R)p_1(T)}{48}-\frac{7p_1(T)^2-4p_2(T)}{5760}
\end{equation}
over $M_4$.

\subsection{A cure for non-compactness: equivariant partition functions}

Consider first non-compact zero modes that originate from 6d hyper multiplets. A nice feature of 6d theories is that they generically have flavor symmetries that act non-trivially on the Higgs branch. Suppose no background flavor symmetry fields are turned on along $M_4$. By computing the partition function on $T^2_\tau \times M_4$ equivariantly with respect to the unbroken flavor symmetry $G_\text{2d}$, one can hope that is will become finite. This will indeed be the case if there are no fixed points under the $G_\text{2d}$-action at the infinite boundary of the Higgs branch.
By the equivariant partition function on $T^2_\tau \times M_4$ we mean the partition function with a non-trivial holonomy of the background $G_\text{2d}$-gauge fields along the time circle of $T^2_\tau$ turned on. In particular, for each $U(1)_i$ subgroup of $G_\text{2d}$ this will give a $U(1)$-valued parameter $x_i$ (which can be naturally analytically continued to a $\C^*$-valued parameter). This will modify the partition function of the 2d theory $T[M_4]$ on the torus from (\ref{EllipticGenus}) to
\begin{equation}
	Z_{T[(M_4,\mu)]}[T^2_\tau](\{x_i\};q) := \Tr_\text{R}(-1)^F q^{L_0}\prod_i x_i^{h_i}
\label{equiv-elliptic-genus}
\end{equation}
where $h_i$ denote the weights of the $U(1)_i$-action on the Hilbert space of 2d theory on a circle. Here and in the rest of the paper, the subscript ``R'' stands for the Ramond sector of the Hilbert space on a circle. When it is well-defined, the right-hand side of (\ref{equiv-elliptic-genus}) gives an element of $\Z[x][[q]]$. Naively, one would expect the result to be a multi-variable weak Jacobi modular form. However, in general this will be spoiled by the fact that the theory $T[M_4]$ should be regarded as a relative theory. We will address this in more detail in section~\ref{sec:relativeness}.

The Hilbert space at each $q$-degree carries a representation of $G_\text{2d}$, and turning on a holonomy $g\in G_\text{2d}$ gives a $q$-series with coefficients in the ring of characters of $G_\text{2d}$. Therefore, the equivariant elliptic genus, when well-defined, can be understood as a map
\begin{equation}
\begin{array}{crcl}
 \text{EG}_{G_\text{2d}}: & \left\{\begin{array}{c} \text{2d $(0,1)$ theories} \\ \text{with $G_\text{2d}$ symmetry}\end{array}\right\} & \longrightarrow & R(G_\text{2d})[[q]] \\
 \\
 & \text{a theory} & \longmapsto  & \Tr_\text{R}(-1)^F g \,q^{L_0} = \sum_{R,m} c_{R,m}\chi_R(g)q^m
\end{array}
\end{equation}
where $R(G)$ is the representation ring of $G_\text{2d}$, $\chi_R$ are characters of irreducible representations of $G_\text{2d}$, and each coefficient $c_{R,m}\in\Z_{\ge0}$ counts the multiplicity of $R$ in the BPS Hilbert space with $q$-degree $m$. Composing it with the map (\ref{flavor-background-map}) we get:
\begin{equation}
\left\{\begin{array}{c}\text{4-manifolds $M_4$}\\ \text{with $G'$-bundle}\end{array}\right\} \stackrel{T}\longrightarrow \left\{\begin{array}{c}\text{2d $(0,1)$ theories $T[M_4]$}\\ \text{with $G_\text{2d}$ symmetry}\end{array} \right\}\xrightarrow{\text{EG}_{G_\text{2d}}}  R(G_\text{2d})[[q]].
\label{equivariant-EG-map}
\end{equation}
We propose that the second map can be refined by replacing the equivariant elliptic genus with an appropriately defined \emph{equivariant topological Witten genus} valued in the ring of $G_\text{2d}$-equivariant topological modular forms. However, in order to make a more precise statement one needs to address the issue of possible relativeness of the 2d theories in the image of the map $T$.

The non-compact zero modes originating from tensor multiplets are more subtle, since the flavor symmetry does not act on them. However, we would like to point out that free tensor multiplets actually are not present in 6d SCFTs, so {\it a priori} it is not obvious if they would contribute or not. One can hope that similarly to the compactification on 6d $(2,0)$ theories they do not actually contribute to supersymmetric configurations, as was argued in  \cite{Vafa:1994tf}.

\subsection{Defect group, relativeness and modularity level}
\label{sec:relativeness}

In this section, we will for a moment ignore the technicalities associated with zero modes and background flavor symmetry bundles
, and instead address a different and, in a sense, completely independent technical complication.
This complication comes from the fact that many 6d $(1,0)$ SCFTs should be understood as \textit{relative} theories, rather than \textit{absolute} ones (see \cite{Freed:2012bs} for a general framework and discussion of such relative theories). A relative theory can be understood as a theory leaving on the boundary of a non-invertible TQFT. The partition function of a $d$-dimensional relative theory on a manifold $M_d$ is not a number, but rather a vector in the Hilbert space of the $d+1$-dimensional TQFT on $M_d$. In the case of a 6d $(1,0)$ theory, the corresponding 7d TQFT is an abelian 3-form Chern--Simons theory with action
\begin{equation}
\int \sum_{ij}\Omega_{ij}C_i d C_j
\end{equation}
where $\Omega$ is the symmetric Dirac pairing matrix on the charge lattice of self-dual strings $\Lambda_\text{string}$. In a way, the relation between 6d theory of self-dual 2-forms to 7d 3-form Chern--Simons theory is analogous to the relation between chiral WZW in 2d and 3d Chern--Simons theory. The essential information about this 7d TQFT is captured by the defect group
\begin{equation}
\mathcal{C}:= \Lambda_\text{string}^*/\Lambda_\text{string} \cong \text{Coker}\,\Omega.
\end{equation}
It was proposed in \cite{DelZotto:2015isa} that, for a 6d SCFT with an F-theory realization, the defect group can be identified with the first cohomology of the three-dimensional link of the singularity in the base in the conformal limit. Note that the defect group comes equipped with a perfect bilinear pairing
\begin{equation}
 \ell k: \mathcal{C} \otimes \mathcal{C} \rightarrow \Q/\Z
\label{defect-pairing}
\end{equation}
which is inherited from the symmetric Dirac pairing ({\it i.e.}~intersection form) on $\Lambda_\text{string}$. It can also be identified with the linking pairing on the first cohomology of the link of the singularity.

On a general 6-manifold $M_6$, $\ell k$ together with the intersection pairing on cohomology defines a non-degenerate antisymmetric form on $H^3(M_6,\CC)$, and the partition function of the 6d SCFT will be labeled by elements of a Lagrangian subgroup of $H^3(M_6,\mathcal{C})$, which, by definition, is a subgroup maximally isotropic with respect to the pairing.
We are interested in the case $M_6=M_4\times T^2_\tau$. Suppose, for simplicity, that $M_4$ is simply-connected and fix a basis in $H^1(T^2_\tau,\Z)$. Then, there is natural choice of a Lagrangian subgroup isomorphic to $H^2(M_4,\mathcal{C})$. And, the partition function of the 6d theory can be defined as a vector labeled by a discrete flux \cite{DelZotto:2015isa,Tachikawa:2013hya,Witten:2009at},
\begin{equation}
 Z_a^\text{4d}[M_4](\tau) := Z^\text{6d}_a[M_4\times T^2_\tau], \qquad a\in H^2(M_4,\mathcal{C}).
\end{equation}
A simple example of this phenomenon is when the 6d SCFT is a $(2,0)$ theory of type $A_1$. Then $\mathcal{C}=\Z_2$ and the the discrete flux $a\in H^2(M_4,\Z_2)$ can be identified with the 't Hooft flux of $SU(2)$ $\CN=4$ 4d SYM on $M_4$ (the second Stiefel-Whitney class $w_2$ of the corresponding  $SU(2)/\Z_2=SO(3)$ principle bundle).

Under the change of basis on $H^1(T^2_\tau,\Z)$ the vector of partition functions transforms as follows (up to an overall extra phase corresponding to gravitational anomaly determined by anomaly polynomials):
\begin{equation}
\begin{array}{rrcl}
 S: & Z_a^\text{4d}[M_4](-1/\tau) & = & \sum\limits_{b\in H^2(M_4,\mathcal{C})}
e^{2\pi i \langle a, b\rangle} Z_a^\text{4d}[M_4](\tau),
\\
\\
T: & Z_a^\text{4d}[M_4](\tau+1) & = & e^{\pi i \langle a, a\rangle} Z_a^\text{4d}[M_4](\tau),
\end{array}
\label{SL2Z-flux-action}
\end{equation}
where
\begin{equation}
 \langle \cdot, \cdot \rangle: H^2(M_4,\mathcal{C}) \otimes H^2(M_4,\mathcal{C}) \rightarrow \Q/\Z
\label{reduced-pairing}
\end{equation}
is defined by composition of intersection form on the second cohomology of $M_4$ with perfect pairing on $\mathcal{C}$ (\ref{defect-pairing}). Note that the diagonal pairing $\langle a , a \rangle$ is well-defined modulo $2\Z$ when $M_4$ is Spin.

Since the partition function of the 6d theory on $M_4\times T^2_\tau$ can be also interpreted as the elliptic genus of the effective 2d theory $T[M_4]$, it follows that the 2d theory is also relative,
\begin{equation}
 Z_a^\text{4d}[M_4](\tau) = Z_{T[M_4],a}[T^2_\tau] := \Tr_{\mathcal{H}_a^R}(-1)^F q^{L_0}.
\end{equation}
The corresponding 3d TQFT is the compactification of 7d 3-form Chern--Simons theory on $M_4$ which is the 3d Abelian Chern--Simons theory with (\ref{reduced-pairing}) being Dirac pairing on the anyons. The relativeness of the 2d $(0,1)$ theory corresponds to the fact the elliptic genus is not a modular form but rather a vector valued modular form transforming under $SL(2,\Z)$ according to (\ref{SL2Z-flux-action}).

Instead of dealing with vector-valued modular forms one can just consider $Z^\text{4d}_0[M_4](\tau)$, the partition function with vanishing discrete flux. For example, when the 6d theory is the $(2,0)$ theory of type $A_1$, this is the partition function of 4d $\CN=4$ SYM on $M_4$ with gauge group $SU(2)$, which is known to be a modular form for the $\Gamma_0(2)$ congruence subgroup of $SL(2,\Z)$. In general, $Z^\text{4d}_0[M_4](\tau)$ is a modular form for $\Gamma_0(N_0)$, where $N_0$ is the smallest positive integer that annihilates $\mathcal{C}$, {\it i.e.}~$N_0\cdot a=0$ for all $a\in\CC$. $N_0$ is also the maximal order of elements in $\CC$ and can be understood more concretely as follows. The defect group is (non-canonically) isomorphic to a product of finite cyclic groups,
\begin{equation}
 \mathcal{C} \cong \prod_i \Z_{p_i}.
\label{defect-group-factorization}
\end{equation}
The perfect pairing (\ref{defect-pairing}) on $\mathcal{C}$ is then zero on a pair of elements from two different cyclic factors, while for elements from the same factor it is given by
\begin{equation}
\begin{array}{rcl}
 \Z_{p_i} \otimes \Z_{p_i} & \longrightarrow & \Q/\Z \\
a \otimes b & \longmapsto & q_iab/p_i \mod 1
\end{array}
\end{equation}
where $q_i$ is coprime with $p_i$. Then,
\begin{equation}
 N_0 = \text{LCM}(\{p_i\}),
\end{equation}
is the least common multiple of all $p_i$'s. It follows that $T^{N_0}$ acts trivially on all components of the vector-valued partition function,
\begin{equation}
\begin{array}{rrcl}
 T^{N_0}: & Z_a^\text{4d}[M_4](\tau) \mapsto Z_a^\text{4d}[M_4](\tau+N_0) & = & e^{\pi i\,N_0 \langle a, a\rangle} Z_a^\text{4d}[M_4](\tau)=Z_a^\text{4d}[M_4](\tau) .
\end{array}
\end{equation}
where we used the fact that the intersection pairing on a spin 4-manifold in even. The zero-flux partition function $Z_0^{(4d)}[M_4](\tau)$ is then invariant under $T$ and $ST^{N_0}S$, the elements generating $\Gamma_0(N_0)\subset SL(2,\Z)$.

To summarize, we get an invariant of $T[M_4]$ under supersymmetry-preserving deformations valued in $\text{MF}(N_0):=\mathrm{ MF}(\Gamma_0(N_0))$, the ring of modular forms of level $N_0$. These are modular forms invariant (up to a factor determined by weight) under the $\Gamma_0(N_0)$ congruence subgroup. Much as for absolute 2d $(0,1)$ theories, where the usual elliptic genus can be refined by the topological elliptic genus valued in $\pi_* \text{TMF}$, for relative $T[M_4]$ we expect to have a topological Witten genus valued in $\pi_* \text{TMF}(N_0)$, where $\text{TMF}(N_0):= \text{TMF}(\Gamma_0(N_0))$ is the spectrum of topological weakly holomorphic modular forms of level $N_0$ (see {\it e.g.}~\cite{mahowald2009topological,hill2016topological}).

Composing it with the map $T$, the compactification of a given 6d SCFT on $M_4$, and ignoring the issues
with non-compactness we get (``naive version''):
\begin{center}
\begin{tikzcd}
\left\{\begin{array}{c} \text{Spin} \\ \text{4-manifolds} \end{array}\right\} \ar[r,"T"] &
\left\{\begin{array}{c} \text{relative} \\ \text{2d $(0,1)$ theories} \end{array}\right\} \ar[r,"\sigma"] \ar[dr,"\text{EG}" ]& \pi_*\text{TMF}(N_0) \ar[d] \\
& & \text{MF}(N_0)\subset \Z[[q]] \\
\end{tikzcd}.
\end{center}
Turning on a non-trivial flavor symmetry background on $M_4$ and replacing the maps by their equivariant version,
we arrive at the refined version of the map (\ref{equivariant-EG-map}):
\begin{center}
\begin{tikzcd}
\left\{\begin{array}{c}  \text{Spin 4-manifolds} \\ \text{with $G'$-bundles} \end{array}\right\}\ar[r,"T"] &
\left\{\begin{array}{c}  \text{relative $(0,1)$ theories} \\
\text{with $G_\text{2d}$ symmetry} \end{array}\right\} \ar[r,"\sigma"] \ar[rd,"\text{EG}_{G_\text{2d}}" ]& \pi_*\text{TMF}_{G_\text{2d}}(N_0) \ar[d] \\
& & \text{MF}_{G_\text{2d}}(N_0)\subset R(G_\text{2d})[[q]] \\
\end{tikzcd}.
\end{center}
This is the diagram announced in the Introduction.

Finally, let us note that theories with the defect group, such that in the decomposition (\ref{defect-group-factorization})
\begin{equation}
 p_i = k_i^2,\; k_i \in \Z_+\quad \text{for all $i$}
\end{equation}
can effectively be made absolute by considering a linear combination
\begin{equation}
 \tilde{Z}^\text{4d}[M_4](\tau) := \sum\limits_{a\in H^2(M_4,\CC')}
Z_a^\text{4d}[M_4](\tau),
\end{equation}
where
\be
\CC'=\bigoplus_{i}\Z_{k_i}
\ee
is a subgroup of $\CC$ on which the induced pairing is trivial. This ensures that $\tilde{Z}$ above is invariant under the full $SL(2,\Z)$. 

More generally, let $p_i=p_i'k_i^2$ where $k_i\in \Z$ and $p_i'$ is an integer with no perfect square factors. Then one can redefine
\begin{equation}
	N_0=\mathrm{LCM}(\{p_i'\}_i)
\end{equation}
and construct a modular form w.r.t. $\Gamma_0(N_0)$.

\subsection{Relativeness and quantization of coefficients of the anomaly polynomials}
\label{sec:anomaly-quantization}

The relative nature of the 6d and 2d theories can also be seen on the level of quantization of coefficients in the anomaly polynomials. Let us first assume that the 6d $(1,0)$ theory is an absolute one. Then the 't Hooft anomalies of such theories should be captured by a 7d invertible TQFT (an SPT in condensed matter language). An invertible TQFT is, in a sense, a classical topological field theory with the action $S^{7d}$, which depends on topology of the space-time manifold $M_7$ and on background gauge fields for the corresponding symmetries. The perturbative anomalies are encoded in a degree-8 anomaly polynomial $I_8$ (\ref{anomaly-6d}), which determines the corresponding part of the action by the descent procedure,
\begin{equation}
 S^{\text{7d}}_{\text{pert}}[M_7] =  \int_{M_8}I_8 ,\qquad \partial M_8=M_7,
 \label{7d-8d-descent}
\end{equation}
which defines a Chern--Simons-like theory in seven dimensions. Given that the action of the invertible TQFT is normalized such that the partition function is $\exp \left(2\pi i\,S^{7d}\right)$, the latter is well-defined by (\ref{7d-8d-descent}) if
\begin{equation}
	I_8[M_8] := \int_{M_8}I_8 \in \Z \;\; \text{for any $M_8$ with}\;\; \partial M_8 =0.
\end{equation}
The manifolds that one can plug in the formulas above are not arbitrary, but should be such that they can be equipped with a certain stable structure --- a lift of the map to $BO$, the classifying space of the stable orthogonal group --- needed to define the boundary 6d theory and the 7d TQFT in the bulk. First of all, one obviously needs orientation and Spin structure, because 6d SCFTs in question do not have time-reversal symmetry and contain fermions. The existence of such structures is equivalent to the conditions $w_1=0$ and $w_2=0$. Then, the Wu formula implies $w_3=0$. Moreover, one can argue that the presence of self-dual 2-form fields also requires $w_4=0$ \cite{Witten:1996md,Sati:2009ic}.  This can be argued as follows. Consider a theory with a single self-dual 2-form field with unit Dirac pairing $\Omega=1$ on $\Lambda_\text{string}\cong \Z$. The corresponding 7d Chern--Simons theory
has a 3-form $C$. From the M-theory realization of such theory one can argue that the quantization
condition for the 4-form flux $G=dC$ in 7d Chern--Simons theory is shifted from the naive one \cite{Witten:1996md,Sati:2009ic}:
\begin{equation}
	G - \frac{p_1(T)}{4} \in H^4(M_7,\Z).
\end{equation}
Note that ${p_1}/{2}$ is a well-defined integer cohomology class for a Spin manifold. Moreover, since
\be
p_1/2\equiv w_4 \pmod 2,
\ee
$p_1/4$  makes sense as an element of integer cohomology when
$w_4=0$. However, in order to actually define $p_1/4$ (given $p_1/2$ is already
defined by a choice of Spin structure) one needs to make some choice, which corresponds to the
ambiguity of lifting $p_1/2$ from the 3rd to 2nd space in the following part of the long exact sequence
\begin{equation}
	\ldots\rightarrow H^3(M_7,\Z_2) \rightarrow H^4(M_7,\Z) \rightarrow  H^4(M_7,\Z) \rightarrow  H^4(M_7,\Z_2) \rightarrow \ldots
\end{equation}
obtained from the short exact sequence of the coefficients
\begin{equation}
	0 \longrightarrow \Z  \stackrel{2\cdot}\longrightarrow \Z  \xrightarrow{\mathrm{mod} 2} \Z_2  \longrightarrow 0.
\end{equation}
This choice is equivalent to a choice of a $O\langle w_1,w_2,w_4\rangle$-structure on $X_7$,\footnote{We use a notation where $O\langle c_1,c_2,\ldots \rangle$-structure requires a trivialization of the characteristic classes $c_1,c_2,\ldots$ of the tangent bundle.} that is the lift of Spin structure with respect to
\begin{equation}
	BO\langle w_1,w_2,w_4\rangle  \longrightarrow B\mathrm{Spin} \stackrel{w_4}\longrightarrow K(\Z_2,4)
\end{equation}
of the classifying spaces. This is sometimes referred to as a Wu-structure. As usual $K(\Z_2,4)\equiv B^4\Z_2$ is the 4-th Eilenberg--Maclane space for $\Z_2$.
As an affine space, the space of such choices is isomorphic to $H^3(M_7,\Z_2)$.
Note that this is the same as the choices of quadratic
refinement of the linking pairing on $\Tor H^3(M_7,\Z)$ ({\it cf.}~\cite{Monnier:2013kna,Monnier:2016jlo}).
However, in the $BO\langle w_1,w_2,w_4 \rangle $ language,
this structure can be understood universally in any dimensions.
In particular, the 6-manifold where 6d theory lives should also
have $O\langle w_1,w_2,w_4\rangle$ structure, which is induced by $O\langle w_1,w_2,w_4\rangle$ structure
of the 7-manifold where the 3-form Chern--Simons theory lives.\footnote{Note that
that $w_4=0$ on any Spin manifold in dimension lower than 8, but not in dimension 8 and higher. This can be seen as follows. There is a general property of Wu classes $v_i\in H^i(M_d,\Z_2)$ on closed $d$-manifolds,
\begin{equation}
	v_i = 0 \text{ for } 2i>d,
\end{equation}
which follows from the Poincar\'e duality and the fact that Steenrod squares $Sq^i$ act
trivially on $H^j(M_d,\Z_2)$ when $i>j$. One also has the relation $v_4=w_4$ when $w_1=w_2=0$.}

The classifying space $BO\langle w_1,w_2,w_4 \rangle$ can be inserted between $B\mathrm{Spin}\simeq BO\langle w_1,w_2\rangle $ and $B\mathrm{String}$ in the
Postnikov tower for $BO$, in the sense that there is a diagram of fibrations,
\begin{center}
\begin{tikzcd}
	B\String \ar[d] \ar[ddd,bend right=80,dashed] & \\
	BO\langle w_1,w_2,w_4 \rangle  \ar[dd] \ar[r,"p_1/4"] & K(\Z,4) \ar[d,"2\cdot"] \\
	& K(\Z,4) \ar[d,"\mathrm{mod}\,2"]\\
	B\Spin   \ar[d] \ar[r,"w_4"] \ar[ur,"p_1/2",dashed] & K(\Z_2,4) \\
	BSO \ar[d] \ar[r,"w_2"] & K(\Z_2,2) \\
		BO \ar[r,"w_1"] & K(\Z_2,1)
\end{tikzcd}
\end{center}

One can explicitly see that the condition $w_4=0$ is indeed required in order to have
\begin{equation}
	I_8^\text{(tensor)}[M_8] \in \Z, \qquad \partial M_8=0
	\label{I-tensor-quantization}
\end{equation}
where $I_8^\text{(tensor)}$ is the anomaly polynomial of a single tensor multiplet (\ref{I-tensor}). Indeed, the gravitational part can be written as
\begin{equation}
	I_8^\text{(tensor)} \; = \; \hat{A}_2 - \frac{L_2}{8} +\ldots
\end{equation}
where
\begin{equation}
	\hat{A}_2=\frac{7p_1^2-4p_2}{5760}
\end{equation}
is the A-roof genus associated with the index of the Dirac operator, and
\begin{equation}
	L_2=\frac{7p_2-p_1^2}{45}
\end{equation}
gives the signature when evaluated on a 8-manifold. On a general Spin 8-manifold (\ref{I-tensor-quantization}) obviously fails ({\it e.g.}~for $M_8=\mathbb{HP}^2$ the signature is 1). But when $w_4=0$ the intersection form of 8-manifold is even
and, therefore, the signature is a multiple of 8.\footnote{This follows from the fact that in general
\begin{equation}
 (a,a) \equiv (a,w_4) \pmod 2,\qquad \forall a \in H^4(M_8,\Z).
\end{equation}
}

Therefore, for absolute theories, $I_8$ belongs to a particular subspace in  (\ref{anomaly-polynomial-6d-bordism}),\footnote{When tensored with $\Q$, bordism groups $\Omega_*^{O\langle w_1,w_2,w_4 \rangle}(\ldots)$ and $\Omega_*^\text{Spin}(\ldots)$ are isomorphic}
\begin{equation}
 I_8 \in  \text{Hom}\left(\Omega_8^{O\langle w_1,w_2,w_4 \rangle}(BSU(2)_R\times BG),\Z\right)
\label{I8-free-bordism}
\end{equation}
which gives the quantization condition on the coefficients in $I_8$. If the 6d theory is absolute, so is the effective 2d theory $T[M_4]$. Its anomaly polynomial $I_4$ therefore should also satisfy a proper quantization constraint:
\begin{equation}
 I_4[X_4]\in \Z, \text{ for Spin }X_4,\text{ with }\partial X_4=0
\label{I4-quantization}
\end{equation}
or, equivalently
\begin{equation}
 I_4 \in  \text{Hom}\left(\Omega_4^\text{Spin}(BG_\text{2d}),\Z\right).
\label{I4-free-bordism}
\end{equation}
In particular, using the explicit formula (\ref{I4-polynomial}) this implies that
\begin{equation}
 2\,(c_\text{R}-c_\text{L})\in \Z
\label{2d-grav-anomaly-quantization}
\end{equation}
because $p_1[X_4]\in 48\,\Z$ on any Spin manifold $X_4$. The anomaly polynomial $I_4$ can be obtained by integrating $I_8$ over $M_4$ as was described in detail in section (\ref{sec:anomaly-reduction}).
The condition (\ref{I4-quantization}) indeed will be automatically satisfied because
\begin{equation}
 I_4[X_4] = I_8[M_4\times X_4] \in \Z
\end{equation}
since $w_4=0$ for any 8-manifold of the form $M_4\times X_4$.

The anomaly polynomials (\ref{I8-free-bordism}) and (\ref{I4-free-bordism}) describe only perturbative anomalies in 6d and 2d respectively. These are anomalies valued in free abelian groups. In general there are also corresponding non-perturbative anomalies valued in torsion abelian groups \cite{Kapustin:2014tfa,Kapustin:2014dxa,Freed:2016rqq},
\begin{equation}
 \mathcal{A}^\text{6d}_\text{tor} \in  \text{Hom}\left(\Tor\Omega_7^{O\langle w_1,w_2,w_4 \rangle}(BSU(2)_R\times BG),U(1)\right)
\label{A6d-bordism}
\end{equation}
and
\begin{equation}
 \mathcal{A}^\text{2d}_\text{tor}  \in  \text{Hom}\left(\Tor\,\Omega_3^\text{Spin}(BG_\text{2d}),U(1)\right).
\label{A2d-bordism}
\end{equation}
They are related by
\begin{equation}
 \mathcal{A}^\text{2d}_\text{tor} = \Theta_{(M_4,\mu)}^*\left(\mathcal{A}^\text{6d}_\text{tor}\right)
\label{bordism-compactification}
\end{equation}
where $\Theta_{(M_4,\mu)}$ is the extension of the map (\ref{bordism-map-simple}) to
\begin{equation}
 \begin{array}{rrcl}
  \Theta_{(M_4,\mu)}: &
\Omega_3^\text{Spin}(BG_\text{2d}) &
\rightarrow &
\Omega_{7}^{O\langle w_1,w_2,w_4 \rangle}(BSU(2)_R\times BG)
\\
& (X_3,\alpha) &
\mapsto &
(X_3 \times M_4, \lambda \times \phi(\alpha \times \mu)).
 \end{array}
\label{bordism-map-tor}
\end{equation}
Here the $O\langle w_1,w_2,w_4 \rangle$-structure on $M_4\times X_3$ is induced by Spin structures on $M_4$ and $X_3$.\footnote{Consider, for simplicity, the case of a simply-connected $M_4$. As was pointed out above, the choice of $O\langle w_1,w_2,w_4 \rangle$ on $M_4\times X_3$ is equivalent to the choice of a quadratic refinement of the linking pairing on $\Tor H^3(M_4\times X_3,\Z)\cong \Tor H^1(X_3,\Z) \otimes H^2(M_4,\Z)$. The Spin structure on $X_3$ fixes a quadratic refinement of the linking pairing on $\Tor H^1(X_3,\Z)$. }

If the 6d theory is relative, then it can not be considered as a boundary theory of an invertible TQFT. The quantization condition in the anomaly polynomial will be in general modified. As we have seen in the section~\ref{sec:relativeness} the relativeness of the theory is effectively measured by an integer $N_0\geq 1$, the maximal order of the elements of the 6d defect group $\mathcal{C}$, with perfect squares removed. In general, the quantization condition on the coefficients of the anomaly polynomials in 6d and 2d will then be modified to
\begin{equation}
 N_0\, I_8[X_8]\in \Z, \text{ for any closed $X_8$ with }w_1=0,\,w_2=0,\,w_4=0
\label{I8-relative-quantization}
\end{equation}
and
\begin{equation}
 N_0\, I_4[X_4]\in \Z, \text{ for any closed $X_4$ with }w_1=0,\,w_2=0.
\label{I4-relative-quantization}
\end{equation}
respectively. In particular, we now have
\begin{equation}
 2N_0\,(c_\text{R}-c_\text{L})\in \Z.
\label{2d-grav-anomaly-relative-quantization}
\end{equation}

By plugging in (\ref{I8-relative-quantization}) the following $O\langle w_1,w_2,w_4\rangle$ 8-manifolds with $SU(2)_R$ bundles:
\begin{itemize}
	\item $T^8\equiv T_{(1)}^2\times T_{(2)}^2\times T_{(3)}^2\times T_{(4)}^2$ with $SU(2)_R$ being associated to the rank 2 complex bundle $V=L\oplus L^{-1}$, where $L$ is a line bundle such that $c_1(L)=[T^2_{(1)}]+[T^2_{(2)}]+[T^2_{(3)}]+[T^2_{(4)}]$.
	\item $S^4 \times K3$ with $SU(2)_R$ bundle given by 1-instanton configuration on
	$S^4$, constant along K3
	\item $K3 \times K3$ with a trivial $SU(2)_R$ bundle.
	\item $Y_8=(\natural^{28} E_8) \cup D^8$, an almost parallelizable\footnote{Being almost parallelizable implies that all but the top Stiefel-Whitney and Pontryagin classes vanish. More details about the construction will be given in section~\ref{sec:TMFfromSigma}.} 8-manifold with signature 224 \cite{kervaire1963groups}, with trivial $SU(2)_R$ bundle.
\end{itemize}
we obtain the explicit conditions
\begin{equation}
\begin{array}{rl}
24\,N_0\, \alpha  & \in \Z\\
48\,N_0\,\beta   & \in \Z\\
2304\,N_0\, (2\gamma+\delta) & \in \Z \\
1440 \,N_0\,\delta & \in \Z
\end{array}
\label{6d-anomaly-coef-quantization}
\end{equation}
on the coefficients of the 6d anomaly polynomial (\ref{anomaly-6d}). The 4 pairs of 8-manifolds with $SU(2)$-bundles above are representatives of $\Omega^{O\langle w_1,w_2,w_4\rangle}_8(BSU(2))$ bordism group and generate it over rational numbers, but are not necessarily generators over integers. So the actual quantization conditions may be stronger. Namely, the actual values of $\alpha,\beta,\gamma,\delta$ for 6d $(1,0)$ theories may always lie in a certain sublattice of the lattice given by (\ref{6d-anomaly-coef-quantization}). We find that in all simple examples a stronger condition, $576\, N_0\,(2\gamma+\delta) \in \Z$, is actually satisfied, instead of the one in the third line in (\ref{6d-anomaly-coef-quantization}). At the same time, there are examples that saturate the other three conditions in (\ref{6d-anomaly-coef-quantization}).

Known examples of absolute 6d SCFTs include E-string theories of arbitrary rank, theories describing $N$ M5-branes probing a $\C^2/\Gamma$ singularity (without decoupling fields associated with center of mass) and minimal conformal matter. We checked that the condition (\ref{I8-relative-quantization}) with $N_0=1$ is indeed satisfied for these theories, and thus (\ref{2d-grav-anomaly-quantization}) and (\ref{I4-quantization}) hold.

On the other hand, the so-called $O(-p)$ theories --- denoted by the base of the corresponding elliptic fibration in F-theory --- with $p=3,4,5,6,7,8,12$ are simple examples of relative 6d (1,0) theories\footnote{We would like to thank Kantaro Ohmori for providing us expressions for anomaly polynomials of $O(-p)$ 6d SCFTs and fruitful discussions on quantization of anomaly coefficients and related topics.}. Their defect group is $\mathcal{C}=\Z_p$. The theory $O(-4)$ is, in a sense, exceptional, because $4=2^2$ and it can be made effectively absolute, as explained in section~\ref{sec:relativeness}. Indeed, one can check that the anomaly polynomial satisfies (\ref{I8-relative-quantization}) with $N_0=1$. For other theories, though, only the weaker quantization condition (\ref{I8-relative-quantization}) with $N_0=p$ generically holds. This is again up to a removal of total squares from $p$. For example, the defect group $\Z_{12}=\Z_{3}\times \Z_{2^2}$ for the $O(-12)$ theory can be effectively turned into $\Z_3$.

After reduction on $M_4$, it may happen that the anomaly polynomial actually satisfies stronger quantization condition than predicted by (\ref{I4-relative-quantization}). This is because products $X_4\times M_4$ generate a quite restricted subgroup in the full 8-dimensional bordism group. Indeed, one can show that, for all $O(-n)$ theories except $O(-5)$ and $O(-7)$, the anomaly polynomial of the effective 2d theory will always satisfy (\ref{I4-quantization}) (and, therefore, (\ref{2d-grav-anomaly-quantization})). For example, in the case of the $O(-12)$ theory, which is expected to have no flavor symmetry, we get
\begin{equation}
 2(c_\text{R}-c_\text{L})=57\sigma+53\chi.
\end{equation}
This is related to the fact that these theories, after reduction on $T^2_\tau$, are self-dual under the full $SL(2,Z)$ duality group.
This indicates that the effective 2d theory $T[M_4]$ in this case can be effectively made absolute, so that the index is a single modular form for the full $SL(2,\Z)$.

On the other hand,
for compactifications of $O(-5)$ and $O(-7)$ theories only (\ref{I4-relative-quantization}) and (\ref{2d-grav-anomaly-relative-quantization}) hold, with $N_0=5,\,7$ and, in fact, these conditions can be saturated.

\subsection{Relation between anomalies in various dimensions}
\label{sec:anomaly-enhancement}

Anomalies play a very important role in our story.
This is already clear from their prominent appearance in the previous subsections,
and will be even more so in section~\ref{sec:TMF2D}.
For this reason, we wish to explain the relation between anomalies of the following theories:
\nobreak
\begin{equation}
\begin{tikzcd}
& \text{6d $(1,0)$ theories} \ar[ldd] \ar[d] \ar[rdd, "\text{on $T^2_{\tau}$}"] & \\
& \begin{array}{c} 5d~\mathcal{N}=1 \\
\text{theories}
\end{array} \ar[ldd, "\text{on $M_4$}"] \ar[rd] & \\
\begin{array}{c} \text{2d $(0,1)$ theories} \\
T[M_4]
\end{array} \ar[d] & &
\begin{array}{c} 4d~\mathcal{N}=2~\text{theories} \\
\text{on $M_4$}
\end{array} \\
\text{1d $\mathcal{N}=1$ SQM} & & 
\end{tikzcd}
\label{dualitycascade}
\end{equation}
where all vertical arrows represent a circle compactification, and the arrow relating 6d and 2d anomalies was already discussed in sections~\ref{sec:anomaly-reduction}, \ref{sec:background-flavor}, and \ref{sec:anomaly-quantization}.

Aside from helping us understand how same anomalies look from different vantage points, \eqref{dualitycascade} will also highlight one key message of this paper: the importance of KK modes and stark contrast with topological twists of ordinary Lagrangian 4d $\mathcal{N}=2$ theories on $M_4$. Recall, that the latter have $U(1)_r$ R-symmetry which plays an important role~\cite{Witten:1988ze} and whose anomaly controls vanishing of the partition function on $M_4$. As we explain below, this is {\it not} the case for the setup of the present paper, even though after reduction on $T^2_\tau$ one finds a 4d theory with $\CN=2$ supersymmetry and the Donaldson--Witten twist on $M_4$, mentioned in section~\ref{sec:twist}.

Before we can see various symmetries and their anomalies related to $U(1)_r$ in other corners of the diagram \eqref{dualitycascade}, let us start with the right corner, {\it i.e.} 6d $(1,0)$ theory reduced on a 2-torus $T^2_\tau$. The anomaly polynomial of the 4d $\CN=2$ theory contains the following terms:
\begin{equation}
I_6 = \frac{1}{12}(n_h-n_v)\,c_1(r)p_1(T)+n_v\,c_1(r)c_2(R)+c_1(r)\,\omega_2^{(4d)}(G)
+\ldots
\label{anomaly-4d}
\end{equation}
where $r$ denotes $U(1)_r$ $r$-symmetry bundle, $n_h$ and $n_v$ are the effective numbers of hyper and vector multiplets, respectively, and $\omega_2^{(4d)}\in H^4(BG,\Q)$.\footnote{Again, when $G=SU(N)$ or $U(N)$ it coincides with the second Chern class of the corresponding rank $N$ complex vector bundle.} In (\ref{anomaly-4d}) we only kept terms linear in $c_1(r)$. Equivalently, instead of $n_h$ and $n_v$ one can use 4d central charges $a$ and $c$,
\begin{equation}
a=\frac{5}{24}n_v+\frac{1}{24}n_h,\qquad c=\frac{1}{6}n_v+\frac{1}{12}n_h.
\end{equation}
Note that, unlike the $SU(2)_R$ symmetry, which is present both in 6d and 4d (before topological twisting), the $U(1)_r$ symmetry appears only in 4d theory in the IR. In particular, the 4d anomaly polynomial above cannot be obtained by simply integrating the 6d anomaly polynomial $I_8$ over the 2-torus. This procedure would obviously give zero, since the tangent bundle of the torus is trivial. Nevertheless, for the so-called ``very-Higgsable'' 6d SCFTs \cite{Ohmori:2015pua}, one can find expression of the coefficients in (\ref{anomaly-4d}) from the coefficients in (\ref{anomaly-6d}). Namely,
\begin{equation}
a=24\alpha-12\beta-18\gamma,\qquad c=64\alpha-12\beta-8\gamma,\qquad
\omega_2^{(4d)}(G) = 96 \,\omega_2^{(2)}(G).
\label{anomaly-4d-very-higgsable}
\end{equation}
The 4d theory on the 4-manifold $M_4$, with vanishing flavor symmetry background, produces an effective 0d theory with the following $r$-charge anomaly,
\begin{equation}
I_2= \Delta r\, c_1(r),
\end{equation}
where
\be
\Delta r = -\frac{n_v}{2}\,\sigma-\frac{n_h+2n_v}{2}\,\chi.
\ee
Such an anomaly can also be referred to as the ``ghost-number anomaly'' if one identifies $r$-charge with the ``ghost-number.''
Using the relations (\ref{anomaly-4d-very-higgsable}), for very Higgsable 6d SCFTs one obtains
\begin{equation}
\Delta r=12 \sigma  (3 \beta +2\delta -16 \gamma )+8 \chi  (3\beta
+4\gamma+7\delta ).
\label{virdim-very-Higgsable}
\end{equation}
In the presence of a non-trivial background for flavor symmetry fields, the $r$-charge anomaly becomes
\begin{equation}
\Delta r = -\frac{n_v}{2}\,\sigma-\frac{n_h+2n_v}{2}\,\chi
+ \Phi\left(\omega_2^{(4d)}(G)\right).
\end{equation}
where the operation $\Phi$ is defined in (\ref{Phi-operation}) in section~\ref{sec:background-flavor}.

In particular, for the three basic 6d $(1,0)$ supermultiplets discussed in section~\ref{sec:KK-reduction}, we find
\bea
\Delta r_\text{(6d tensor)} & = & -1/2(\chi+\sigma) = -b_0+b_1-b_2^+ \\
\Delta r_\text{(6d hyper)} & = & -\sigma/4 \\
\Delta r_\text{(6d vector)} & = & -1/2(\chi+\sigma) = -b_0+b_1-b_2^+
\eea
As in~\cite{Dedushenko:2017tdw}, these values have interpretation in 2d $(0,1)$ theory $T[M_4]$:
$$
\#\text{left-moving fermions} - \#\text{right-moving fermions}
\; = \;
\begin{cases}
\Delta r \,, & \text{6d tensor} \\
-\Delta r \,, & \text{6d hyper} \\
-\Delta r \,, & \text{6d vector}
\end{cases}
$$
which suggest that the total $r$-charge anomaly $\Delta r$ can be related to a 't Hooft anomaly in 2d. Namely, it should be a global symmetry such that the fermions arising from compactification of hyper and vector multiplets are in conjugate representation compared to the fermions arising from compactification of a tensor multiplet. We will further elaborate on this later in the present subsection.

Suppose for a moment that the 4d theory in question has an $\CN=2$ Lagrangian description with a gauge group $H$ and matter in the representation $R_H$. In other words, it has a vector multiplet valued in the Lie algebra $\mathrm{Lie}(H)$ and a matter hypermultiplet valued in $R_H$. Then, the localization procedure tells us that the Feynman path integral can be reduced to the finite-dimensional integral over the moduli space $\CM$ of solutions to certain generalized first-order differential equations on $M_4$. Such equations are non-abelian monopole equations studied in {\it e.g.}~\cite{Labastida:1995zj}, where the focus was on the $SU(N)$ gauge group. In section~\ref{sec:BPS-equations}, we discuss in more detail the moduli spaces of solutions to such equations, as well as their five-dimensional lifts.

After localization, the partition function (\ref{Z6d})-(\ref{EllipticGenus}) naively reduces to the following sum of finite-dimensional integrals
\begin{equation}
Z_\text{6d}[M_4\times T^2_\tau] = Z_\text{4d}[M_4](\tau) =
\sum_{n}q^n \int_{\CM_n} 1
\label{Z4d-localization}
\end{equation}
where $\CM=\sqcup_n \CM_n$ and $\CM_n$ are subspaces of solutions (modulo gauge transformations) with a constraint on the homotopy type of the gauge bundle. The number $n$ is the integral over $M_4$ of a certain degree-4 characteristic class of the $H$-bundle ({\it i.e.}~the pullback of some element $\omega_2(H)\in H^4(BH,\Z)$ via the map $M_4\rightarrow BH$). It has a meaning of the ``total instanton number.'' When $H$ is a product of (special) unitary groups, the class $\omega_2(H)$ is a degree-4 polynomial of the corresponding first and second Chern classes, with $\deg c_k=2k$:
\begin{equation}
q^n=\exp\{2\pi i \tau \int_{M_4}\omega_2(H)\}=\exp\{2\pi i\int_{M_4\times S^1} A_\text{KK} \wedge \omega_2(H)\}
\end{equation}

In topological twists of ordinary (Lagrangian) 4d $\CN=2$ theories, the virtual dimensions of the moduli spaces $\CM_n$, {\it i.e.}~the naive dimensions obtained by subtracting the number of degrees of freedom modulo gauge transformations minus the number of constraints, can be shown to coincide with the $r$-charge anomaly,
\begin{equation}
\mathrm{virdim}\,\CM_n = \Delta r .
\end{equation}
A simple argument for this is that in both cases the result is given by the index of a twisted Dirac operator, the total Dirac operator acting on all the spinor fields of the 4d theory on $M_4$.
Suppose the virtual dimension coincides with the actual dimension of $\CM_n$. Then, from the point of view of localization, the partition function (without any insertions) vanishes (or is ill-defined) because in (\ref{Z4d-localization}) one integrates a degree-zero form over a manifold of non-zero dimension. In the case $\mathrm{dim}\,\CM_n \neq \mathrm{virdim}\, \CM_n$ the situation is more subtle. Similar to other localization scenarios ({\it e.g.}~in Gromov-Witten theory), one can argue that one should actually replace $\CM_n$ in (\ref{Z4d-localization}) by the virtual fundamental class $[\CM_n]^\text{vir}$, which is a homology class of degree $\mathrm{virdim}\, \CM_n$.

Note, that the 4d theory being superconformal means, among other things, the absence of the mixed gauge--$U(1)_r$ anomaly. From the relation between the $r$-charge anomaly and the virtual dimension, this is equivalent to the statement that the virtual dimensions of the moduli spaces $\CM_n$ are independent of the instanton number $n$. This is very different from the case of the familiar Donaldson--Witten theory (topologically twisted pure $\CN=2$ $SU(2)$ SYM) or Seiberg--Witten theory (topologically twisted pure $\CN=2$ $U(1)$ SQED with one flavor), where the virtual dimension depends on the instanton number. In those theories one saturates the anomaly by inserting an appropriate operator in the path integral, where the choice of the operator depends on the instanton number. This corresponds to replacing the integrand in \eqref{Z4d-localization} with a non-trivial characteristic class on the moduli space, and pairing it with the virtual fundamental class of $\CM_n$.

Our situation is quite different in a number of ways. Not only the virtual dimension is independent on $n$, but the vanishing of the $r$-charge anomaly is not required for the partition function \eqref{Z6d} to be well defined. Indeed, the crucial contribution to \eqref{Z6d} comes from the tower of Kaluza-Klein modes, which break the $U(1)_r$ symmetry to a discrete subgroup that we discuss next.
Another crucial difference is that we are computing equivariant integrals over moduli spaces, which in general regularize both bosonic and fermionic zero-modes.

While the $SU(2)_R$ subgroup of the 4d $\CN=2$ R-symmetry $U(1)_r\times SU(2)_R$ is already manifest in 6d, the $U(1)_r$ part is not. In particular there is no direct way to see the corresponding anomaly captured by (\ref{anomaly-4d}) in 6d, or in the effective 2d theory $T[M_4]$. Nevertheless, we would like to argue that a remnant of this anomaly can still be seen in a certain way.

First, let us try to understand if we can see the 4d $U(1)_r$ anomaly in 1d $\CN=1$ supersymmetric Quantum Mechanics (SQM) obtained by compactifying 5d $\CN=1$ theory on $M_4$ or equivalently, by reducing $T[M_4]$ on a circle $S^1$.
In 5d, there is still only $SU(2)$ R-symmetry.
Moreover, we know that there are no anomalies of any continuous symmetries in 5d.
However, if we reduce 5d theory on a circle and take the strict limit of zero size $S^1$,
we get a 4d theory which has emergent $U(1)_r$ that does have 't Hooft anomaly,
and one can still ask whether some part of this $U(1)_r$ and its anomaly is manifest in 5d.

Indeed, in 5d there is a discrete symmetry that enhances to $U(1)_r$ in 4d, and this discrete symmetry can be anomalous.
Let us choose a normalization of $r$-charges, such that 4d fermions have $r = \pm 1$
(the choice of sign depends on their chirality and is different for vector and hyper multiplets),
the scalars in the 4d vector multiplets have $r=+2$ and all other fields in 4d vector and
hyper multiplets have zero $r$-charge.
We claim that in 5d (and in Euclidean signature, to be precise) we can see a $\Z_4$ subgroup of $U(1)_r$.
The generator of this subgroup acts on the complex scalar in the 4d $\CN=2$ vector by
multiplying it by $-1$. In terms of 5d fields, this scalar is
\begin{equation}
 \phi_\C = \phi_\R + i \int_{S^1} A^{(5d)}
\end{equation}
where $\phi_\R$ and $A^{5d}$ are real scalar and vector bosonic fields in 5d $\CN=1$ vector multiplet.
Because of large gauge transformations, the gauge invariant combination is $\exp(\phi_\C)\in \C^*$.
The $U(1)_r$ then emerges as the (double cover of) rotation symmetry of $\C$ in the limit of zero radius,
when $\C^*$ effectively turns into $\C$. (It corresponds to the local rotation around 1 of $\C^*$.)
However, the symmetry
\begin{equation}
 \phi_\C \rightarrow - \phi_\C
\end{equation}
is still present for finite radius.
In particular, it involves the sign change of the 5-th component of the gauge field,
\begin{equation}
 A_5 \rightarrow -A_5.
\end{equation}
Therefore, it is a time-reversal 
symmetry $T$
in the effective 1d QM (if time is the coordinate along the circle $S^1$).
Note that, in principle, there is some choice in how the time-reversal symmetry acts
on fields\footnote{These choices correspond to different ways of putting the
theory on non-orientable manifolds.}, and only the action on the gauge fields is unambiguous.
One also needs to commit to the condition $T^2=1$ or $T^2=(-1)^F$,
where $F$ is the fermion number.
(Again, this is in Euclidean signature, and in Minkowski signature the conditions are reversed).
We are interested in the second choice, so that $T$ acts as $\pm i$ on fermions,
which is precisely the action of the generator of the $\Z_4$ subgroup of $U(1)_r$ in the 4d limit.

Thus, we have identified a $\Z_4$ subgroup of $U(1)_r$ with a $\Z_4$ symmetry
generated by $T$ in 1d SQM. What about anomaly? The anomalies of time-reversal symmetry in 1d
are classified by $\Z$ in the free case, which is reduced to \cite{fidkowski2011topological,Kapustin:2014dxa,Freed:2016rqq}
\begin{equation}
 \text{Hom}\left(\Omega_2^{\text{Pin}^-}(\text{pt}),U(1)\right) \cong \Z_8
\end{equation}
in the
interacting case. This comes from the bordism group of 2-manifolds $\Omega_2^{\text{Pin}^-}(\text{pt})\cong \Z_8$. Massless
1d fermions contribute $\pm 1$ to this anomaly depending on whether $T$
acts as $+i$ or $-i$ on them.

Consider 1d fields that are obtained by KK reduction of 5d vector and hyper multiplets. Equivalently, these fields can be obtained by dimensional reduction of 2d theory described in section~\ref{sec:KK-reduction}. Then one can see that the $U(1)_r$ anomaly is indeed the same as the time-reversal anomaly, up to the fact that the latter is valued modulo 8 in the interacting case. Note that 1d time-reversal anomaly can be understood as the 4-manifold compactification of 5d time-reversal anomaly, which is valued in
\begin{equation}
 \text{Hom}\left(\Omega_6^{\text{Pin}^-}(\text{pt}),U(1)\right) \cong \Z_{16}.
\end{equation}
However, we can only see the $\Z_8$ subgroup of $\Z_{16}$ in supersymmetic theories.
This compactification is done similar to (\ref{bordism-compactification})-(\ref{bordism-map-tor}).

From the 2d/6d point of view, the time-reversal symmetry anomaly in 1d/5d can be understood
as the anomaly of a certain $\Z_2$ global symmetry. Note that there is a
general correspondence between time-reversal anomaly in $d$ dimensions
and $\Z_2$ global symmetry anomaly in $d+1$, which can be explained, for
example, by the Smith isomorphism between (non-trivial factors of)
$\Omega_d^{\text{Pin}^-}(\text{pt})$ and $\Omega_{d+1}^{\text{Spin}}(B\Z_2)$ bordism
groups \cite{Kapustin:2014dxa}. In particular, the classification of anomalies of a $\Z_2$ global
symmetries in 2d is also either $\Z$ or $\Z_8$ in the free or interacting cases,
respectively. And this $\Z_2$ acts on the 2d fermions
either as $(-1)^{F_L}$ or as $(-1)^{F_R}$, where the choice depends on whether
the eigenvalue of $T$ in the 1d system is either $+i$ or $-i$.

\subsection{3-manifolds}
\label{sec:3-manifolds}

Given an invariant of 4-manifolds, the most straightforward way to produce invariants of 3-manifolds is to consider 4-manifolds of the form $M_3\times S^1$. As was pointed out in section~\ref{sec:special-holonomy}, without any flavor symmetry background, this invariant is not very interesting because the 2d supersymmetry is enhanced to $\mathcal{N} = (1,1)$. However, for a general flavor symmetry background, that is a map
\begin{equation}
	M_3\times S^1 \rightarrow BG'
	\label{3man-to-BG}
\end{equation}
the supersymmetry will still be $(0,1)$ and the (topological) Witten genus will be non-trivial. In particular, this will be the case when the map above is not homotopic to a product of maps $M_3\rightarrow BG_1$ and $S^1\rightarrow BG_2$, where $G_{1,2}$ are two commuting subgroups of $G'$. Therefore, for a fixed 6d $(1,0)$ theory with flavor symmetry $G$ one produces invariants of 3-manifolds equipped with the map (\ref{3man-to-BG}) valued in $G_\text{2d}:=\text{Centralizer}_G(G')$-equivariant TMF.

Another way to produce non-trivial invariants is to consider 4-manifolds associated to 3-manifolds via
\begin{equation}
	M_4 = (M_3 \times S^1) \# Z_4
\end{equation}
where $Z_4$ is some fixed ``canonical'' 4-manifold, for example $Z_4=\mathbb{CP}^2$, or $\bar{\mathbb{CP}}^2$, or $S^2\times S^2$. Then, even with a trivial flavor symmetry background the effective 2d theory will have generically $(0,1)$ supersymmetry and the corresponding (topological) Witten genus is expected to be non-trivial. 

In fact, under mild assumptions, the connected sum defines a commutative algebra inside $\TMF_*$ and each $M_3$ gives rise to a module, which should be a rather powerful invariant for $M_3$. The multiplication in this algebra is almost always different from the ring multiplication, as the latter is realized by taking the disjoint union of 4-manifolds, not the connected sum.\footnote{In the toy model discussed in section~\ref{sec:toy-model}, the two actually coincide, but the toy model doesn't come from any 6d theory. Instead, it is obtained by taking a \textit{subsector} of the 6d free tensor multiplet on $M_4$.}

Instead of considering invariants of standalone 3-manifolds, one can also consider invariants of 4-manifolds with 3-manifold boundaries. Compactification of a 6d theory on a 4-manifold with boundary then produces a 3d effective $\CN=1$ theory with a boundary condition which breaks supersymmetry to $(1,0)$ in 2d. Different 4-manifolds with the same boundary correspond to different boundary conditions in the same 3d theory $T[M_3]$. This is analogous to the setup considered in \cite{Gadde:2013sca} where compactification of a 6d $(2,0)$ theory on a 4-manifold with boundary gives an effective 3d $\CN=2$ theory with $(0,2)$ boundary condition. As in \cite{Feigin:2018bkf} and {\it op.~cit.}~one can use this point of view to interpret gluing of 4-manifolds along a common boundary in terms of ``sandwiching'' 3-dimensional $\CN=1$ theories. In particular, it means there must exist many non-trivial 2d $(0,1)$ dualities which correspond to 4d Kirby moves. 

Another interesting question is whether the cutting-and-gluing mentioned above is \textit{funtorial}. In other words, given a 6d $(1,0)$ theory, can one upgrade the map
\be
T:\quad \{\text{4-manifolds}\} \rightarrow \pi_*(\TMF)
\ee 
into a functor
\be
\CT:\quad \mathrm{Cob_4} \rightarrow \pi_*(\TMF)\mathrm{-mod},
\ee
where the left-hand side is the category of smooth 4-dimensional cobordisms and the right-hand side is the category of modules over the ring $\pi_*(\TMF)$? This is a ``4d TQFT over the ring $\pi_*(\TMF)$'' that associates each 3-manifold a $\pi_*(\TMF)$-module, each cobordism a map between modules, and each closed 4-manifolds an elements in $\pi_*(\TMF)$.


\section{2d (0,1) theories and topological modular forms}
\label{sec:TMF2D}

The theory of topological modular forms, usually abbreviated as ``tmf,'' was originally constructed by Hopkins and Miller to be the ``universal elliptic cohomology theory.'' The name for this cohomology theory is motivated by the fact that the cohomology ring of a point is rationally isomorphic to the ring of integral modular forms (for the full modular group $SL (2,\Z)$). The most important facts about tmf relevant for the present paper are gathered in appendix~\ref{app:TMF}. See also \cite{goerss2009topological, lurie2009survey} for an introduction and \cite{douglas2014topological} for a comprehensive overview. 

There are three different versions of closely related generalized cohomology theories, $\mathrm{tmf}^*$, $\mathrm{Tmf}^*$ and $\mathrm{TMF}^*$. Most relevant to us is the last one, $\TMF^*$, which gives a 576-periodic generalized cohomology theory. The original construction of topological modular forms uses homotopy theory, and it has been an important unsolved problem to give a geometric interpretation of the TMF cocycles. Stolz and Teichner proposed that the TMF cohomology classes could be represented by supersymmetric theories in two dimensions \cite{ST1,ST2}. We interpret their proposal as follows. Given a topological space $X$,
\be
\TMF^*(X)\cong \left\{\text{families of 2d $(0,1)$ theories parametrized by $X$}\right\}/\text{deformations}.
\ee
Here all continuous SUSY-preserving deformations are allowed, and as will be made clear later, the right-hand side is also naturally a graded abelian group. In other words, if we denote the space of all 2d $(0,1)$ theories as $\CT$, we expect
\be\label{TMFSpace}
\TMF^{*}(X)\cong[X,\CT]
\ee
where the right-hand side stands for homotopy class of maps from $X$ to $\CT$.\footnote{In this section, by a 2d $(0,1)$ theory we always mean a Euclidean theory on a torus with an odd Spin structure ({\it i.e.}~periodic boundary conditions for fermions on both a- and b-cycles). Although we only discuss absolute theories with well-defined partition functions, much can be generalized to relative theories by replacing $\TMF$ with $\TMF(N_0)$.} The first step toward understand this proposal is to establish an isomorphism when $X$ is a point
\be\label{TMF=Theory}
\TMF^{*}(\mathrm{pt})\cong\pi_0(\CT),
\ee
which is what we will explore in the first half of this section. In the second half, we will discuss the equivariant version of TMF, which we need for applications to 4-manifold invariants.

\subsection{Space of theories as a ring spectrum}

The left-hand side of \eqref{TMF=Theory} can be identified with the homotopy groups of the TMF spectrum, which is commonly denoted as $\pi_*(\TMF)$.\footnote{More precisely, we have $\pi_*(\TMF)=\TMF^{-*}(\mathrm{pt})$.} It forms a ring and has a delicate structure, of which we already had a glimpse in section~\ref{sec:toy-model}. On the right-hand side, $\CT$ is a sequence of spaces $\CT_d$, each defined as the space of 2d $(0,1)$ theories with central charges satisfying $2(c_R-c_L)=d$. At a given degree $d\in\Z$, we have
\be\label{TMF=Theory2}
\pi_d(\TMF)=\pi_0(\CT_d).
\ee

The proposed relations \eqref{TMFSpace} and \eqref{TMF=Theory} have many interesting implications for the theory space $\CT$. First of all, as TMF is an $E_\infty$-ring spectrum, for \eqref{TMFSpace} to hold, the space of theories $\CT$ also needs to be an $E_\infty$-ring spectrum. The ring structure on $\CT$ is given by the following three operations:
\begin{itemize}
\item The sum ``+'': taking the sum $T_1+T_2$ of two QFTs. The new theory will have two disconnected branches; going onto one of them recovers $T_1$ or $T_2$. The moduli space of the new theory is the disjoint union of the moduli spaces of $T_1$ and $T_2$. If $T_1$ and $T_2$ are sigma models onto $M_1$ and $M_2$, the new theory is the sigma model onto $M_1$\rotatebox[origin=c]{180}{$\Pi$}$M_2$. In general, $T_1$ and $T_2$ need not to have the same degree.
\item The multiplication ``$\times$'': taking tensor product $T_1\otimes T_2$ of two QFTs. The new theory is obtained by stacking $T_1$ and $T_2$ together without interaction. When $T_1$ and $T_2$ are sigma models, the new theory is the sigma model onto the Cartesian product $M_1\times M_2$.
\item The reversal ``$-$'': flipping the parity of a QFT. The new theory has the same dynamics as the old theory, but the parity of the vacuum is flipped, from fermionic to bosonic or vice versa. For a sigma model, this operation corresponds to reversing the orientation of the target.
\end{itemize}

The multiplicative unit $\bf 1$ is given by a theory with a single gapped vacuum and no degrees of freedoms. Tensor product of this theory and another theory leaves that other theory unaffected, ${\bf 1}\otimes T\simeq T\otimes {\bf 1} \simeq T$. The additive unit $\bf 0$ is the trivial theory with no vacuum and no degrees of freedom. It can be viewed as the sigma model to the empty manifold. Like the empty manifold, which should be viewed as a manifold of all dimensions simultaneously, the trivial theory should be viewed as a theory living in all possible degrees. Multiplying the trivial theory with another theory gives a trivial theory, ${\bf 0}\otimes T \simeq T\otimes {\bf 0}\simeq{\bf 0}$.

What is more interesting beside the ring structure is that the space $\CT$ of 2d $(0,1)$ theories also has to be an $\Omega$-spectrum. This means that we must have a structure map\footnote{The structure map is in this direction and not the opposite because the TMF spectrum is given by $\TMF_d\simeq \CT_{-d}$. This ensures that $\pi_d(\TMF)=\pi_0(\TMF_{-d})=\pi_0(\CT_d)$. }
\be
s_d: \CT_d \ra \Omega\CT_{d-1}
\ee
for each $d$, which is a weak homotopy equivalence. In other words, the induced map on homotopy groups is an isomorphism
\be
(s_d)_*: \pi_*(\CT_d) \ra \pi_*(\Omega\CT_{d-1})\cong\pi_{*+1}(\CT_{d-1}).
\ee
It would be interesting to understand the above maps from physics. However, quantum field theory indeed predicts a map in the opposite direction that should be the homotopy inverse of $s_d$, 
\be
t_d: \Omega\CT_{d-1} \ra \CT_d 
\ee
given by viewing an $S^1$-family of 2d $(0,1)$ theories of degree $d-1$ as a single theory of degree $d$. 

\subsubsection{A dictionary}

One might consider establishing \eqref{TMF=Theory2} in three steps:
\begin{enumerate}
\item Construct a 2d $(0,1)$ theory for each TMF class.
\item Use invariants to show that theories representing different TMF classes cannot be deformed into each other.
\item Show that theories representing the same TMF class can always be deformed into each other.
\end{enumerate}
We will partially complete the first step, and we hope to further explore the rest in future work.

The spectra $\pi_*(\TMF)$ form a graded ring, which can be viewed as a graded module over $\pi_0(\TMF)=\Z[x]$. Since we expect the relation $\TMF(\mathrm{pt})=\pi_*(\CT)$ to be a ring homomorphism, we need to first identify the generators, and then verify that the ring structures are the same. A nearly complete dictionary for generators in degrees $d\in[-24,24]$ is summarized in Table~\ref{tab:TMF}, 
with the notations and details to be explained in the following subsections.
\begin{table}[ht]
\begin{center}
\resizebox{!}{0.62\textwidth}{%
\begin{tabular}{c|c|c}
$d$ & $\pi_d(\TMF)$ & \text{Generators in terms of $(0,1)$ theory}\\
\hline
0 & $\Z[x]$   &   massive theory, $T_{E_8}\otimes \sigma(Y_8\times Y_8)$ \\
1 & $\Z_2[x]$ &  $\sigma(S^1)$ \\
2 & $\Z_2[x]$ & $\sigma(T^2)$ \\
3 & $\Z_{24}$ & $\sigma(S^3)$ \\
4 & $\Z[x]$   & $T_{E_8}\otimes\sigma(Y_8\times Y_{12})$ \\
5 & 0 &  $-$ \\
6 & $\Z_2$&  $\sigma(S^3\times S^3)$ \\
7 & 0 & $-$ \\
8 & $\Z[x]\oplus \Z_2$ & $\sigma(SU(3))$, $\sigma(Y_8)$
\\
9 &  $\Z_2[x]\oplus \Z_2$  & $\sigma(U(3))$, $\sigma(S^1 \times Y_8)$ 
\\
10 &  $\Z_2[x]\oplus \Z_3$  & $\sigma(Sp(2))$, $\sigma(T^2\times Y_8)$ 
\\
11 & 0 & $-$ \\
12 & $\Z[x]$  & $\sigma(Y_{12})$ \\
13 &$\Z_3$ & $\sigma(S^3\times Sp(2))$\\
14 & $\Z_2$ & $\sigma(G_2)$\\
15 & $\Z_2$ & $\sigma(U(1)\times G_2)$ \\
\vdots & \vdots & \vdots \\
24 & $24\Z + x\Z[x]$ & $\sigma(Y_8^3)$, $\sigma(Y_{24})$ \\
\hline
$-1$, $-2$, $-3$ & 0 & $-$ \\
$-4$ & $\Z[x]$  & $T_{E_8}\otimes \sigma(Y_{12})$ \\
$-5$ & 0 & $-$ \\
$-6$ & $\Z_2[x]$ & $\sigma(T^2)\otimes T_{E_8}\otimes \sigma(Y_{8})$ \\
$-7$ & $\Z_2[x]$ & $\sigma(S^1)\otimes T_{E_8}\otimes \sigma(Y_{8})$ \\
$-8$ & $\Z[x]$ & $T_{E_8}\otimes \sigma(Y_{8})$\\
$-9$, $-10$, $-11$ & 0 & $-$ \\
$-12$ & $\Z[x]$ & $T_{E_8}^{\otimes 2}\otimes\sigma(Y_8\times Y_{12})$ generates the submodule $x\Z[x]$\\
$-13$ & $0$ & $-$ \\
$-14$ & $\Z_2[x]$ & $\sigma(T^2)\otimes T_{E_8}$ \\
$-15$ & $\Z_2[x]$ & $\sigma(S^1)\otimes T_{E_8}$ \\
$-16$ & $\Z[x]$ & $T_{E_8}$ \\
$-17$, $-18$, $-19$ & 0 & $-$ \\
$-20$ & $\Z[x] $ & $T_{E_8}^{\otimes 2}\otimes\sigma(Y_{12})$ \\
\vdots & \vdots & \vdots \\
$-24$ & $24\Z + x\Z[x]$ & $(T_{E_8})^{\otimes 2}\otimes \sigma(Y_{8})$, ``twisted $SO(24)_1$''  \\
\end{tabular}
}
\end{center}\caption{2d $(0,1)$ representatives for important generators of $\pi_*\TMF$ in degrees from $-24$ to $24$. In degree zero, $\pi_0(\TMF)=\Z[x]$ and we list the theories corresponding to 1 and $x$. In other degrees, we only list the generator of $\pi_*(\TMF)$ as a $\Z[x]$-module. In the table, $T_{E_8}$ is the chiral $E_8$ WZW at level 1, $\sigma(M)$ denotes the sigma model with target $M$, while $Y_8$, $Y_{12}$ and $Y_{24}$ are certain 8-, 12- and 24-manifolds. Details will be explained in later subsections.}\label{tab:TMF}
\end{table}

\subsubsection{$E_4$, $2E_6$ and $24\Delta$ from sigma models}\label{sec:TMFfromSigma}

The free part of $\pi_*(\TMF)$ can be represented by modular forms in the image of the map
\be\label{TMFtoMF}
	\pi_*(\TMF)\ra \mathrm{MF}_*.
\ee
Here $\mathrm{MF}_*\cong\Z\left[E_4,E_6,\Delta^{\pm 1}\right]/(E_4^3-E^2_6-1728\Delta)$ denotes the ring of integral weakly holomorphic modular forms, with $E_4,E_6$ and $\Delta$ being the fourth and the sixth Eisentein series and the modular discriminant, placed in degrees 8, 12, and 24, respectively. The map \eqref{TMFtoMF} is zero on torsion elements of $\pi_*(\TMF)$, and non-zero on the free part. The image of $x$ is the $j$-invariant $j:=E_4^3/\Delta$. From degree 0 up to degree 28, the image is generated as a $\Z[x]$-module by 1, $2E_4^2E_6/\Delta$, $E_4$, $2E_6$, $E_4^2$, $2E_4E_6$, $24\Delta$, $E_4^3$, $2E_4^2E_6$. And we will find the 2d $(0,1)$ theories for these generators first.

Let us remind that the usual physical quantity $\mathrm{Tr}_\text{R} (-1)^F q^{L_0}$, with trace taken over the Ramond sector of the Hilbert space on the torus, is not a modular form, but rather a modular function transforming under $SL(2,\Z)$ with a certain multiplier system that depends on the gravitational anomaly $c_\text{R}-c_\text{L}=d/2$. Multiplying it by $\eta^{d}$ makes it into a true modular form 
\begin{equation}
	\eta^{d}\,\mathrm{Tr}_\text{R} (-1)^F q^{L_0}
\end{equation}
of weight $d/2$, which represents a free element of $\pi_{d}(\mathrm{TMF}$).

Many TMF classes in positive degrees can be represented by string manifolds, and (0,1) sigma models onto these manifolds give a (tautological) way to construct the corresponding quantum field theories. Since the topological Witten genus map $\mathrm{MString}_*\ra \mathrm{tmf}_*$ is surjective \cite{hopkins2002algebraic,mahowald293structure}, all TMF classes coming from tmf can be represented this way.

To find the string manifold corresponding to $E_4$, consider the following construction \cite{kervaire1963groups}. The open 8-manifold $X_8$ obtained by $-E_8$-plumbing has an exotic 7-sphere as its boundary. Then, the boundary sum $\natural^{28} X_8$ gives a manifold whose boundary is $S^7$. Capping it off with an 8-ball gives a closed 8-manifold $Y_8$ with signature $28\times 8=224$. $X_8$ is parallelizable by construction. As a consequence, the tangent bundle of $Y_8$ is almost parallelizable with only the top Pontryagin class being non-trivial. This in particular implies that $Y_8$ is string. Then, it follows from Hirzebruch signature theorem that
\be
p_2(Y_8)=\frac{45}{7}\sigma(Y_8)=-1440.
\ee
Hence
\be
\hat{A}(Y_8)=1
\ee
and its Witten genus is given by $E_4$.

Similarly, one can find a manifold that corresponds to $2E_6$. One now takes the open 12-manifold $X_{12}$ obtained by $-E_8$-plumbing. As the group of exotic spheres in dimension 11 is $\Z_{992}$, one can now take 992 copies of $X_{12}$, form the boundary sum, and glue a 12-ball to get a closed 12-manifold. The resulting string manifold $Y_{12}$ has signature $-992\times 8$, $p_3=-120960$ and has $\hat{A}(Y_{12})=2$. The Witten genus of $Y_{12}$ is given by $2E_6$. Due to Rokhlin's theorem, the $\hat{A}$-genus of a 12-manifold is even, so it is impossible to obtain $E_6$ from a sigma model.\footnote{As an aside, there is a generalization of Rokhlin's theorem for general $(0,1)$ theories, which states that when the degree $d\equiv 4 \bmod 8$, the Witten index is even. This is because the BPS Hilbert space in this case is equipped with a quaternionic structure, making the dimension always even. Similar results also hold for $\CN=1$ quantum mechanics in 0+1 dimensions.}

This construction also works for higher dimension, but we don't expect to get anything new, as all these $\TMF$ classes can be expressed in terms of $[E_4]$ and $[2E_6]$. For example, in dimension 16, we can take 8128 copies of $X_{16}$.\footnote{For $X_d$ obtained in this way, $\partial X_d$ is a generator of $bP_{d+1}$, which is a subgroup of $\Theta_d$ represented by exotic sphere that bound parallelizable manifolds. For $d=16$, $bP_{d+1}\cong\Z_{8128}$ while $\Theta_d\cong bP_{d+1}\oplus \Z_2$.} Capping it off again gives $Y_{16}$ with $\hat{A}(Y_{16})=1$. So $Y_{16}$ gives $E_4^2$, which can also be obtained by taking $Y_8\times Y_8$.

There also exist a 24-dimensional manifold $Y_{24}$ whose Witten genus is $24\Delta$. An explicit construction of such a $Y_{24}$ is in \cite[Sec.~9]{mahowald293structure}. Further, there is also a more recent construction using ternary code giving a different field theory representative of this TMF class \cite{Gaiotto:2018ypj}. They also constructed representatives for 12$\Delta^2$, 8$\Delta^3$, $6\Delta^4$ and $24\Delta^5$. As these theories are $\CN=1$ holomorphic CFTs, one can put them on the left-moving sector, giving representatives for 12$\Delta^{-2}$, 8$\Delta^{-3}$, $6\Delta^{-4}$ and $24\Delta^{-5}$ as well.

For the other sigma models we described here, it would also be interesting to find equivalent description at the IR conformal fixed points.

\subsubsection{Left-moving compact bosons}

In negative degrees, TMF classes can no longer be represented by sigma models, and this is where we hope QFT can help to gain insight into the TMF spectrum. We first consider the simplest chiral CFT given by compact bosons.

Left-moving bosons are absolute when they propagate on $\R^n/\Lambda_n$ with $\Lambda_n$ being an even self-dual lattice. The simplest case is when $n=8$ and $\Lambda_8$ is the $E_8$ root lattice. Then, the theory $T_{E_8}$ is a chiral $E_8$ WZW theory with level 1. This gives a degree-$(-16)$ TMF class which corresponds to the modular form $E_4/\Delta$. Tensor product of this theory with a sigma model to $Y_8\times Y_8$ gives the sought after generator $x$ in $\pi_0(\TMF)$.

Using instead the Leech lattice gives a TMF class in degree $-48$ corresponding to $\frac{j-720}{\Delta^2}$. One can consider a $\Z_2$-orbifold of this theory, which leads to $\frac{j-744}{\Delta^2}$. The difference $24\Delta^{-2}$ is twice the generator $12\Delta^{-2}$ in this degree. One cannot hope to do better using lattices, as the coefficient of the $\Delta^k$ term of the theta series of any even self-dual lattice of dimension $24k$ is divisible by 24, {\it cf.} \cite[Thm.~12.1]{borcherds1995automorphic}.

\subsubsection{Affine Lie algebras}

The chiral $E_8$ WZW theory at level 1 has generalizations: one can consider modular invariant modules of more general affine Lie algebras. Here, we will only give an example related to free fermions.

$2r$ free left-moving fermions transform under the vector representation of $\hat{so}(2r)_1$ and provide an example of a theory that is not absolute. However, in some cases, we can add a spinor module to get a modular invariant chiral CFT. We will consider the case of $r=12$. The characters are
\be
\chi_v=\frac{1}{2}\left(\frac{\theta_3^{12}-\theta_4^{12}}{\eta^{12}}\right)
\ee
and
\be
\chi_s=\frac{1}{2}\frac{\theta_2^{12}}{\eta^{12}}.
\ee
Then, the following combination
\be
\chi_s-\chi_v=24
\ee
gives a absolute chiral theory whose elliptic genus is simply 24, without any $q$-dependence. Therefore, this theory defines a class in degree $-24$, which maps into the modular form $24\Delta^{-1}$.

By using other affine Lie algebras, we hope to obtain multiples of $\Delta^{-d}$ in similar ways for other $d$. The most important one is $\Delta^{-24}$ which gives the periodicity of TMF.

\subsubsection{Lie groups, torsion elements and ring structure}

Another class of interesting manifolds consists of Lie groups with the left-invariant framing (\textit{cf.} \cite{HenriquesST}). In fact, such manifolds give all the torsion elements of low positive degrees in Table~\ref{tab:TMF}. Important ones are $U(1)$, $SU(2)$, $SU(3)$, $Sp(2)$, $G_2$ and their products.

Apart from identifying the generators, we also wish to understand the ring structure. One of the most intriguing features is that $x$ has a non-trivial product only with the generators of $\TMF$ that descend to generators of real K-theory. It is not clear why this has to be the case. However, one can see that $x$ indeed multiplies trivially on $\sigma(S^3)$, because $Y_8\times S^3$ is string-cobordant to the empty set. In fact, $\Omega_{11}^{\mathrm{string}}=0$.

\subsubsection{Generalizations to higher dimensions}

As field theories with minimal supersymmetry in dimension one and two are classified by generalized cohomology theories --- KO-theory and TMF respectively --- one may ask what happens in higher dimensions or with enhenced supersymmetry. Since in any dimensions, theories can still be added or multiplied, connected components of the space of theories will always form a ring. However, it is no longer obvious that whether the space of theories still form a spectrum, as it is unclear how to make sense of the ``pushforward operation'' (viewing a family of theories parametrized by a compact topological space as a single theory) with extended supersymmetry or in higher dimensions.

As KO and TMF are closely related to the Thom spectra MSpin and MString, it is tempting to ask what is the role played by MFivebrane, and whether there is a spectrum FB related to 6d physics, whose connective version fb has a ``Fivebrane orientation,'' given by an $E_\infty$ map
\be
{\rm MFivebrane \rightarrow fb}.
\ee
Here MFivebrane, sometimes denoted as M$O\langle9\rangle$, is the Thom spectrum of the 8-connected cover of $BO$ that fits in the Whitehead tower 
\be
\cdots\rightarrow B{\rm Fivebrane} \rightarrow B\String \rightarrow B{\Spin} \rightarrow BSO \rightarrow BO,
\ee
and $B{\rm Fivebrane}$ is obtained by killing all homotopy groups of $BO$ upto $\pi_8(BO)=\Z$. A manifold $M$ admits a Fivebrane structure if the classifying map $M\rightarrow BO$ for the tangent bundle can be lifted to a map $M\rightarrow B{\rm Fivebrane}$. Some evidence that the Fivebrane structure is relevant for 6d theories was found in \cite{Sati:2008kz}.

\subsection{What we do and do not know about ``equivariant TMF''}
\label{sec:equivariant-TMF}

The coefficient ring $\pi_*\text{TMF}$ is where the invariants of 2d $(0,1)$ theories under supersymmetry-preserving deformations are valued. Similarly, we expect that there exists an equivariant version $\pi_*\text{TMF}_{G}$ to be the home for invariants under deformations preserving both supersymmetry and flavor symmetry $G$.\footnote{In this section, we denote $G_\text{2d}=G$ to simplify the notations.} Similar to $\TMF^*$, we expect $\TMF_G^*$ to be a generalized cohomology theory represented by $\CT_G$, the space of 2d $(0,1)$ theories with flavor symmetry $G$, which is expected to be also an $E_\infty$-ring spectrum.
Unfortunately, the equivariant version of TMF has not been fully developed yet in the mathematical literature.  In this section we collect various predictions for equivariant TMF that follow from its proposed physical interpretation. 

\subsubsection{Gradings}
\label{sec:equiv-TMF-grading}
Let us first describe the expected grading structure on $\pi_*\text{TMF}_{G}$. There are well-known invariants of quantum field theories: 't Hooft anomalies for global symmetries (including space-time symmetries). They are valued in a discrete abelian group and additive under stacking ({\it i.e.}~taking tensor products) of theories. Therefore, such an abelian group can play the role of an additional grading. Using the fact that 't Hooft anomalies in 2d are in one-to-one correspondence with equivariant invertible TQFTs in 3d, it follows that for 2d theories with fermions and global symmetry $G$ this group is given by \cite{Kapustin:2014tfa,Kapustin:2014dxa,Freed:2016rqq}:
\begin{equation}
	\CA_\text{2d}^{G}:= \text{Hom}\left(\Omega_4^\text{Spin}(BG),\Z\right)\oplus\text{Hom}\left(\text{Tor}\,\Omega_3^\text{Spin}(BG),U(1)\right)
	\label{anomalies-2d-G}
\end{equation}
where the first term is a free abelian group and the second one is torsion. Elements of the first term are usually referred to as anomaly polynomials in the physics literature.

Let us check that this agrees with the familiar structure in the non-equivariant case. When $G=1$, we have
\begin{equation}
	\CA_\text{2d}^{1}=  \text{Hom}(\Omega_4^\text{Spin}(\text{pt}),\Z) \cong \Z
	\label{anomalies-2d-noG}
\end{equation}
with the generator
\begin{equation}
	\begin{array}{rrcl}
	p_1/48:\quad & \Omega_4^\text{Spin}(\text{pt}) & \longrightarrow & \Z, \\
	& [K3] & \longmapsto & 1.
	\end{array}
\end{equation}
A general element of $\CA$ is then given by
\begin{equation}
	d\cdot\frac{p_1}{48} \in \CA,\qquad d\in \Z.
\end{equation}
The coefficient $d= 2(c_\text{R}-c_\text{L})$, the gravitational anomaly of the 2d theory, is indeed the grading $d$ on $\pi_d\text{TMF}_{G}$.

\subsubsection{Relation to the equivariant $\KO$-theory}
\label{sec:equiv-KO}

In the non-equivariant case, there are maps between spectra (see {\it e.g.}~\cite{ando2010multiplicative,hill2016topological,bunke2009secondary}):
\begin{equation}
	\text{tmf} \longrightarrow \KO[[q]]
\end{equation}
and
\begin{equation}
	\text{TMF} \longrightarrow \KO((q)),
	\label{TMF-KOq}
\end{equation}
where $\KO$ is the spectrum of the real K-theory. The Witten genus map (understood as $q$-series with coefficients valued in the KO-groups of a point) factors through the above maps at the level of stable homotopy groups. Namely,
\begin{equation}
\begin{tikzcd}
 \Omega^{\text{String}}_d(\text{pt}) \ar[r] \ar[rr,bend right=30,"\text{Witten genus}"]&
 \pi_d\text{tmf} \ar[r] &
 \pi_d \KO[[q]] & \cong \Z[[q]]\text{ for }d\equiv 0\pmod 4.
\end{tikzcd}
\end{equation}
Physically, such maps have the following meaning. A compactification of 2d $(0,1)$ theory on a circle produces an infinite tower of supersymmetric $\CN=1$ quantum mechanical systems labeled by spin ({\it i.e.}~eigenvalue of the generator of the $U(1)$ rotation of the circle).  Then, for each quantum mechanics, one can calculate an invariant valued in $\pi_*KO$ \cite{atiyah1964clifford,AlvarezGaume:1983at,Witten:1986bf,stolz2004elliptic,stolz2011supersymmetric} so that the powers of $q$ in (\ref{TMF-KOq}) correspond to the spins. Let us focus on the coefficient in front of the lowest power of $q$, that is $\CN=1$ quantum mechanics obtained by dimensional reduction of 2d $(0,1)$ theory.

The stable homotopy groups $\pi_d \KO$ have periodicity 8 and are listed in the second column of Table \ref{table:equivariant-KO}. The grading $d$ has the following geometric meaning: when the quantum mechanics is an $\CN=1$ sigma model with a $d$-dimensional target manifold, it has invariant valued in $\pi_d \KO$. Mathematically it is realized by Atiyah-Bott-Shapiro orientation map of spectra \cite{atiyah1964clifford},
\begin{equation}
	M\Spin \longrightarrow \KO
\end{equation}
where the target manifold represents an element in $M\Spin$. Note, that in order to define $\CN=1$ quantum mechanics one needs to introduce Spin structure on the target \cite{AlvarezGaume:1983at}. The Hilbert space $\CH$ can then be identified with the space of sections of the spinor bundle and the supercharge $Q$ with the Dirac operator. The fermion number $F$, when non-anomalous ({\it i.e.}~when $d\equiv0\bmod 2$), can be identified with the chirality of the spinors. When $d\equiv0 \bmod 4$ the corresponding invariant of the supersymmetric quantum mechanics is the usual $\Z$-valued Witten index, which can be defined as
\begin{equation}
	\Tr_\CH (-1)^F e^{-\beta H} = \mathrm{dim}\,\mathrm{ker}\, Q|_{F=0} -  \mathrm{dim}\,\mathrm{ker}\, Q|_{F=1}\in \Z
\end{equation}
over the Hilbert space. When $d=1$ one can define the invariant as
\begin{equation}
	(\mathrm{dim}\, \mathrm{ker}\, Q) \bmod 2 \in \Z_2.
\end{equation}
where $Q$ is the supercharge. When $d=2$, one can define the invariant as
\begin{equation}
	\left(\mathrm{dim}\,\mathrm{ker}\, Q|_{F=0}\right) \bmod 2 \in \Z_2.
\end{equation}

As in the case of equivariant TMF (see section \ref{sec:equiv-TMF-grading}) we expect the gradings to be identified with 't Hooft anomalies. We propose that the $\Z_8$ grading of $\pi_*\KO$ can be identified with the anomalies of $\text{Pin}^-$ spacetime symmetry valued in
\begin{equation}
	\CA_{1d}^{1}:=\text{Hom}\left(\text{Tor}\,\Omega_2^{\text{Pin}^-}(\text{pt}),U(1)\right)\cong \Z_8.
	\label{Pin-Z8-anomaly}
\end{equation}
Note, that 1d Euclidean quantum mechanics obtained by dimensional reduction of a 2d Euclidean theory has time-reversal ({\it i.e.}~orientation-reversal) symmetry because time-reversal transformation can be lifted to the space-time rotational symmetry. To be precise, by a reduction from 2d to 1d we mean a compactification on a circle and taking the strict zero radius limit. The unit of the anomaly (\ref{Pin-Z8-anomaly}) can be produced by a free real fermion. The sign of the contribution depends on whether the time-reversal transformation acts as $i$ or $-i$ on it. (Both are consistent with $T^2=(-1)^F$ condition.) If the fermion is understood as the dimensional reduction of a 2d fermion, then the choice of sign corresponds to the choice of chirality, which can be seen from the fact that the time-reversal transformation in 1d is lifted to a rotation by $\pi$ in 2d. Therefore the $\Z_8$-valued anomaly in 2d can be understood as the remnant of the $\Z$-valued gravitational anomaly in 2d. It follows that both definitions of the grading $d$ coincide at least for a sigma model with $T^d$ target.

There is also a way to see it at the level of reduction of the corresponding invertible TQFTs from 3d to 2d. The unit of the gravitational anomaly in 4d corresponds to level-1 Spin-gravitational 3d Chern--Simons theory that can be understood as the boundary term to $p_1(TM)/48$. Its value of the action on a torus $T^3$ with an odd Spin structure along all 3 circles is equal to $1/2\bmod 1$, which is easy to see from cutting a K3 surface in half in a symmetric way along such $T^3$ and using the fact that $p_1[\text{K3}]=48$. The value of the partition function of the invertible TQFT is then $e^{2\pi i (1/2)}=-1$.

On the other hand the effective action of the 2d invertible TQFT generating (\ref{Pin-Z8-anomaly}) is the Arf-Brown-Kervaire invariant \cite{Kapustin:2014dxa}. Compactification of the $T^3$ down to 2d gives $T^2$ with odd Spin structure. The value of the invariant is $4 \bmod 8$, and value of the partition function of the invertible TQFT is then $e^{2\pi i (4/8)}=-1$, the same as the above.

Verifying the agreement of the generator of $\Omega_2^{\text{Pin}^-}(\text{pt})$ is, however, more involved since one has to properly lift a 2d non-orientable Pin$^{-}$-manifold to a 3-dimensional oriented Spin manifold. In the 1d boundary of a 2d TQFT the orientation reversal should be lifted to a rotation by $\pi$ in the 2d boundary of the 3d TQFT. Therefore, it is natural to lift a non-orientable 2-manifold $\Sigma$ to $(\tilde{\Sigma}\times S^1)/\Z_2$ where $\tilde{\Sigma}$ is the orientation double cover of $\Sigma$ and $\Z_2$ acts as the deck transformation on $\tilde{\Sigma}$ and as reflection on $S^1$.

Consider the 3-manifold $\mathbb{RP}^3\#\mathbb{RP}^3$ which is known to be a circle bundle over $\mathbb{RP}^2$, obtained as a quotient $(S^2\times S^1)/\Z_2$ where $\Z_2$ acts as described above (see {\it e.g.}~\cite{hatcher2000notes}). Each $\mathbb{RP}^3$ can be understood as the boundary of the unit disk bundle of $O(2)$ complex line bundle over $\mathbb{CP}^1$. Gluing two copies of such 4-manifolds together along the common boundary produces $\mathbb{CP}^2 \# \mathbb{CP}^2$.
Therefore, $\mathbb{RP}^3\#\mathbb{RP}^3$ can be obtained by cutting $4\#\mathbb{CP}^2$ in half in a symmetric way. Since $p_1[4\#\mathbb{CP}^2]/48=1/4$, the value of the Spin-gravitational Chern--Simons invariant on $\mathbb{RP}^3\#\mathbb{RP}^3$ is then $1/8 \bmod 1$. This is in agreement with the value of the Arf--Brown--Kervaire invariant on $\mathbb{RP}^2$, which is $1\bmod 8$.

One should expect that there is an equivariant version of the map (\ref{TMF-KOq}). Therefore, part of the information about the equivariant TMF can be captured by the equivariant $\KO$-theory. The coefficient ring of the latter has representation theoretic description \cite{atiyah1969equivariant,bruner2010connective,fok2014real}. Similarly to the non-equivariant case \cite{atiyah1964clifford}, The $d$-th group $G$-equivariant KO-group of a point $\KO^{d}_{G}(\text{pt})=\KO^{d,0}_{G}(\text{pt})$ can be obtained as the quotient of the Groethendieck groups of $G$-equivariant modules of the Clifford algebras of dimensions $d$ and $d+1$. 
The result is summarized in the second column in Table \ref{table:equivariant-KO}, where $R(G)$, $RSp(G)$ and $RO(G)$ are representation rings of complex, symplectic and real representations respectively.

\begin{table}[t!]
\begin{center}
    \begin{tabular}{| c | c| c| c |c| c|}
    \hline
	$d$ & $\KO^{d,0}_G(pt)$ &  $G=1$ & $G=\Z_2$ & $G=U(1)$ & $G=SU(2)$
\\ \hline \hline
$0$ & $RO(G)$ & $\Z$ & $\Z\oplus \Z t$ &
$\Z\oplus \Z(t+t^{-1})\oplus\ldots$ &
$\Z\mathbf{1}\oplus 2\Z\mathbf{2} \oplus \Z\mathbf{3} \oplus \ldots$
\\ \hline
$1$ & $RO(G)/R(G)$ & $\Z_2$ & $\Z_2\oplus \Z_2 t$ &
$\Z_2$ & $\Z_2\mathbf{1}\oplus \Z_2\mathbf{3} \oplus \Z_2\mathbf{5}\oplus \ldots$
\\ \hline
$2$ & $R(G)/RSp(G)$ & $\Z_2$ & $\Z_2\oplus \Z_2t$ & $\Z_2\oplus t\Z[t]$ &
$\Z_2\mathbf{1}\oplus \Z_2\mathbf{3} \oplus \Z_2\mathbf{5}\oplus \ldots$
\\ \hline
$3$ & $0$ & $0$ & 0 & 0 & 0
\\ \hline
$4$ & $RSp(G)$ & $\Z$ & $\Z\oplus \Z t$ & $2\Z \oplus \Z(t+t^{-1})\oplus \ldots$ & $2\Z\mathbf{1}\oplus \Z\mathbf{2} \oplus 2\Z\mathbf{3} \oplus \ldots$
\\ \hline
$5$ & $RSp(G)/R(G)$ & $0$ & 0 & 0 &
$\Z_2\mathbf{2}\oplus \Z_2\mathbf{4} \oplus \Z_2\mathbf{6}\oplus \ldots$
\\ \hline
$6$ & $R(G)/RO(G)$ & $0$ & 0 & $t\Z[t]$ &
$\Z_2\mathbf{2}\oplus \Z_2\mathbf{4} \oplus \Z_2\mathbf{6}\oplus \ldots$
\\ \hline
$7$ & $0$ & $0$ & 0 & 0 & 0
\\ \hline
    \end{tabular}
    \end{center}
\caption{Coefficient ring of $G$-equivariant $\KO$-theory and some examples.}
\label{table:equivariant-KO}
\end{table}

Now let us elaborate on the meaning of the additional superscript ``$0$'' in $\KO^{d,0}(BG)$. As was previously mentioned, we expect invariants of supersymmetic quantum mechanics to be valued in a ring graded by the group of 't Hooft anomalies of the protected symmetries (or, equivalently, equivariant invertible TQFTs in 2d). The generalization of (\ref{anomalies-2d-noG}) for non-trivial $G$ is given by\footnote{In principle there is also a free part similar to the one in \ref{anomalies-2d-G}, but in this case it is always trivial, as
$
 \text{Hom}\left(\text{Tor}\,\Omega_3^{\text{Pin}^-}(BG),\Z\right)\cong 0.
$
}
\begin{equation}
	\CA_{1d}^{G}:=\text{Hom}\left(\text{Tor}\,\Omega_2^{\text{Pin}^-}(BG),U(1)\right).
\end{equation}
The forgetful map $\Omega_2^{\text{Pin}^-}(BG)\rightarrow \Omega_2^{\text{Pin}^-}(\text{pt})$ induces the first map in the short exact sequence
\begin{equation}
 \CA_{1d}^{1}\longrightarrow \CA_{1d}^G \longrightarrow \CA_{1d}^G/\CA_{1d}^{1},
 \label{anomalies-1d-1-G}
\end{equation}
which splits in all simple examples. The index $0$ in $\KO^{d,0}(BG)$ then indicates that we restricted our attention to the grading ({\it i.e.}~anomaly, valued in $\CA_{1d}^G$) in the image of the first map in (\ref{anomalies-1d-1-G}). In other words, the 't Hooft anomaly for the symmetry $G$ itself is trivial. This is the situation when the Hilbert space forms an honest representation of $G$, rather than a projective one, and one can indeed define invariants valued in the equivairant $\KO$-groups listed in the Table \ref{table:equivariant-KO}. Namely, for $d=0\,(\text{or 4}) \bmod 8$ it is $\dim\Ker\,Q|_{F=0}-\dim\Ker\,Q|_{F=1}$ considered as an element of the representation ring $RO(G)$ (or $RSp(G)$). For $d=1\,(\text{or 5}) \bmod 8$, it is $\dim\Ker\,Q$ considered as an element of $RO(G)$ (or $RSp(G)$) modulo $R(G)$. And, for $d=2\,(\text{or 6}) \bmod 8$, it is $\dim\Ker\,Q|_{F=0}$ considered as an element of $R(G)$ modulo $RSp(G)$ (or $RO(G)$), {\it cf.}~\cite{furuta2014equivariant}.

For more general grading with non-vanishing class in $\CA_{1d}^G/\CA_{1d}^1$, the Hilbert space in general forms a projective representation of $G$, and one should replace the representation rings $R(G),\,RO(G),\,RSp(G)$ with the appropriate abelian groups of projective representations (altogether they still form a ring with gradings in $\CA_{1d}^G$).

\subsubsection{The free part}

As reviewed in section~\ref{sec:TMF2D}, the ``free part'' of the coefficient ring of topological modular forms $\pi_*\text{TMF}$ can be explicitly described as a subring of the ring of weakly holomorphic modular forms,
\begin{equation}
 \pi_*\text{TMF}/I_{\Tor}
 \subset \text{MF}_* = \Z[E_4,E_6,\Delta^{\pm 1}]/(E_4^3-E_6^2-1728\Delta).
\end{equation}
Here $I_{\Tor}$ is the ideal of torsion elements. The cokernel of the embedding map is explicitly described in \cite{henriques2007homotopy}. In particular, both rings are rationally isomorphic,
\begin{equation}
 \pi_*\text{TMF}\otimes \Q \cong \text{MF}_*\otimes \Q.
\end{equation}

We expect that a similar statement should hold in the equivariant case. Consider first the case when $G$ is a simply-connected simple Lie group or $U(1)$. Then it is natural to propose that
\begin{equation}
 \pi_*\text{TMF}_{G}/\Tor \subset J_{*,*},\qquad
 \pi_*\text{TMF}_G\otimes \Q \cong J_{*,*}\otimes \Q.
\end{equation}
For $G=U(1)$ the right-hand side, $J_{*,*}$, is the ring of meromorphic Jacobi forms \cite{eichler1985theory}. In the case of simple Lie group $G$, it is a natural generalization, the ring of meromorphic Jacobi forms associated to the root system of type $G$ \cite{wirthmuller1992root}. This ring has two integral gradings: the weight (the direct anolog of a weight of a modular form) and the index. This is in agreement with the general proposal for the grading group \eqref{anomalies-2d-G},
\begin{equation}
	\CA_\text{2d}^{G}= \text{Hom}\left(\text{Tor}\,\Omega_3^\text{Spin}(BG),U(1)\right)\oplus \text{Hom}\left(\Omega_4^\text{Spin}(BG),\Z\right)\cong 0\oplus \Z^2
\end{equation}
with the generators being $p_1/48$ and a characteristic class $\omega_2(G)\in H^4(BG,\Z)$ that correspond to weight and index, respectively. (When $G=U(1)$ or $SU(N)$, $\omega_2$ is given by the second Chern character, and we have $\omega_2(G)=c_1^2(G)/2$ or $\omega_2(G)=c_2(G)$, respectively.) Note, that here we assume that weight and index are properly quantized. For example, when $G=U(1)$ one needs to consider half-integer weights and indices according to their usual normalization so that, in particular, the ordinary Jacobi theta function (and its inverse) is included. This is the value of the elliptic genus of the pair of $(0,1)$ Fermi (chiral) multiplets forming an irreducible representation of $G=U(1)$.

Let us consider a slightly more general example for the case $G=U(1)$.
Take free (or, possibly, with a superpotential) $2n_f$ $(0,1)$ Fermi and $2n_c$ $(0,1)$ chiral multiplets. Let us combine them into pairs forming $U(1)$ representations of charges $n_i$ and $m_i$ respectively. The equivariant index, rescaled by $\eta(q)^d$ (see  Appendix~\ref{app:TMF} and Section~\ref{sec:TMF2D} for details) should be an element of the free part of $\text{TMF}^{d,*}_G(\text{pt})$ with $d=2n_c-2n_f$, where $d$ is the weight grading. The rescaled index (in the Ramond sector) is given by
\begin{equation}
G(x;q) := I(x;q)\cdot \eta(q)^d =
\frac{\prod_{i}^{n_f} \theta(x^{n_i};q)}{\prod_{i=1}^{n_c}
\theta(x^{m_i};q)} \cdot (q;q)_\infty^d
\end{equation}
where
\begin{equation}
\theta(x;q) :=(x^{1/2}-x^{-1/2})\prod_{n\geq1} (1-q^nx)(1-q^n/x).
\end{equation}
Now, using the fact that
\begin{equation}
\theta(e^u;q)/(q;q)_\infty^2= u\,\exp\left(-2\sum_{k\geq 1}\frac{u^{2k}}{(2k)!}\, E_{2k}(q)\right)
\end{equation}
we get
\begin{equation}
G(e^u;q) = u^{n_f-n_c}\,\left(1+\sum_{k>=1} u^{2k} P_{4k}(E_2, E_4, E_6,\ldots) \right)
\end{equation}
where $P_{4k}$ is a certain polynomial of degree $4k$ of Eisenstein series
$E_{2m}$ (given $\text{deg}\,E_{2m}=4m$) with rational coefficients. Note that $E_2$ does not appear if and only if
\begin{equation}
\sum_{i=1}^{n_c}
m_i^2 = \sum_{i=1}^{n_f}n_i^2,
\end{equation}
{\it i.e.}~there is no 't Hooft anomaly for $G=U(1)$. Restricting to this case corresponds to taking the subring $\text{TMF}^{d,0}_G(\text{pt})\subset \text{TMF}^{d,*}_G(\text{pt})$. This is analogous to considering the subring $\KO^{d,0}_{BG}(\text{pt})\subset \KO^{d,*}_{BG}(\text{pt})$. As was explained in section~\ref{sec:equiv-KO}, $\KO^{d,0}_{BG}(\text{pt})$ is the ordinary coefficient ring of the equivariant $\KO$-theory. In this case $P_{4k}(\ldots)$ can be interpreted as an element of $\pi_{4k}\text{TMF}\otimes \Q$. Taking into account that $u$ can be interpreted as the generator of $H^*(BU(1),\Z) = \Z[u]$, we have
\begin{equation}
G(e^u;q) \in  \bigoplus_{k} H^{d-k}(BG, \pi_{k}\text{TMF})\otimes \Q.
\label{index-BU1}
\end{equation}
The right-hand side can be understood as the result of applying the Atiyah-Hirzebruch spectral sequence to $\text{TMF}^{d}(BG)$, assuming free elements survive to $E_\infty$,
\begin{equation}
E_2^{p,q} = H^p(BG, \pi_q\text{TMF} ) \Rightarrow \text{TMF}^{p-q}(BG).     \end{equation} 
This suggests that, similarly to the Atiyah--Segal completion \cite{atiyah1969equivariant} that connects $\mathrm{K}_G(\text{pt})$ with $\mathrm{K}(BG)$ (see also \cite{bruner2010connective} for a detailed discussion of such relation for KO-theory), there might exist a close relation between $\text{TMF}^{d,0}_G(\text{pt})$ and $\text{TMF}^{d}(BG)$.

Note that the example above with $G=U(1)$ can be straightforwardly generalized to any semisimple Lie group $G$ by considering its maximal torus $\cong U(1)^{\text{rank}\,G}$ and imposing Weyl invariance.

Consider now the other extreme, when $G$ is a finite group. Then, the proposed grading group is
\begin{multline}
	\CA_\text{2d}^{G}= \text{Hom}\left(\text{Tor}\,\Omega_3^\text{Spin}(BG),U(1)\right)\oplus \text{Hom}\left(\Omega_4^\text{Spin}(BG),\Z\right)\\
	\cong \text{Hom}\left(\Omega_3^\text{Spin}(BG),U(1)\right)\oplus \Z
\end{multline}
where the first term is torsion. The forgetful map $\Omega_3^\text{Spin}(BG)\rightarrow \Omega_3^{SO}(BG)$ induces the injective map
\begin{equation}
 H^3(BG,U(1))\cong \text{Hom}\left(\Omega_3^\text{SO}(BG),U(1)\right)\longrightarrow
 \text{Hom}\left(\Omega_3^\text{Spin}(BG),U(1)\right).
\end{equation}
When the grading is restricted to the image of this map, we propose that $\text{TMF}^{*,*}(BG)\otimes \Q$ can be identified with the ring of meromorphic functions in $q= e^{2\pi i\tau}$ and a commuting pair $h,g\in G$, transforming under $SL(2,\Z)$ as
\begin{equation}
 F_{g,h}(\tau)=\epsilon_{g,h}
 \left(\begin{array}{cc}
        a & b \\
        c & d
       \end{array}
\right)\,
F_{g^dh^{-c},g^{-b}h^a}\left(\frac{a\tau+b}{c\tau+d}\right)
\end{equation}
where $\epsilon_{g,h}$ is a $U(1)$-valued multiplier system that can be explicitly constructed from a given 3-cocycle in $H^3(BG,U(1))$, see {\it e.g.}~\cite{Bantay:1990yr,Coste:2000tq,Gaberdiel:2013nya,Persson:2013xpa}.

\subsubsection{Equivariance and level structure}

Equivariance and level structure can be combined in essentially independent manner.
In the diagram below we list the hierarchy of 2d $(0,1)$ theories and their invariants:
\nobreak
\begin{equation}
\begin{tikzcd}
	\pi_* \mathrm{M}\String =\Omega_*^\text{String}(\text{pt}) \ar[rd]\ar[rrd] &  & &\\
	\left\{\begin{array}{c} \text{$(0,1)$ sigma-models}
	\end{array} \right\}
	\ar[r] \ar[d,hook] \ar[u,"\text{[Target]}"] &
	\pi_*\text{tmf} \ar[r] \ar[d,hook] & \text{mf} \ar[r,hook] \ar[d,hook] & \Z[[q]] \ar[d,equal]
	\\
	\left\{\begin{array}{c} \text{$(0,1)$ theories} \\ \text{w/o non-compact} \\ \text{zero-modes}
	\end{array} \right\}
	\ar[r]  \ar[d,hook] \ar[dd,hook, bend right = 60] &
	\pi_*\text{TMF} \ar[r]  \ar[d,hook] \ar[dd,hook, bend right = 60] & \text{MF}  \ar[d,hook] \ar[dd,hook, bend right = 60] \ar[r,hook] & \Z[[q]]  \ar[d,equal] \ar[dd,hook, bend right = 60]
	\\
	\left\{\begin{array}{c} \text{relative} \\ \text{$(0,1)$ theories} \\ \text{w/o non-compact} \\ \text{zero-modes}
	\end{array} \right\}
	\ar[r]  \ar[dd,hook, bend right = 60] &
	\pi_*\text{TMF}(\Gamma) \ar[r]  \ar[dd,hook, bend right = 60] & \text{MF}(\Gamma)  \ar[dd,hook, bend right = 60] \ar[r,hook] & \Z[[q]]  \ar[dd,hook, bend right = 60]
	\\
	\left\{\begin{array}{c} \text{$(0,1)$ theories} \\ \text{w/ $H$-action}
	\end{array} \right\}
	\ar[r]  \ar[d,hook] &
	\pi_*\text{TMF}_H \ar[r]  \ar[d,hook] & \text{MF}_H  \ar[d,hook] \ar[r,hook] & R(H)[[q]]  \ar[d,equal]
	\\
	\left\{\begin{array}{c} \text{relative} \\ \text{$(0,1)$ theories} \\ \text{w/ $H$-action} 
	\end{array} \right\}
	\ar[r] &
	\pi_*\text{TMF}_H(\Gamma) \ar[r] & \text{MF}_H(\Gamma) \ar[r,hook] & R(H)[[q]]
\end{tikzcd}
\label{TMF-hierarchy}
\end{equation}
where mf (MF) stands for the ring of (weakly holomorphic) modular forms, $\Gamma\subset SL(2,\Z)$, $R(H)$ is representation ring of $H$. In the table above, 2d $(0,1)$ sigma models are with compact target space and have no left-moving fermions, and $H$-actions are assumed to have compact fixed loci.

\section{Examples}
\label{sec:examples}
In this section, we give some examples illustrating various aspects of the general program outline in previous sections.

\subsection{ $N$ M5-branes probing $\C^2/\Z_k$ singularity }
\label{sec:Nk-theories}

While more general 6d $(1,0)$ SCFTs can be constructed via F-theory, we will first focus on a two-parameter family that can be realized as the world-volume theory of $N$ M5-branes probing a $\R\times \C^2/\Z_k$ singularity (here $\Z_k$ acts as rotations by opposite phases on $\C^2$). The topologically twisted compactification of the 6d theory on a spin 4-manifold $M_4$ then can be realized geometrically in M-theory as follows
\be
\begin{matrix}
{\mbox {\textrm{$N$ fivebranes:}}}~~~\qquad & \R^2 & \times & M_4& \\
&   \cap &  & \cap \\
{\mbox{\rm space-time:}}~~~\qquad & \R^3&  \times  & X_8 & .  \\ \qquad \end{matrix}
\label{BraneGeom1}
\ee
$X_8$ here is a local Spin(7)-holonomy space given by the total space of a fibration
\be
\begin{matrix}
\C^2/\Z_k & \ra & X_8  \\
&      & \da \\
& &  M_4  &  \\ \qquad \end{matrix}
\ee
obtained by identifying the $SU(2)_+$ factor in Spin(4)$_{M_4}$ $=SU(2)_+\times SU(2)_-$ holonomy group of $M_4$ with the $SU(2)$ isometry group of $\C^2/\Z_k$. The 4-manifold $M_4$ then is Cayley cycle in $X_8$. 

The geometry involving a spin 3-manifold $M_3$ is similar --- $M_3$ will now be embedded in a 7-dimensional $G_2$-manifold as an associative cycle, with $\Z_k$ singularity fibered along it by identifying $Spin(3)_{M_3}=SU(2)$ holonomy of $M_3$ with $SU(2)$ isometry of $\C^2/\Z_k$:
\be
\begin{matrix}
\C^2/\Z_k & \ra & X_7 \\
&      & \da \\
& &  M_3 &  \\ \qquad \end{matrix}.
\ee
For $k=1$ such setup which provides a correspondence between 3-manifolds and 3d $\CN=2$ theories was considered in \cite{Eckhard:2018raj}.

 To set up the convention, we now describe more explicitly the twisting procedure. We use $\phi_1,\ldots,\phi_5$ to parametrize the $\C^2\times \R_5$ transverse space and to denote the corresponding five scalars coming on the M5-brane world-volume. The $\R_5$ direction is not used for topological twist, while the isometry group of $\C^2=\R^4$ is $SO(4)$ with the double cover being $\mathrm{Spin}(4)=SU(2)_+\times SU(2)_-$. The two copies of $SU(2)$ act on
\be
\left(\begin{matrix}
\phi_1+i\phi_2 & i\phi_3+\phi_4\\
i\phi_3-\phi_4 & \phi_1-i \phi_2
\end{matrix}\right)
\ee
by multiplication of the left and right. And $\Z_k$ is generated by the left-multiplication of
\be
\left(\begin{matrix}
e^{2\pi i/k} & 0\\
0 & e^{-2\pi i/k}
\end{matrix}\right).
\ee
For generic values of $k$, the commutant of $\Z_k$ in $\mathrm{Spin}(4)$ is $SU(2)_+$,\footnote{Notice that for $k>1$, the commutant of $\Z_k$ in $\mathrm{Spin}(5)$ lives inside the $\mathrm{Spin}(4)$ subgroup, this ensures that the $\R_5$ direction is not needed for the topological twist.} which is identified with the R-symmetry group of the 6d $(1,0)$ theory. After the topological twist, $\phi_{1,\ldots,4}$ will transform under a complex two dimensional (real 4 dimensional) representation of $SU(2)_+$ factor in Spin(4)$_{M_4}$ (or $SU(2)=\mathrm{Spin}(3)_{M_3}$), while $\phi_5$ will remain a singlet. 

Notice that even for $k=1$ this M-theory setup is different (and generically preserves half as much supersymmetry) from the one usually used to describe 6d (2,0) theories topologically twisted on 4- and 3-manifolds, where 5-branes wrap coassociative cycle in a $G_2$-manifold and a Lagrangian cycle in a Calabi-Yau 3-fold respectively. 

\subsubsection{$(N,k)=(2,1)$ theory on $M_4=\Sigma_1\times \Sigma_2$}
When $k=1$ the supersymmetry is actually $(2,0)$. However, effectively one can consider the $(2,0)$ theory as $(1,0)$ with flavor symmetry $G=SU(2)_f$. This follows from decomposition of the $(2,0)$ R-symmetry group as $\Spin(5)_R \supset SU(2)_R\times SU(2)_f$ where $SU(2)_R$ is $(1,0)$ R-symmetry.

In other words, we consider a 6d $(2,0)$ theory on $M_4 \times T^2_\tau$, with Donaldson--Witten twist that uses only $SU(2)_R$ subgroup, {\it i.e.}~the one that can be generalized to any 6d $(1,0)$ theory. Also, when this twist is applied to $(2,0)$ the resulting effective 2d theory $T[M_4]$
has the same amount of SUSY as for compactifications of $(1,0)$, so, in
a sense, for this twist there is no qualitative difference between
6d $(2,0)$ and $(1,0)$ theories.

Using the general formula for the anomaly polynomial of $(N,k)$
theories, one finds the central charges with trivial $SU(2)_f$ flavor symmetry background on $M_4=\Sigma_1 \times
\Sigma_2$ to be
\begin{equation}
	\begin{array}{c}
	c_\text{L} = 41(g_1-1)(g_2-1), \\
	c_\text{R} = 42(g_1-1)(g_2-1).
	\end{array}
\end{equation}

The case of $M_4=\Sigma_1 \times
\Sigma_2$ can be studied by first compactifying it on $\Sigma_1$ and then computing $\Sigma_2 \times
T^2$ index of the effective theory. The
effective 4d $\CN=1$ is a particular member of the family of 4d $\CN=1$ theories
obtained by wrapping M5-branes on the zero
section of $L_p \times L_q$ bundle over $\Sigma_1$ such that $c_1(L_n)=n$ and
$p+q=2-2g_1$. This has been considered in the literature
and in certain
cases there is a Lagrangian description. The case when $p=q=1-g_1$ corresponds to the case when the $SU(2)_f$ background on $\Sigma_1$ is trivial. If it also trivial along $\Sigma_2$ it means that $G'=1$, $G_\text{2d}=G=SU(2)_f$ in term of the general setup. When $p\neq q$, this corresponds to turning on non-trivial flux of $G'=U(1)_f \subset G=SU(2)_f$ along $\Sigma_1$. If the flavor symmetry background along $\Sigma_2$ remains inside the same subgroup, then the unbroken 2d flavor symmetry group is $G_\text{2d}=G'=U(1)_f$.

The case of two M5-branes gives the $A_1$ $(2,0)$ theory after decoupling the center of mass motion.
As was mentioned before, one can first compactify the 6d $(2,0)$ theory
on $\Sigma_1$ to get an effective 4d $\CN=1$ theory described, for example, in \cite{Bah:2012dg}. Compared to the usual class
$\CS$ theories, here one has trinions colored by $\pm$. The trinions of the
same type are glued by $\CN=2$ vector multiplet and trinions of different
types are glued by $\CN=1$ vector multiplets. In general the resulting
theory has $U(1)_r$ and $U(1)_f$ non-anomalous flavor symmetry and there is
a way to identify the correct IR R-symmetry ({\it i.e.}~how it mixes with
$U(1)_f$). For the twist corresponding to the trivial flux along $\Sigma_1$ there are equal number of $\pm$
trinions in the generalized quiver. In this case $U(1)_f$ enhances to
$SU(2)_f$. The case of having different numbers of $\pm$
trinions corresponds to having a non-trivial $U(1)_f$ flux along $\Sigma_1$ given by the difference of these numbers.

Denote the flux of  $G'=U(1)_f\subset G=SU(2)_f$ along $\Sigma_i$ as $n_i\in \Z$. Let us first pick
\begin{equation}
 n_1= (g_1-1), \qquad
n_2= -(g_2-1)
\label{fluxes-KW}
\end{equation}

For the manifold of this particular type
($M_4=\Sigma_1 \times \Sigma_2$) the resulting twist (for R- and flavor
symmetries together) can be also interpreted as Kapustin-Witten (aka
"Langlands") twist of 4d $N=4$ SYM, i.e. the one where we twist the full
$SO(4)$ on $M_4$.

The result for the partition function is the following (Note that the symmetry $g_1\leftrightarrow g_2$ is a
non-trivial self-consistency test since the calculation treats
$\Sigma_1$ and $\Sigma_2$ in a very asymmetric fashion):
\begin{equation}
Z^\text{6d}[\Sigma_1\times \Sigma_2 \times T^2_\tau] = A(q,v)^{3(g_1-1)(g_2-1)}
\;\;\in R(G_\text{2d})[[q]]
\end{equation}
where
\begin{multline}
A(q,v) := (v^2-1/v^2)\cdot
( 1 - 8q + (26-1/v^4-v^4)q^2 + 8(-6+1/v^4+v^4)q^3 +
\\
(78-27/v^4-27 v^4)q^4 + 8(-20+7/v^4+7v^4)q^5 + ... )
\end{multline}
and $v$ is the $U(1)_f=G_\text{2d}$ fugacity. Note
that with the choice of fluxes (\ref{fluxes-KW}) 2d theory actually has (2,2) symmetry, but
turning on $v$ breaks it to (0,2)).
\begin{equation}
\left.
\frac{A(q,v)}{(v^2-1/v^2)}
\right|_{v\rightarrow 1} = \eta(q)^8/\eta(q^2)^4
\end{equation}

Consider now a different background:
\begin{equation}
n_i=+(g_i-1),\quad \text{for $i=1,2$.}
\end{equation}
The unbroken 2d flavor symmetry is again $G_\text{2d}=G'=U(1)_f\subset SU(2)_f=G$. The result for the partition function reads
\begin{equation}
Z^\text{6d}[\Sigma_1\times \Sigma_2 \times T^2_\tau] = B(q,v)^{(g_1-1)(g_2-1)}
\;\;\in R(G_\text{2d})[[q]]
\end{equation}
where
\begin{multline}
B(q,v) = \frac{(v-1/v)^3}{(v+1/v)}\cdot
(1 - 12 (v^2+2+1/v^2) q + \\
(414+75/v^4+284/v^2+284 v^2+75 v^4) q^2 + O(q^3) ).
\end{multline}
Note that in this case the overall factor
\begin{equation}
 \frac{(v-1/v)^3}{v+1/v}
\end{equation}
has infinite series in $v$, unlike for $n_1=(g_1-1), n_2=-(g_2-1)$ case.

Finally consider the case $g_1=2$ with zero flux. One possible way to
realize the corresponding 4d $\CN=1$ theory is by the following content:
\begin{itemize}
	\item
$\CN=1$ $SU(2)_{i=1,2,3}$ vector multiplets
\item
Chirals in $(2,2,2,2,+1/2)$ representation of
$SU(2)_+ \times SU(2)_- \times SU(2)_3 \times SU(2)_f \times U(1)_r$
\end{itemize}
Here the normalization of R-charge is such that
supercharges have charge 1. The fact that it is half-integer for chirals gives some
technical difficulties, so that later we have to consider condition
$(g_2-1) \in 2\Z$.
Note that naively $SU(2)_f$ is part of $U(2)_f =(U(1)_t \times SU(2)_f)/\Z_2$
global symmetry, but its diagonal $U(1)_t$ is anomalous (there are
$c_1(U(1)_t)c_2(SU(2)_i)$ terms in the anomaly polynomial).
If one tries to calculate $\Sigma_2 \times T^2_\tau$ partition function of this theory by the same method, the following problem arises: the
corresponding Bethe equations are degenerate, i.e. have infinite number of solutions.

To circumvent this problem one can choose some deformation. Consider the following way to lift the degeneracy:
formally turn on the fugacity corresponding to the anomalous $U(1)_t$.
Even though $U(1)_t$ is anomalous symmetry in 2d, it
is a valid symmetry of effective quantum mechanics obtained by
compactification on $S^1$, even if we keep all KK modes. In particular,
it is a valid $U(1)_t$ global symmetry of 3d $\CN=2$ theory obtained by
putting the above 4d $\CN=1$ on $S^1$. Mathematically, this
means that there is a corresponding grading on the vector space, but
not on the chiral algebra (i.e. no corresponding 2d $U(1)_t$ currents). For $g_1=2$, $g_2=3$ the partition function with this deformation has the following $q$-expansion:
\begin{multline}
	Z^\text{6d}[\Sigma_1 \times \Sigma_2 \times T^2_\tau]/2^8 = \\
	[729\cdot\mathbf{1} + t^2 (3898 \cdot\mathbf{1} + 3990 \cdot\mathbf{3} )
	+t^4(11978 \cdot\mathbf{1} + 24495 \cdot\mathbf{3} +8713 \cdot\mathbf{5}) + ...]
	\\
	+q [-(13832 \cdot\mathbf{1} +11889 \cdot\mathbf{3})
	+ t^2(4287704 \cdot\mathbf{1} + 6395508 \cdot\mathbf{3} + 2108956 \cdot\mathbf{5}) + ...]
	\\
	+ q^2 [ t^{-2}(25250 \cdot\mathbf{1} + 25230 \cdot\mathbf{3}) + ...]
	\\
	+... \in R(G_\text{2d}\times U(1)_t)[[q]]
\end{multline}
where $\mathbf{d}$ is the character of $G=G_\text{2d}=SU(2)_f$ representation of dimension $d$.

\subsubsection{ $(N,k)=(2,2)$ theory on $M_4=\Sigma_1\times \Sigma_2$ }

As in the case of $(N,k)=(2,1)$ theory, to calculate the partition function of 6d $(N,k)=(2,2)$ theory on $\Sigma_1\times \Sigma_2\times T^2_\tau$ one can first reduce the theory on $\Sigma_1$ and then calculate $\Sigma_2\times T^2_\tau$ topologically twisted index \cite{Benini:2016hjo} of the effective  4d $\CN=1$ theory. The effective 4d theory is not Lagrangian per se, but can be constructed from pieces that have Lagrangian description \cite{Razamat:2016dpl}. In particular, such construction involves taking certain couplings to infinity and gauging on a global symmetries which are not present in UV but appear in the IR. Still this description is sufficient for calculation of the index using localization. In other words, one needs to generalize the calculation of $S^3\times S^1$ superconformal index done in \cite{Razamat:2016dpl} to the case of $\Sigma_2\times T^2_\tau$ index. The latter case is technically more involved. In particular it requires solving a system of rather complicated algebraic (at finite order in $q$) Bethe ansatz equations. Again, one has to require $(g_2-1)\in 2\Z$ in order to apply localization.

For closed $\Sigma_{1,2}$ with no flavor symmetry fluxes the reduction of the anomaly polynomial gives the following formula for the central charges:
\begin{equation}
	\begin{array}{c}
	c_\text{L} = 134(g_1-1)(g_2-1), \\
	c_\text{R} = 132(g_1-1)(g_2-1).
	\end{array}
\end{equation}

In general we would like to turn on some fluxes along $\Sigma_1 \times \Sigma_2$. The 6d theory has flavor symmetry $SO(7)
\supset SU(2) \times SU(2) \times U(1)$ with maximal torus $U(1)_\beta \times
U(1)_\gamma \times U(1)_t$ (in the notations of
\cite{Razamat:2016dpl}). The effective 4d theory obtained by compactification of the 6d theory on $\Sigma_1$ with fluxes w.r.t. $U(1)_\beta \times
U(1)_\gamma \times U(1)_t$ can be again described by gluing together certain trinion theories $T^{\pm}_{A}$ and $T^{\pm}_{B}$. Each trinion corresponds to a sphere with three (``maximal'') punctures. Each puncture breaks $SO(7)$ flavor symmetry down to a certain $SU(2)\times U(1)^2$ subgroup. There are punctures of two different ``colors'' corresponding to different embeddings. The trinions $T^{\pm}_{A}$ have all three punctures of the same color while $T^{\pm}_{B}$ have punctures of different color. The theories $T_A^+$ and $T_B^+$ correspond to spheres supporting fluxes (1/4,1/4,1) and (-1/4,1/4,1) respectively. The theories $T_A^-$ and $T_B^-$ correspond to spheres with opposite fluxes and differ by charge conjugation. Each puncture also introduces $SU(2)^2$ global symmetry, so that gluing punctures together corresponds to gauging the diagonal of $SU(2)^2\times SU(2)^2$ symmetry (after introducing extra matter depending on the type of punctures). The trinion theories $T^{\pm}_{A}$ and $T^{\pm}_{B}$ can be build from building blocks that have Lagrangian description. The description involves three copies of $SU(2)$ gauge groups. We direct the reader to \cite{Razamat:2016dpl} for the details.

Consider for example genera $g_1=2$ and $g_2=3$. If we have zero fluxes $(0,0,0)$ ({\it i.e.} $G'=1$) along $\Sigma_1$, then one encounters again the problem that the Bethe equations, that arise in
calculation of the partition function of the effective 4d $\CN=1$ on
$\Sigma_2 \times T^2_\tau$, are degenerate.

If we instead take fluxes $(1/2, 1/2, 2)$ along $\Sigma_1$ (this preserves
$SU(2)_\text{diag} \times U(1)^2$ flavor symmetry in 4d and can be realized, for example,
by gluing two copies of $T_A^+$ theories), the Bethe equations are
non-degenerate, and one can get a finite answer when the fugacities
$(\beta,\gamma,t)$ of unbroken part of $SO(7)$ are turned on.

Suppose the
fluxes along $\Sigma_2$ are $(\Beta, \Gamma, 0)$.
For generic $\Beta$, $\Gamma$ the unbroken symmetry is $G_\text{2d}=U(1)^3$ but when $\Beta=\Gamma$, it is
 $G_\text{2d}=SU(2)_\text{diag} \times U(1)^2$, with $SU(2)_\text{diag}$ fugacity being
$\sqrt{\beta/\gamma}$). Then (by summing over
$2^{12}$ solutions of Bethe equations) we get the answer of the
following form :
\begin{multline}
Z^\text{6d}[\Sigma_1 \times \Sigma_2 \times T^2_\tau] =
\\
t^{16}
\frac{(t^4\gamma^8)^\Gamma
(t^4\beta^8)^{\Beta}}{(1-\beta^2\gamma^2)^8}
\times
(
(\beta\gamma)^4 + O(t))
\\
+ q (
4(\beta ^8 \gamma ^4-3 \beta ^6 \gamma ^6
-2 \beta ^6 \gamma ^2+\beta ^4 \gamma ^8
-8 \beta ^4 \gamma ^4+\beta ^4-2 \beta ^2 \gamma ^6
-3 \beta ^2 \gamma ^2+\gamma ^4 )
+O(t)
)
+O(q^2) )
\\ \in R(G_\text{2d})[[q]]
\end{multline}

One can also consider $\Sigma_1$ to be just a basic building block, i.e. $\Sigma_1=S^2\setminus {3\text{pt}}$, a pair of pants. Suppose that all three punctures are of the same color so that the effective 4d theory is $T_A^+$. The result for the partition function of the 6d theory, depends on $q$ the nome of 2-torus, $t,\beta,\gamma$, the fugacities of $SU(2)\times U(1)^2\subset SO(7)$ global symmetry, and on $SU(2)^2$ fugacities $u_{1,2}$, $v_{1,2}$, $z_{1,2}$ associated to each puncture. In particular we have:
\begin{equation}
	Z^\text{6d}[(S^2\setminus {3\text{pt}}) \times \Sigma_2 \times T^2_\tau](q;u_{1,2}, v_{1,2}, z_{1,2};t,\beta,\gamma)
	=\frac{t^8\beta^2\gamma^2}{(1-\beta^2\gamma^2)}+O(t^{10})+O(q)
\end{equation}
for $g_2=3$ and zero flavor fluxes along $\Sigma_2$.

\subsection{E-string theory}

\subsubsection{ E-string on $M_4=\Sigma_1\times \Sigma_2$ }

The calcuation of the partition function on $M_4 =\Sigma_1\times \Sigma_2$ with possible fluxes can be done similarly to the case of $(N,k)=(2,2)$ theory, by using the results of \cite{Kim:2017toz}.
Consider for example the case $g_1=1$, $g_2=2$ with one unit of flux for $G'=U(1) \subset G=E_8$ along $\Sigma_1$. The flux breaks $E_8$ flavor symmetry down to $E_7\times U(1)$. The compactification on $\Sigma_1$ first produces the 4d $\CN=1$ theory that has Lagrangian description with $SU(2)^2$ gauge symmetry, $SU(8)\times U(1)$ flavor symmetry and the chiral multiplets in the following representations:
\begin{equation}
	\begin{tabular}{c|cccc}
	\hline
	 & $SU(2)$ & $SU(2)$ & $SU(8)$ & $U(1)$ \\ 
	\hline 
	$\Phi_1$ & $\mathbf{2}$ & $\mathbf{1}$ & $\mathbf{8}$ & $-1/2$ \\
	$\Phi_2$ & $\mathbf{1}$ & $\mathbf{2}$ & $\bar{\mathbf{8}}$ & $-1/2$ \\
	$B_{1,2}$ & $\mathbf{2}$ & $\mathbf{2}$ & $\mathbf{1}$ & $+1$ \\
	$F_{1,2}$ & $\mathbf{1}$ & $\mathbf{1}$ & $\mathbf{1}$ & $-2$ \\
	\hline 
	\end{tabular}
\end{equation} 
There are also the following terms turned on in the superpotential: $\Phi_1\Phi_2 B_1$, $\Phi_1\Phi_2 B_2$, $B_1^2F_1$, $B_2^2F_2$ with obvious projections on the trivial representation. The $SU(8)$ flavor symmetry is enhanced to $E_7$ in the IR.

Calculation of the $T^2_\tau \times \Sigma_2$ topologically twisted index of this theory yields:
\begin{multline}
Z^\text{6d}[\Sigma_1\times \Sigma_2\times T^2_\tau] = (\mathbf{1}\,t^4 + \mathbf{56}\,t^6 + ...) \\
- (2\cdot\mathbf{1}+2\cdot\mathbf{56}\,t^2+...)q
+(\mathbf{1}\,t^{-4}+\mathbf{56}\,t^2+...)q^2 +... \in R(G_\text{2d})[[q]]
\end{multline}
where $t$ is the $U(1)$ flavor fugacity and $\mathbf{d}$ denotes representation of $E_7$ of dimension $d$.

\section{Twisted indices of 5d theories}
\label{sec:BPS-equations}

While our starting point in 6d involves mostly non-Lagrangian theories, many of their Kaluza--Klein reductions to 5d admit Lagrangian descriptions that are conjectured to capture all information in the BPS sector of the 6d theory. This allows to define (and compute) topologically twisted indices of the resulting 5d gauge theories on $S^1 \times M_4$, which recover the elliptic genera of $T[M_4]$, the effective theory obtained by partially twisting the 6d theory on a 4-manifold.

Therefore, compared to our previous discussion, in this section we decompactify one circle in $T^2_{\tau}$ and instead consider the geometry
\be
\text{6d $(1,0)$ theory on }M_4\times S^1 \times \R.
\ee
After reducing on the $S^1$ factor, one obtains a 5d $\CN=1$ theory on $M_4 \times \R$ with a partial twist on $M_4$. Then supersymmetric vacua of the effective 1d $\CN=1$ quantum mechanics on $\R$ can be identified with solutions to a system of BPS equations. As these equations are PDEs on $M_4$, we will refer to them as the ``4d BPS equations.'' They will depend on the flavor symmetry background $\mu$, and the moduli space of solutions $\CM_{\text{BPS}}(M_4,\mu)$ will have disconnected components labeled by topological types of the gauge bundle. They correspond to mutually non-interacting sectors of the quantum mechanics on $\R$, which we will loosely refer to as ``instanton number sectors.''  Using intersection theory on the moduli space $\CM_{\text{BPS}}(M_4,\mu)$, one can hope to define numerical invariants of $M_4$ for a given $\mu$ and a given instanton number sector. However, the physical system also predicts that there is  a vector space associated to the pair $(M_4,\mu)$, given by the BPS Hilbert space $\CH_\text{BPS}$ of the 1d $\CN=1$ quantum mechanics. 
To obtain such BPS Hilbert space it is usually not sufficient to view the 1d theory as a sigma model onto $\CM_{\text{BPS}}$ because there can be 1) singularities on $\CM_{\text{BPS}}$ and 2) quantum tunneling effects (sometime referred to as ``worldline instantons''). When $M_4$ has $b_2^+>1$, $\CM_{\text{BPS}}$ is expected to be smooth for a generic choice of the metric, so here we will not worry about singularities. To cure the second problem, one needs to consider configurations that can also vary along the $\R$ direction.

However, {\it a priori} we actually do not expect that the BPS Hilbert space itself is the invariant of $(M_4,\mu)$ since it can change under continuous deformations, including deformations of the metric on $M_4$. To be more specific, assume that the fermionic parity is non-anomalous\footnote{For absolute theories ($N_0=1$), this is the case when $d=2(c_\text{R}-c_\text{L})$ is even.} then the vector space is $\Z_2$ graded ($\CH_\text{BPS}=\CH^0\oplus \CH^1$) and can be understood as $\Z_2$ graded complex on which the differential, physically realized as the supercharge, acts trivially.  The individual components $\CH^0$ and $\CH^1$ can vary under small deformations,\footnote{Later in this section we briefly discuss a possible scenario when the individual components $\CH^0$ and $\CH^1$ are actually invariant under deformations of the metric on $M_4$.} in particular deformations of the metric on the 4-manifold. This can happen already on the level of an ordinary 1d $\CN=1$ sigma-model, where $\CH^0$ and $\CH^1$ are vector spaces of chiral harmonic spinors on the target, and their dimensions can depend on the metric of the target. The equivalence class of the formal difference $\CH^0-\CH^1$, considered as the virtual vector space (moreover, a virtual representation of the 5d global symmetries), is, however, an invariant. The difference should be also immune to any tunneling effects. Therefore one can identify $\CH^0-\CH^1$ with the index of the Dirac operator on $\CM_\text{BPS}(M_4,\mu)$. The isomorphism class of the virtual vector space $\CH^0-\CH^1$ is completely captured by the equivariant Euler characteristics of $\CH_\text{BPS}$, which can be also understood as the twisted index of the 5d theory, with holonomies for global symmetries turned on along the $S^1$ circle of $S^1\times M_4$. It can be expressed as an equivariant $\hat{A}$-genus on $\CM_{\text{BPS}}$ that can be computed via a fixed point formula.\footnote{
When the dimension of $\CM_{\text{BPS}}$ is not even, instead of the (equivariant) index of the Dirac operator, one can define torsion-valued equivariant KO-classes via the equivariant Atiyah--Bott--Shapiro map (see {\it e.g.}~\cite{joachim2004higher}). Similarly, when the dimension is 2 mod 4, there will also be torsion classes supplementing the equivariant index.} If the 5d theory is a reduction of a 6d theory, then it will have a canonical symmetry $U(1)_q$ coming from the rotation of the 6d-to-5d circle, {\it i.e.}~the graviphoton of KK compactification. In the 5d gauge theory description, $U(1)_q$ is in general a mixture of instanton symmetry and flavor symmetry, and thus acts on the moduli space non-trivially. Therefore, the twisted partition function, with the fugacity for $U(1)_q$ turned on, has good chance of being well-defined even if the 6d theory has no flavor symmetry. In general, apart from $U(1)$, the 5d theory has global symmetry $G_{2d}\subset G$ unbroken by the background $\mu$.\footnote{More generally one may consider 5d theory obtained by compactification of 6d theory on $S^1$ with a non-trivial holonomy $h\in G_\text{2d}\subset G$. Then the actual global symmetry of 5d theory on $M_4$ would be $G_\text{1d}:=\text{Centralizer}_{G_\text{2d}}(h)\subset G_\text{2d}$ and one would need to replace $G_\text{2d}$ with $G_\text{1d}$ where appropriate in the formulas below.} The symmetry $G_{2d}$ can be also realized as the mixture of instanton and flavor symmetries. An example of such scenario is given by the E-string theory on a circle, where the effective 5d theory is an $SU(2)$ gauge theory with 8 flavors and a $U(1)$ subgroup of $G=E_8$ in the 6d theory is the $U(1)$ instanton symmetry in 5d.

Let us be more explicit. Suppose the 6d $(1,0)$ theory on $S^1$ can be described in terms of a 5d $\CN=1$ theory with the gauge group $H=\prod_{i=1}^LH_i$ (where $H_i$ is a simple Lie group or $U(1)$) and hypermultiplets in some quaternionic representation $R_H$ of $H$. The 5d theory (in background $\mu$) has $U(1)_q\times G_\text{2d}$ global symmetry. Let the generator of $U(1)_q$ be a linear combination of the generators of $U(1)_i$ instanton symmetries corresponding to gauge groups $H_i$, whose currents are $  \star\frac{1}{8\pi^2} \Tr (F_i\wedge F_i)$, with coefficients $c_i$ and an ordinary $U(1)$ flavor symmetry acting on hypermultiplets. This flavor symmetry action on hypermultiplets induces action of $U(1)_q$ on $\CM_{\text{BPS}}$. Similarly, $G_\text{2d}$ both acts on the hypermultiplets as a usual flavor symmetry, which induces an action of $G_\text{2d}$ on $\CM_\text{BPS}$, and as combination of instanton symmetries. 
The latter action is realized by certain homomorphisms
\begin{equation}
	\xi_i:G_{\text{2d}}\rightarrow U(1)_i.
\end{equation}
The partition function of the 5d theory on $S^1\times M_4$ with holonomy $g\in G_{2d}$ around $S^1$ and (complexified) holonomy $q$ of $U(1)_q$ then reads
		\begin{equation}
		Z^\text{6d}[M_4\times T^2_\tau] =Z^\text{5d}[M_4\times S^1] =
		\sum_{n_1,\ldots, n_L}\;\;\; \prod_{i=1}^Lq^{c_in_i}\xi_i(g)^{n_i}
		\hspace{-6ex}\int\limits_{\scriptsize\begin{array}{c} \CM_\text{BPS}^{n_1,\ldots,n_L} \\ U(1)_q\times G_\text{2d}\text{-equivariant}   
		\end{array}}
		\hspace{-9ex}
		\hat{A} 
		\;\;\;\in R(G_\text{2d})((q))
		\label{Z5d-A-roof}
	\end{equation}
	where $\CM^{n_1,\ldots,n_L}_\text{BPS}$ is the component of the moduli space of solutions of BPS equations $\CM_\text{BPS}$ with fixed instanton numbers
	\begin{equation}
		n_i=\frac{1}{8\pi^2}\int_{M_4}\Tr F_i\wedge F_i.
	\end{equation}
For each term in the sum (\ref{Z5d-A-roof}) the integral $\int \hat{A}$ can be understood either as $U(1)_q\times G_\text{2d}$-equivariant index of the Dirac operator on $\CM_\text{BPS}^{n_1,\ldots,n_L}$, or, when $g$ is in the maximal torus of $G_\text{2d}$, as the equivariant integral of the $\hat{A}$-genus characteristic class. Both interpretations provide an element of $R(G_\text{2d})[q,q^{-1}]$. When $\CM_\text{BPS}^{n_1,\ldots,n_L}$ is non-compact, one should consider a certain completion of this ring, allowing infinite series in positive powers of $q$ and also infinite (virtual) sums of representations of $G_\text{2d}$. The coefficients in the sum over instanton numbers $n_i$ belong to the same ring.

The formula (\ref{Z5d-A-roof}) assumes that the actual dimension of $\CM_\text{BPS}^{n_1,\ldots,n_L}$ coincides with its virtual one, given by the dimension of the space of gauge and hypermultiplet fields, minus the dimension of the target space of the first order differential operators that specify BPS equations, minus the dimension of the space of gauge transformations. In general this is not the case. The mismatch will result in a real vector bundle $E$ over $\CM_\text{BPS}$ of zero modes of fermionic fields\footnote{The simplest example of this is the theory of a singe 5d hypermultiplet and no gauge field. The virtual dimension of $\CM_{\text{BPS}}$ is $-\sigma/4$. For a generic metric on $M_4$  we have:
\begin{equation}
	(\CM_\text{BPS},E)=\left\{ \begin{array}{cl}
	(\R^{-\sigma/4},\bar{\text{pt}}), & \sigma \leq 0, \\
	(\text{pt},\bar{\R^{\sigma/4}}), & \sigma >0,
	\end{array}
	\right.
\end{equation}
where $\bar{V}$ denotes a trivial bundle with constant fiber $V$.
 } 
 (not paired with the bosonic zero modes by supersymmetry in $\CN=1$ effective 1d quantum mechanics). The rank of the bundle is equal to the difference between the actual and the virtual dimensions of $\CM_\text{BPS}$. The index of the ordinary Dirac operator then should be replaced by the index of the Dirac operator coupled to $\CS(E)$, the spinor bundle of $E$. Equivalently, in the formula (\ref{Z5d-A-roof}) one should replace
 \begin{equation}
	 \int\limits_{\scriptsize\begin{array}{c} \CM_\text{BPS}^{n_1,\ldots,n_L} \\ U(1)_q\times G_\text{2d}\text{-equivariant}   
	 		\end{array}}
	 		\hspace{-9ex}
	 		\hat{A}
	 		\qquad\qquad\longrightarrow
	 			 \int\limits_{\scriptsize\begin{array}{c} \CM_\text{BPS}^{n_1,\ldots,n_L} \\ U(1)_q\times G_\text{2d}\text{-equivariant}   
	 			\end{array}}
	 		\hspace{-9ex}
	 		\hat{A}\, \mathrm{ch}\, \CS(E).
	\end{equation}
Equivalently, one can understand the original integral in the virtual sense. However, let us assume, for simplicity of the discussion, that the BPS equations are not degenerate and the bundle of fermion zero modes $E$ is trivial.
	
The right-hand side of (\ref{Z5d-A-roof}) in principle can be defined for any 5d gauge theory, with some global symmetries $U(1)_q$ and $G_\text{2d}$ that act as combinations of instanton and flavor symmetries. However, only for 5d theories that arise from compactification of 6d theories on a circle, one expects (\ref{Z5d-A-roof}) to be modular in $\tau$, with a multiplier system determined by $c_\text{R}-c_\text{L}$ and 2d 't Hooft anomaly of $G_\text{2d}$. 
Also, let us emphasize that although the partition function (\ref{Z5d-A-roof}) is written in terms of moduli spaces of solutions of 4d BPS equations on $M_4$, it is \text{not} the same as the partition function of the 4d $\CN=2$ theory obtained by dimensional reduction. In particular, the latter would involve an integration of a different characteristic class (the equivariant Eular class $e(E)$ instead of $\hat{A}\,\mathrm{ch}\CS(E)$) over $\CM_\text{BPS}$. The result of equivariant integration then would be a rational function in logarithms of fugacities of global symmetries, instead of series in fugacities with integral coefficients. The invariants of smooth 4-manifolds produced by 5d theory can be considered as K-theoretic lift of the usual Seiberg--Witten and Donaldson-like invariants. In certain special cases, however, it may happen that there is a 4d gauge theory that captures 6d theory on $T^2_\tau$ with all KK modes. A possible candidate for this is 4d $SU(2)$ gauge theory with 4 flavors describing the  6d $O(-4)$ theory on a torus.

Let us point out a possibility of having a more refined invariant than just equivariant Euler characteristic of $\CH_\text{BPS}=\CH^0\oplus \CH^1$, or, equivalently, equivariant index of the Dirac operator on $\CM_\text{BPS}$. Without taking into account of tunneling effects, $\CH^0$ and $\CH^1$ are given by the vector spaces of harmonic spinors on $\CM_\text{BPS}$ (assuming it is a manifold). The individual components may vary under generic deformations of metric on $\CM_\text{BPS}$. However, it may happen that even for a generic metric on $M_4$ the metric on $\CM_\text{BPS}$ is not a generic one, but belongs to some particular family such that the individual spaces $\CH^0$ and $\CH^1$ remain the same along the family. Taking into account tunneling effects then can be done by a construction similar to Floer homology. Solutions of 5d BPS equations on $M_4\times \R$ should define a (real) supercharge $Q$ that acts non-trivially on $\CH_\text{BPS}$ obtained from sigma-model on $\CM_\text{BPS}$. Then $\CH_\text{BPS}^{\text{(corrected)}}=\ker Q|_{\CH_\text{BPS}}$.

Now it is a good point to pause and summarize the benefits of working with 6d theories (or their 5d avatars) instead of their 4d reductions:
\begin{itemize} 
\item The 6d theory always give rise to modular forms, which in general cannot be recovered from the 4d theory obtained by dimensional reduction. In fact, the 4d theory often contains strict less information and cannot even be used to obtain a $q$-series. An example is the E-string theory in 6d which has Minahan--Nemeschansky's $E_8$ theory as a 4d limit. The latter theory has no marginal deformation and its twisted partition function on a 4-manifold is not naturally a $q$-series. On the other hand, the E-string theory can give rise to $q$-series, which was explicitly computed in the previous section for a class of 4-manifolds.  

\item The 6d and 5d theories often contain other interesting information that is not visible in the 4d theory. For example, the 5d $\CN=1$ $SU(2)$ gauge theory with $N_F=4$ has enhanced $D_5$ symmetry \cite{Seiberg:1996bd}, while the 4d reduction only captures a $D_4$ subgroup. Therefore, there is no computation in the purely 4d point of view that can reproduce an equivariant partition function with $D_5$ symmetry. Also, the partition functions of 6d and 5d theories are naturally series with integer coefficients, while the partition function of the 4d theory is only expected to be a series with rational coefficients.   

\item Using higher dimensional theories, one can obtain stronger invariants of 4-manifolds that are topological modular forms, (virtual) vector spaces, or modules over the ring of topological modular forms. These are only available if one studies 6d or 5d theories.

\end{itemize}

One can also run the above program on 3-manifolds. As we have concluded in section~\ref{sec:3-manifolds}, for the 2d $(1,1)$ theory $T[M_3\times S^1]$, only the sector with zero instanton number can give non-trivial invariants for $M_3$. Such invariants, viewed as TMF classes, take values in the subring $\Z\subset \pi_*\TMF$, and one might regard them as uninteresting. However, due to enhanced supersymmetry, they can be categorified to homological invariants, one for each inequivalent flavor background $\mu$, and carries an action of $G_{\text{2d}}$. The collection of these homological invariants might be powerful at distinguishing 3-manifolds. 

When $M_4=M_3\times S^1$, it is possible to further reduce the 4d BPS equations by assuming that the fields are constant along the circle. The resulting system of PDEs on $M_3$ will be referred to as the ``3d BPS equations.''
We expect to have the relation
\be
\CM_{\text{BPS}}(M_3)=\CM_{\text{vac}}(T[M_3])
\ee
between the moduli space of solutions to BPS equations on $M_3$ and the moduli space of vacua of the 3d $\CN=1$ theory $T[M_3]$ on $S^1\times \R^2$, both are labeled by the 6d $(1,0)$ theory and depend on the flavor symmetry background. This gives a way to test proposals for $T[M_3]$.

As the 3d equations are simpler while having all the essential ingredients, we will start from there and then move to 4d and 5d. We will consider general 5d $\CN=1$ gauge theories, which may or may not come from a 6d $(1,0)$ SCFT. This would enable one to categorify 4-manifold invariants such as the Donaldson and Seiberg--Witten invariants whose corresponding 4d $\CN=2$ theory cannot be lifted to be a 6d SCFT.

\subsection{3d BPS equations}

To start, we consider a general 5d $\CN=1$ theory with gauge group $H$ and matters transforming in a quaternionic $H$-representation $R_H$. The vector-multiplet consists of a gauge connection $A$, a gaugino $\lambda$ and a real scalar $\phi$ together with a auxiliary field $\CD$, all of which are valued in $\frak{h}=\mathrm{Lie}\,H$. The hyper-multiplet consists of complex scalars $q$ and complex fermions $\psi$. The way they transform under $\mathrm{Spin}(1,4)\times SU(2)_R$ is listed in the table below.\footnote{We work with Lorentzian signature in this section.}
\be
\begin{aligned}
\text{$A$:} & \ {\bf(5,1)}\ra {\bf 3}^0\oplus {\bf 1}^{\pm 2},\\
\text{$\lambda$:} & \ {\bf(4,2)}\ra {\bf 1}^{\pm 1}\oplus {\bf 3}^{\pm 1},\\
\text{$\phi$:} & \ {\bf(1,1)}\ra {\bf 1}^{0},\\
\text{$\CD$:} & \ {\bf(1,3)}\ra {\bf 3}^{0},\\
\text{$q$:} & \ {\bf(1,2)}\ra {\bf 2}^0, \\
\text{$\psi$:} & \ {\bf(4,1)}\ra {\bf 2}^{\pm 1}.
\end{aligned}
\ee
The last column of the table indicates how the fields transform under $\mathrm{Spin}(3)'_{M_3}\times SO(1,1)$ after the topological twist. The gauginos, as well as the supercharges, are symplectic-Majorana fermions satisfying
\be
\bar{\lambda}_A=\left(\lambda^T\right)^B\epsilon_{BA}\Omega,
\ee
where $\Omega$ is the symplectic form on the $\bf 4$ of $\mathrm{Spin}(1,4)$ and $A,B$ are indices for the $\bf 2$ of $SU(2)_R$. For the two scalar supercharges, we denote the two corresponding SUSY variation parameters as $\varepsilon_{+}$ and $\varepsilon_-$, where the subscripts denote helicity in $\R^{1,1}$.

The BPS equations are obtained from the variation of fermionic fields, given by
\be\label{FermionVar}
\begin{aligned}
\delta\lambda&=\frac{1}{2}F_{\mu\nu}\gamma^{\mu\nu}\varepsilon-iD_\mu\phi\gamma^\mu\varepsilon+i\CD^I\sigma^I\varepsilon,\\
\delta\psi&=\sqrt{2}\left(D_\mu q^A\cdot\gamma^\mu \varepsilon_A+\phi q^A\varepsilon_A\right),\\
\delta\bar{\psi}&=\sqrt{2}\left(\bar{\varepsilon}^A\gamma^\mu\cdot D_\mu\bar{q}_A+\bar{\varepsilon}^A\bar{q}_A\phi\right).
\end{aligned}
\ee
Here $I=1,2,3$ is the index for $\bf 3$ of $SU(2)_R$. One can choose the $\varepsilon$ above to be any linear combination $\varepsilon=a\varepsilon_{+}+b\varepsilon_{-}$ and the BPS equation will in principal depend on $a,b$. We first consider the case of $a=1$ and $b=0$.

Notice that $\gamma^{\mu}$ with $\mu=1,2,3$ will flip the helicity in $\R^{1,1}$, and, as a consequence, each equation in \eqref{FermionVar} will give rise to two BPS equations coming from terms with even/odd number of $\gamma$'s. Define $\mu(q)$ as the moment map for the $H$-action on $R_H$, it is valued in $\frak{h}$ times $\bf 3$ of $SU(2)_R$. After the twist it defines an $\frak{h}$-valued ``squaring map'' from the spinor bundle $\CS(M_3)$ to $\Omega^1(M_3)$ which can be expressed in component language as $\mu(q)=\bar{q}_i\sigma^I T^a_{ij}q_jdx^I$ locally. Then the equation of motion for $\CD$ is
\be
\CD=\mu(q).
\ee
Now the BPS equations are given by
\be
\begin{aligned}\label{BPSEq}
\star F_A-\mu(q)&=0,\\
d_A\phi &=0,\\
d_A q=d_A\tilde{q}&=0,\\
\phi q=\phi \tilde{q}&=0.
\end{aligned}
\ee
For general $a,b$ we have\footnote{The exact equations will depends on conventions for $\varepsilon_\pm$.}
\be
\begin{aligned}\label{BPSEq'}
a\left(\star F_A-\mu(q)\right)+b\cdot d_A\phi &=0,\\
b\left(\star F_A-\mu(q)\right)+a\cdot d_A\phi &=0,\\
a {\slashed{D}}_A q+b\cdot\phi q&=0,\\
b {\slashed{D}}_A q+a\cdot\phi q&=0,
\end{aligned}
\ee
where ${\slashed{D}}_A$ is the Dirac operator and equations for $\tilde{q}$ are omitted. For generic $a$ and $b$, this set of equations is equivalent to \eqref{BPSEq}, but when $a=b$, since $\varepsilon_+ + \varepsilon_-$ is invariant under the flipping of helicity, we have only one-half of the equations,
\be
\begin{aligned}\label{BPSEq+}
\star F_A-\mu(q)+ d_A\phi &=0,\\
 {\slashed{D}}_A q+\phi q&=0.
\end{aligned}
\ee
Similar phenomenon happens for $a=-b$, for which $\varepsilon$ is anti-invariant under the flipping of helicity in $\R^{1,1}$ and we have
\be
\begin{aligned}\label{BPSEq-}
\star F_A-\mu(q)-d_A\phi &=0,\\
 {\slashed{D}}_A q-\phi q&=0.
\end{aligned}
\ee

\subsection{4d and 5d BPS equations}

Before further analyzing the above equations, we derive a set of related equations on 4-manifolds. The holonomy group of a general 4-manifold $M_4$ is $SO(4)$. We consider its double cover $\mathrm{Spin}(4)=SU(2)_+\times SU(2)_-$. A general 5d $\CN=1$ theory can be twisted over $M_4$ by replacing $SU(2)_+$ with the diagonal subgroup $SU(2)'_+$ of $SU(2)_+\times SU(2)_R$. The supercharges transform under the $SU(2)_+\times SU(2)_-\times SU(2)_R$ and $SU(2)'_+\times SU(2)_-$ respectively as
\be
Q: ({\bf 1, 2,2})\oplus({\bf 2, 1,2}) \ra ({\bf 1,1})\oplus  ({\bf 3,1}) \oplus ({\bf 2,2}).
\ee
We will denote the scalar supercharge as $Q_0$ and the corresponding SUSY parameter $\varepsilon_0$. The fields will transform in the following way after the twist,
\be
\begin{tabular}{c|c|c}
fields & $SU(2)_+\times SU(2)_-\times SU(2)_R$ & $SU(2)'_+\times SU(2)_-$\\ \hline
$A$: & ${\bf(2,2,1)}\oplus {\bf (1,1,1)}$ & ${\bf (2,2)} \oplus {\bf(1,1)}$\\ \hline
$\lambda$ & $({\bf 1, 2,2})\oplus({\bf 2, 1,2})$ & $({\bf 1,1})\oplus({\bf 3,1}) \oplus ({\bf 2,2})$\\ \hline
$\phi$ & ${\bf(1,1)}$ & ${\bf(1,1)}$ \\ \hline
$\CD$ & ${\bf(1,1,3)}$ & ${\bf (3,1)}$\\ \hline
$q$ & ${\bf(1,1,2)}$ & ${\bf (2,1)}$ \\ \hline
$\psi$ &  $({\bf 1, 2,1})\oplus({\bf 2, 1,1})$& ${\bf (1,2)\oplus (2,1)}$
\end{tabular}
\ee

Now $q$ and $\tilde{q}$ are sections of the spinor bundle $\CS^+(M_4)$,\footnote{When $M_4$ is not Spin, one needs to twist the $H$-bundle to make $\CS^+\otimes R_H$ well defined. For the 6d theory obtained by $N$ M5-branes probing $\C^2/\Gamma$ singularity, there is always a $U(1)$ factor in $H$, which can be used to deal with this problem, as all oriented 4-manifolds are Spin$^{c}$.} and $\phi':=-iA_5$ is a $\frak{h}$-valued real scalar over $M_4$ satisfying $\phi'=\phi'^\dagger$. The BPS equations are now straightforward to derive from the variation of $\lambda$ and $\psi$. Since $\varepsilon_0$ has definite helicity under $\mathrm{Spin}(4)$ and a $\gamma$ will flip helicity, there will be at least two equations coming from variations of either $\lambda$ and $\psi$. In fact, there is an additional one from the ``trace part'' of the variation of $\lambda$. Another way to see that there are three equation coming out of $\delta\lambda$ is to notice that this variation is also decomposed as $({\bf 1,1})\oplus({\bf 3,1}) \oplus ({\bf 2,2})$ under $SU(2)'_+\times SU(2)_-$, and each piece gives rise to an independent equation. In total, we have
\be
\begin{aligned}\label{BPSEq4D}
F_A^+ -\mu^+(q)&=0,\\
d_A(\phi+\phi') &=0,\\
\slashed{D}_A q=\slashed{D}_A \tilde{q}&=0,\\
\varphi \cdot q=\bar{\varphi}\cdot \tilde{q}&=0,\\
[\varphi,\varphi^\dagger]&=0.
\end{aligned}
\ee
Here $\varphi=\phi+i\phi'$,\footnote{We use Euclidean signature the above equations. Also, one can choose the gauge with $\phi'=0$.} and $\mu^+$ is now the squaring map from sections of $\CS^+(M_4)\otimes R_H$ to $\Omega^{2,+}(M_4,\mathrm{ad}(H))$ is again coming from the equation of motion for $\CD$.

One can also allow fields to have non-trivial dependence on the time direction. This leads to a system of flow equations:
\be
\begin{aligned}\label{BPSEq5D}
F_A^+ -\mu^+(q)&=0,\\
F_0-d_A \phi &=0,\\
\slashed{D}_A q=\slashed{D}_A \tilde{q}&=0,\\
-iD_0 q+\phi \cdot q=iD_0\tilde{q}+\phi\cdot \tilde{q}&=0,\\
D_0\phi&=0,
\end{aligned}
\ee
where $F_0$ is a one-form on $M_4$ that satisfies $(F_0)_\mu=F_{0\mu}$ locally. 

Now we have all the 3d/4d/5d BPS equations \eqref{BPSEq}, \eqref{BPSEq4D} and \eqref{BPSEq5D}. When flavor backgrounds are turned on, one only needs to make obvious modifications to the equations by making the covariant derivative also dependent on the background flavor connections. Supersymmetry further demands that the connections for flavor symmetries are self-dual for the non-abelian part and may have constant curvature for the abelian part.  

\subsection{Vanishing theorems}

We now analyze the BPS equations in greater detail. The first step is to prove a vanishing theorems.

We start with the three equations
\be
E_1=E_2=E_3=0
\ee
with $E_1=F_A^+ -\mu^+(q)$, $E_2= \slashed{D}_A q$ and $E_3= \slashed{D}_A \tilde{q}$, which are shared between the 4d and 5d BPS equations.

Using the Lichnerowicz--Weitzenböck formula
\be
\slashed{D}_A^\dagger\slashed{D}_A=\nabla_A^\dagger\nabla_A +\frac{1}{4}\CR + \slashed{F}^+,
\ee
where $\nabla_A$ is the covariant derivative assoicated with $A$ and the Levi-Civita connection, $\CR$ is the scalar curvature and $\slashed{F}^+: \Gamma(\CS^+\otimes R_H)\ra \Gamma(\CS^+\otimes R_H)$ the Clifford action, we have
\be
|E_1|^2+|E_2|^2+|E_2|^2=|F_A^+|^2+|\nabla_A q|^2+|\nabla_A \tilde{q}|^2+\frac{1}{4}\CR(|q|^2+|\tilde{q}|^2).
\ee
This leads to
\paragraph{Vanishing Theorem.}If the metric on $M_4$ has positive scalar curvature, then we must have
$$
q=\tilde{q}=0
$$
and
$$
F_A^+=0
$$
for the 4d and 5d BPS equations to hold.
\bigskip

Analogous theorem also holds for the 3d equations, which can be viewed as a direct corollary of the above theorem applied to $S^1$-invariant configurations on $M_4=M_3\times S^1$.

\subsection{$\CM_{\text{BPS}}$ for $L(p,1)\times S^1$}
As an application of the vanishing theorem, we determine $\CM_\text{BPS}$ for $M_4=L(p,1)\times S^1$. The vanishing theorem ensures that $q=\tilde{q}=0$, which leaves only two non-trivial equations
\be
F^+_A=0,
\ee
and
\be
d_A\phi=0.
\ee
$\CM_{\text{BPS}}$ will be a disjoint union of $\CM_i$ with $i\geq 0$ being the instanton number. When the instanton number is zero, $F^-_A=0$ also holds. So $\CM_0$ projects to the moduli space of flat connections on $L(p,1)\times S^1$, which further projects onto $\CM_{\text{flat}}(L(p,1))$. Denote a flat connections on $L(p,1)$ as $a$, then the fiber of $\CM_{\text{flat}}(L(p,1)\times S^1)\ra\CM_{\text{flat}}(L(p,1))$ is $H_a:=\mathrm{Stab}_H(a)$. The fiber of the first projection $\CM_0\ra \CM_{\text{flat}}(L(p,1)\times S^1)$ at $(a,h_0)$ is the vector space $\mathrm{Stab}_{\frak{h}_a}(h_0)$. So we see that the moduli space has interesting structure.

\subsection{Abelian theories}

We now consider the case where we have an abelian theory in 5d. It could come from the 6d (1,0) theory obtained by a M5-brane probing $\C^2/\Gamma$ singularity, or it could be a theory without a 6d lift, such as the $U(1)$ theory with a fundamental hypermultiplet. As it turns out, in abelian theories, there won't be tunneling effects and therefore the exact BPS Hilbert space $\CH_\text{BPS}$ is given by just by counting solutions of 4-dimensional equation, as described in the beginning of the section. 

To see this, consider a general $t$-dependent solution to the 5d BPS equations \eqref{BPSEq5D} $(A,A_0,\phi,q,\tilde{q})$. Since $H$ is abelian, the field $\phi$, transforming in the adjoint of $H$, is now a collection of decoupled free fields. And we also have $D_0=\partial_0$. As a consequence, \eqref{BPSEq5D} implies
\be
\partial_0\phi=\partial_0q=\partial_0\tilde{q}=0,
\ee
which in turn implies
\be
\partial_0F_0=0
\ee
and
\be
\partial_0F^+_A=0.
\ee
In fact, since $F_0=d\phi$ is exact on any slice $M_4\times \{t\}$, from Bianchi identity we have
\be
\partial_0F_A=-dF_0=0.
\ee
Now we only needs to show that we can make $A$ and $A_0$ time-independent by doing a gauge transformation
\be
A\mapsto A+df,\quad A_0\mapsto A_0+\partial f
\ee
with a function $f$ on $M_4\times \R_t$.

Since we have
\be
\partial_0A=F_0+dA_0=d(\phi+A_0),
\ee
it is easy to see that the gauge transformation given by the globally-defined function
\be
f=\int_0^tdt(A_0+\phi)
\ee
will make $\partial_0A=0$. After this gauge transformation, we will also have
\be
A_0=-\phi
\ee
becoming independent of time. This leads to

\paragraph{Vanishing Theorem for $Q$.} When $H$ is abelian, an arbitrary solutions of the 5d BPS equation will be gauge equivalent to a time-independent solution.
\bigskip

So modulo the field $A_0=-\phi$, solutions of the 5d equations are in one-to-one correspondence with the solutions of 4d multi-monopole equations. In other words, the homology groups are generated by solutions of the 4d equations with no differentials. For the non-abelian case, we expect $\CH_\text{BPS}$ to be in general $\Z_2$-graded. However, since the abelian case has no differential, it will now be $\Z$-graded. The grading is given by the index of an elliptic operator
\be
T: \Omega^1\oplus (\CS^+\otimes P(R_H))\ra \Omega^0\oplus\Omega^{2,+}\oplus(\CS^-\oplus P(R_H))
\ee
given by the multi-monopole equations, with $P(R_H)$ being the vector bundle associated with the gauge principal $H$-bundle $P$.

For each factor of $U(1)$ in $H=U(1)^r$, there will be an abelian instanton symmetry in the 5d theory. Therefore, one can make the homology $\Z^{r+1}$-graded. We also observe ``flavor-homology interlocking''---the homological degree can be written as a linear combination of instanton numbers of $U(1)$'s in $H$.

\subsection{An alternative viewpoints on the BPS equations}

The number of solutions to the 3d BPS equations are the same as the number of supersymmetric vacua of $T[M_3]$, or equivalently, its Witten index. If one replaces $\R^3$ in \eqref{BraneGeom1} with $T^3$ and perform compactification on it first, one would obtain a 3d $\CN=4$ (or 3d $\CN=8$ for $k=1$) theory $T[T^3]$ twisted on $M_3$. It is not hard to see that the twist, using $SU(2)_H$ of the $SU(2)_C\times SU(2)_H$ R-symmetry of the 3d $\CN=4$ theory, exactly gives the Rozansky--Witten theory of the Coulomb branch.

\section*{Acknowledgments}
We would like to thank Ali Daemi, Mike Freedman, Mike Hopkins, Anton Kapustin, Ciprian Manolescu,  Kantaro Ohmori, Shlomo Razamat, Peter Teichner, Edward Witten, Ida Zadeh, Gabi Zafrir, and Michele del Zotto for fruitful discussions. We especially would like to thank Peter Teichner for extensive discusions on topological modular forms and Edward Witten for his suggestion to view the 4-manifold invariants we obtain from 6d $(1,0)$ theories as defining a class in TMF. The work of S.G. is supported by the U.S. Department of Energy, Office of Science, Office of High Energy Physics, under Award No. DE-SC0011632, and by the National Science Foundation under Grant No. NSF DMS 1664240.
 The work of D.P. is supported by the Walter Burke Institute for Theoretical Physics, the U.S. Department of Energy, Office of Science, Office of High Energy Physics, under Award No.\ DE{-}SC0011632, and in part by the center of excellence grant ``Center for Quantum Geometry of Moduli Space" from the Danish National Research Foundation (DNRF95).
P.P. gratefully acknowledges the support from Marvin L. Goldberger Fellowship and the DOE Grant 51 DE-SC0009988 during his affiliation with IAS. The work of C.V. is supported in part
by NSF grant PHY-1067976. We would like to thank the hospitality of Simons Center for Geometry
and Physics, Kavli Institute for Theoretical Physics, International Center for Theoretical Physics, and Max Planck Institute for Mathematics where parts of this work were done.

\appendix

\section{Some relevant facts about TMF}
\label{app:TMF}

In this appendix, we gather several relevant facts about TMF, starting with its definition.

\subsection{$\Omega$-spectrum and elliptic cohomology}
An $\Omega$-spectrum $E$ is a sequence of topological spaces $E_n$ with $n\in\Z$ equipped with structure maps
\be
s_n: \quad E_n \ra \Omega E_{n+1} \quad \forall n\in\Z
\ee
from $E_n$ to the based loop space of $E_{n+1}$ such that all $s_n$'s are weak homotopy equivalences. In other words, $s_n$ induces isomorphisms between all homotopy groups
\be
s_n^*: \quad \pi_*(E_n) \xrightarrow{\sim} \pi_*(\Omega E_{n+1})\cong\pi_{*+1}(E_{n+1}). 
\ee

By Brown's representability theorem, $E$ defines a reduced generalized cohomology theory, denoted also as $E$, which assign to a topological space $X$ the cohomology groups\footnote{Here and below, there are two possible conventions of the sign of the degree.}
\be
E^{n}(X)=\lim_{k\ra\infty}[\Sigma^kX,E_{n+k}] \quad \forall n\in\Z,
\ee
where the expression on the left-hand side inside the limit stands for homotopy classes of maps from $\Sigma^k X$, the $k$-th suspension of $X$, to $E_{n+k}$. Conversely, any reduced generalized cohomology theory comes from an $\Omega$-spectrum. 

The homotopy groups of the spectrum $E$ is defined as
\be
\pi_n (E) := \lim_{k\ra\infty}\pi_{k+n} (E_k) =
E^{-n} (\text{pt}).
\ee

For an abelian group $G$, the singular cohomology $H^{*}(-,G)$ is given by the  Eilenberg--Maclane spectrum 
\be
E_{n}=K(n,G).
\ee
Other well-known examples include the spectra of complex and real K-theory, given by
\be
\mathrm{K}_{2n}=BU \times \Z,\quad \mathrm{K}_{2n+1}=U,
\ee
and
\bea
\KO_{8n}=&BO \times \Z, &\KO_{8n+1}=O,\\
\KO_{8n+2}=&O/U,  &\KO_{8n+3}=U/Sp,\\
\KO_{8n+4}=&BSp \times \Z, &\KO_{8n+5}=Sp,\\
\KO_{8n+6}=&Sp/U, &\KO_{8n+7}=U/O
\eea
where $U$, $Sp$ and $O$ stand for $U(\infty)$, $Sp(\infty)$ and $O(\infty)$ respectively.
Both spectra are periodic with period 2 and 8. Further, K-theory is even, meaning that $\mathrm{K}^{2n+1}(\mathrm{pt})=0$. 

Another even periodic cohomology theory, closely related to topological modular forms, 
 is the \textit{elliptic cohomology} $\mathrm{Ell}_{C/R}$. We won't detail its construction and only will remark that it depends on a choice of an elliptic curve $C$ over a commutative ring $R$. 

As an elliptic curve over $R$ is the same as a map 
\be
\mathrm{Spec}(R)\ra\CM_{\rm ell}
\ee
where $\CM_{\rm ell}$ is the \textit{moduli stack} of elliptic curves,\footnote{The moduli stack $\CM_{\rm ell}$ contains slightly more information than the moduli space. In particular, it remembers that certain elliptic curves can have automorphisms, which lead to the torsion classes in $\TMF_*$.} the elliptic cohomologies define a presheaf of cohomology theories over $\CM_{\rm ell}$. However, this presheaf cannot be made into a sheaf and it has no global section. The insight of Goerss, Hopkins and Miller is that the category of spectra is better behaved, and one can ask for a sheaf of spectra $\CO^{\rm top}$ over $\CM_{\rm ell}$. They proved that this sheaf indeed exists. More precisely, there is a sheaf of $E_\infty$-ring spectra $\CO^{\rm top}$ on $\CM_{\rm ell}$ in the \'etale topology, 
\be
\CO^{\rm top}: \quad \{\text{\'etale maps to $\CM_{\rm ell}$}\} \ra \{\text{$E_\infty$-ring spectra}\},
\ee
whose associated presheaf of cohomology theories is that of the elliptic cohomologies.

Then one can talk about the global section of $\CO^{\rm top}$,
\be
\TMF:=\CO^{\rm top}(\CM_{\rm ell}),
\ee 
which is a $E_\infty$-ring spectrum. The associated cohomology theory is no longer an elliptic cohomology. Instead, it should be viewed as the \textit{universal elliptic cohomology theory}. 

\subsection{Topological modular forms}

There are several variants of TMF. One is
\be
\mathrm{Tmf}:=\CO^{\rm top}(\bar{\CM}_{\rm ell}).
\ee
Here $\bar{\CM}_{\rm ell}$ is the Deligne--Mumford compactification of $\CM_{\rm ell}$ by allowing nodal singularities, which $\CO^{\rm top}$ extends over. The connective cover of $\Tmf$ is known as $\tmf$, and has the property
\be
\pi_{<0}(\tmf)=0.
\ee 
The version $\TMF$ is periodic with the ``periodicity element'' commonly denoted as $\Delta^{-24}\in\pi_{-576}(\TMF)$. It is related to the other versions via
\be
\TMF=\tmf\left[\Delta^{-24}\right]=\Tmf\left[\Delta^{-24}\right].
\ee
At the level of homotopy groups, we have the isomorphism between rings,
\be
\pi_*(\TMF)=\pi_*(\tmf)\left[\Delta^{-24}\right].
\ee
Therefore, to understand $\pi_*(\TMF)$, it suffices to describe $\pi_*(\tmf)$. There are three ring homomorphisms involving the latter, and when combined, can shed much light on its structure.

\begin{itemize}
\item The first is the Hurewicz homomorphism from the homotopy groups of spheres
\be
\pi_*(\mathbb{S}) \ra \pi_*(\tmf).
\ee
This map is an isomorphism upto degree 6 and allows us to identify $\tmf$ classes in low degrees with elements in $\pi_*(\mathbb{S})$. Recall that the generator of the latter in low degrees are the unit $1\in\pi_0(\mathbb{S})$ and the two Hopf invariants $\eta\in \pi_1(\mathbb{S})$ and $\nu\in \pi_3(\mathbb{S})$, which satisfy the relations 
\be
2\eta=24\nu=\eta^4=\nu^4=2\nu^2=0
\ee 
and 
\be 
\eta^3=12\nu.
\ee

\item The second is the \textit{topological Witten genus} map 
\be
\Omega^{\String}_* \ra \pi_*(\tmf)
\ee
coming from the $\String$-orientation of $\tmf$, analogous to the Atiyah--Bott--Shapiro map $\Omega^{\Spin}_* \ra \pi_*(\KO)$ given by the Spin orientation of $\KO$. In fact, the Hurewicz homomorphism factorizes through the topological Witten genus map, as it comes from the fact that the sphere spectrum $\mathbb{S}$ is the unit object in the category of $E_\infty$-spectra. Further, this map is surjective, which means that every element in $\tmf$ can be represented by a String manifold.

\item The third is the elliptic genus map
\be
\pi_{*}(\tmf) \ra \mathrm{MF}_{*/2},
\ee
which factorize the Witten genus map
\be
\pi_{*}({\rm MString}) \ra \pi_{*}(\tmf) \ra \mathrm{MF}_{*/2}.
\ee
Over rational numbers, this is an isomorphism
\be
\pi_{*}(\tmf) \otimes \Q \xrightarrow{\sim} \mathrm{MF}_{*/2} \otimes \Q,
\ee
but it has both kernel and cokernel over $\Z$. The kernel is the ideal generated by all torsion classes, while
\be
\text{coker} \left( \pi_n (\tmf) \to \mathrm{MF}_{\frac{n}{2}} \right)
= \begin{cases}
	\Z \slash \frac{24}{\text{gcd} (k,24)} \Z  & ~\text{if}~ n=24k, \\
	(\Z / 2\Z)^{\lceil \frac{n-8}{24} \rceil} & ~\text{if}~ n \equiv 4 \pmod 8, \\
	0 & ~\text{otherwise.}~
\end{cases}
\ee
For example, when $n=24$, $\mathrm{MF}_{12}\simeq \Z\oplus\Z$ is generated by $E_4^3$ and $\Delta$, but $\pi_n (\tmf)\simeq \Z\oplus\Z$ is generated by the pre-image of $E_4^3$ and $24\Delta$.

\end{itemize}

\section{Bauer--Furuta invariants and their generalizations} \label{app:BF}

As was briefly mentioned at the end of Section \ref{sec:twist}, one can expect that invariants of 4-manifolds valued in the torsion subgroups of $\pi_*\text{TMF}$ are closely related to Bauer--Furuta invariants \cite{bauer2004stable,bauer2004stable2} and their possible generalizations. 

In particular, take 6d $(0,1)$ theory that consists of a vector multiplet and a charged hypermultiplet and take the strict 4d limit (that is, forget all KK modes). As we discussed earlier, on a closed simply-connected 4-manifold $M_4$, after the topological twist, bosonic modes of the hypermultiplet are sections of the spinor bundle $\CS^+$, while that of a vector multiplet take values in $\CA$, the space of connections on a line bundle~$L$. The Seiberg--Witten equations define a $U(1)$-equivariant map between these spaces:
\be
\mu: \quad \Gamma (\CS^+ \otimes L) \times \CA \; \longrightarrow \; \Gamma (\CS^- \otimes L) \times \Gamma (\Lambda^{+})
\ee
where the $U(1)$ circle action can be thought of as the framing at a base point on $M_4$. The moduli space of solutions to Seiberg--Witten equations is a quotient by this circle action,
\be
\CM_{SW} \; = \; \mu^{-1} (0) / S^1
\ee
In this setting, the Bauer--Furuta invariant is a stable homotopy class of the $U(1)$-equivariant map of spheres,
\be
\mu_{\text{fin}}: \quad 
S (\mathbb{C}^{c_+} \oplus \R^{d_+}) \; \longrightarrow \; S (\mathbb{C}^{c_-} \oplus \R^{d_-})
\ee
where $\mu_{\text{fin}}:~ \mathbb{C}^{c_+} \oplus \R^{d_+} \to \mathbb{C}^{c_-} \oplus \R^{d_-}$ is a finite-dimensional approximation to $\mu$. When 4-manifold $M_4$ is Spin, this map is, in fact, a Pin$_2$-equivariant proper map
\be
\mu_{\text{fin}}: \quad 
\mathbb{H}^{c_+} \oplus \tilde \R^{d_+} \; \longrightarrow \; \mathbb{H}^{c_-} \oplus \tilde \R^{d_-}
\ee
where $\tilde R$ is the non-trivial 1-dimensional representation of Pin$_2$ defined by the multiplication by Pin$_2/S^1 = \{ \pm 1 \}$, and non-negative integers $c_i$ and $d_i$ satisfy
\be
c_+ - c_- = - \frac{\sigma}{16}
\,,\qquad \qquad
d_+ - d_- = b_2^+
\label{ccddcond}
\ee
The group Pin$_2$ here is the centralizer of $S^1$ in $Sp(1)$, where $Sp(1)$ is the group of unit quaternions and $S^1$ is the intersection of $Sp(1)$ with $\mathbb{C} \subset \mathbb{H}$.
To summarize, one can distinguish three kinds of Bauer--Furuta invariants:

$i)$ non-equivariant ones with values in $\pi_{d+1} (\mathbb{S})$, where $d =  \text{dim} \CM_{SW} = \frac{\lambda^2 - \sigma}{8}$;

$ii)$ $U(1)$-eqiuvariant invariants that take values in $\{ \mathbb{R}^{b_2^+} , \mathbb{C}^{d} \}_{U(1)} \cong \pi^{b_2^+ - 1} (\mathbb{C}{\bf P}^{d-1})$;

$iii)$ and $\text{Pin}_2$-equivariant Bauer--Furuta invariants, valued in $\{ \tilde{\mathbb{R}}^{b_2^+} ,  \mathbb{H}^{\frac{\sigma}{32}} \}_{U(1)}$.

\noindent
For example, the case of $d=0$ corresponds to $\pi_1 (\mathbb{S}) \cong \Z_2$, and the corresponding (non-equivariant) Bauer--Furuta invariant is a mod 2 reduction of the Seiberg--Witten invariant. However, already in the next case of $d=1$, we see that numerical Seiberg--Witten invariants vanish, whereas Bauer--Furuta invariants take values in $\pi_2 (\mathbb{S}) \cong \Z_2$. This Bauer--Furuta invariant is non-trivial {\it e.g.} for the connected sum $K3 \# K3$.

The discussion in the present section suggests a promising avenue for constructing new 4-manifold invariants, in particular, when homotopy theoretic and equivariant techniques are combined together. The simplest example of a promising direction that, to the best of our knowledge, has not been explored so far is the ``multi-monopole version of the Bauer--Furuta invariant'' based on the generalization of Seiberg-Witten equations with $N_f$ spinor fields $\Psi_i$:
\begin{eqnarray}
F_A^+ & = & i \sum_{i=1}^{N_f} \Psi_i \bar \Psi_i, \label{NfSWeq} \\
\partial \!\!\! \slash \, \Psi_i & = & 0, \qquad i = 1, \ldots, N_f, \nonumber
\end{eqnarray}
Unlike the standard Seiberg-Witten equations (which correspond to $N_f=1$), when $N_f > 1$ the moduli space $\mathcal{M}_{N_f}$ of solutions to these multi-monopole equations is non-compact. This problem, however, can be circumvented by considering $SU(N_f)$-equivariant integrals over $\mathcal{M}_{N_f}$ since, as shown in \cite{Dedushenko:2017tdw}, the set of fixed points is compact. Although this gives a finite result for the ``$SU(N_f)$-equivariant multi-monopole invariants,'' they all can be expressed in terms of Seiberg-Witten invariants. A natural way to extract further information is to consider $SU(N_f)$-equivariant Bauer--Furuta invariants based on \eqref{NfSWeq}.

\section{Landscape of 6d SCFTs and geography of 4-manifolds}
\label{sec:geography}

In the study of 6d $(1,0)$ theories, there is a ``landscape problem'': What are the constraints on the coefficients of the anomaly polynomials? Similarly, in the study of 4-manifolds, there is a ``geography problem'': For what values of topological invariants, there exist a simply-connected smooth 4-manifold? In this concluding section, we will discuss the interplay between the landscape of 6d $(1,0)$ theories and the geography of 4-manifolds.

For simplicity, we will focus on the case when $M_4$ is a minimal complex surfaces of general type,\footnote{Minimal surface is a surface that cannot be obtained by blowing up any other surface at a point. Topologically this means that it is not a connected sum of any other surface with $\bar{\mathbb{CP}}^2$. Surfaces of general type on the formal level means that the Kodaira dimension is 2.} then it admits a K\"aher metric, and we consider compactification of 6d $(1,0)$ theories without non-trivial flavor symmetry backgrounds. The supersymmetry of the effective two-dimensional theory in this case will be enhanced to $\CN=(0,2)$. And, as was described in section~\ref{sec:anomaly-reduction}, assuming that the true 2d $U(1)$ R-symmetry for the IR SCFT can be identified with the maximal torus of the 6d $SU(2)_R$ symmetry, one can obtain expressions for left- and right-moving central charges separately. The unitarity of the 2d theory then imposes the following inequalities
\begin{equation}
\begin{array}{l}
c_\text{R}=18\cdot(3 \alpha -2\beta)\sigma+36 \alpha \chi \geq 0, \\
c_\text{L}=18\cdot(3 \alpha -3 \beta +8 \gamma +4 \delta) \sigma+12\cdot( 3\alpha - \beta) \chi \geq 0.
\end{array}
\label{2d-unitarity-bounds}
\end{equation}

The coefficients $\alpha,\beta,\gamma,\delta$ of the 6d anomaly polynomial are also expected to satisfy certain information theory bounds \cite{Yankielowicz:2017xkf} and Hofman-Maldacena bounds:
\begin{equation}
\begin{array}{l}
P_1:=\alpha-4(\beta-4\gamma) \geq 0, \\
P_2:=3\alpha-2\beta \geq 0, \\
P_3:= 8\alpha-6\beta+4\gamma+7\delta \geq 0, \\
P_4:= 2\alpha-9\beta+16\gamma-2\delta \geq 0.
\end{array}
\label{6d-bounds}
\end{equation}
Note that the condition of the positivity of the 6d central charge \cite{Cordova:2015fha}
\begin{equation}
a_\text{6d}=8/3(\alpha-\beta+\gamma)+\delta \geq 0
\label{a6d-bound}
\end{equation}
follows from (\ref{6d-bounds}).

On the other hand, there are known inequalities on the Euler number $\chi$ and signature $\sigma$ for minimal complex surfaces of general type:
\begin{equation}
 \begin{array}{rcl}
(2\chi+3\sigma) &\geq & 0, \\
(2\chi+3\sigma) &\leq &9\cdot\frac{\chi+\sigma}{4}, \\
(2\chi+3\sigma) &\geq &2\cdot\frac{\chi+\sigma}{4}-6,
 \end{array}
\label{min-surface-bounds}
\end{equation}
where $2\chi+3\sigma= c_1^2$ is the square of the canonical class and $(\chi+\sigma)/4= \chi_h$ is the holomorphic Euler number. The first inequality in (\ref{min-surface-bounds}) just means the positivity of the square of the canonical class, the second inequality is known as Bogomolov-Miyaoka-Yau bound and the third one is the Noether bound.

It is very tempting to ask if the geometrical bounds (\ref{min-surface-bounds}) are related to the physical bounds (\ref{2d-unitarity-bounds}), (\ref{a6d-bound}) and (\ref{6d-bounds}). Unfortunately, the answer seems to be negative.

\begin{figure}[tbp]
	\centering
	\includegraphics[scale=0.7]{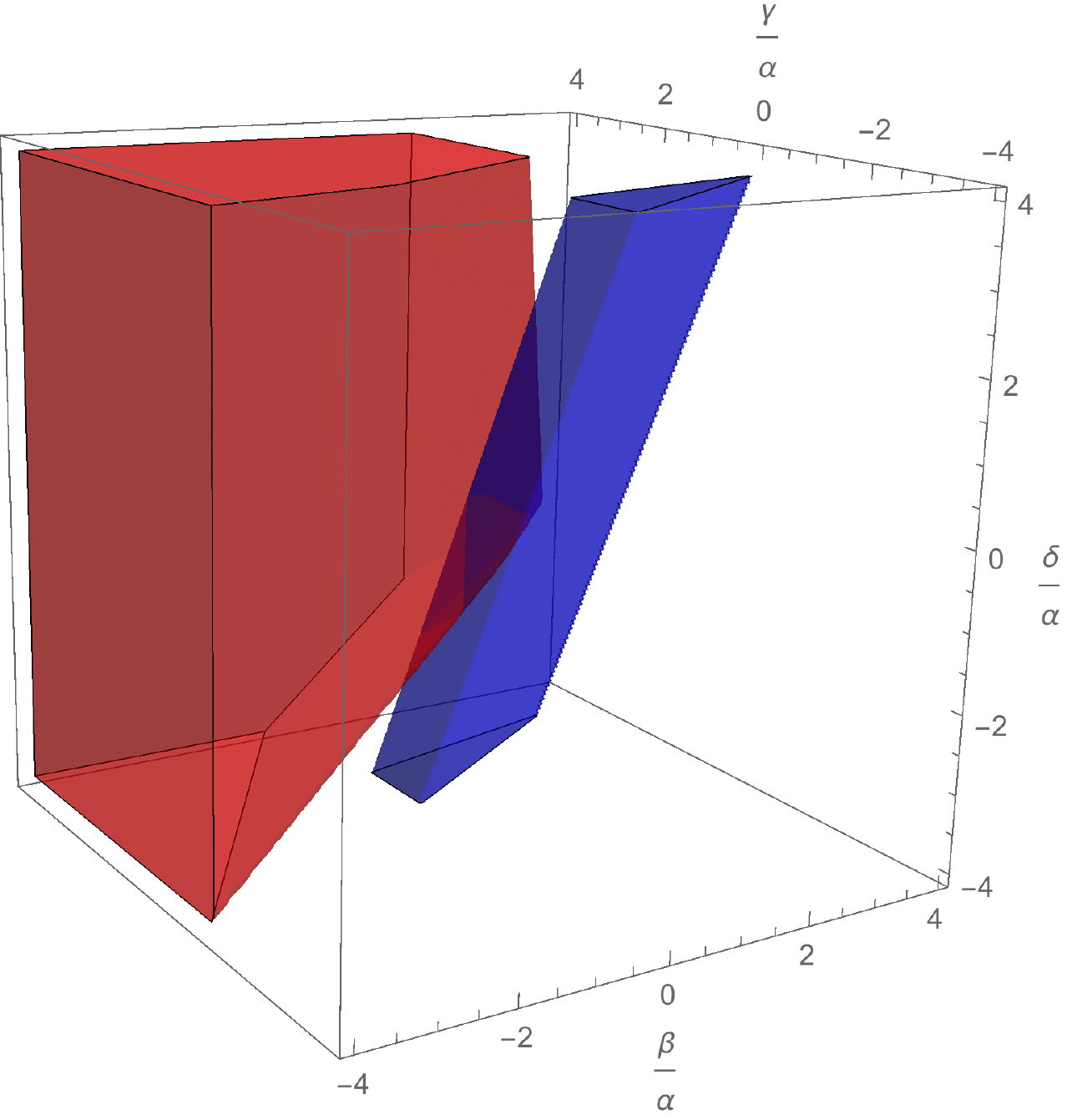}
	\caption{\label{fig:geography-landscape} The figure displays two regions in the 3d projective space of 6d anomaly polynomials in the affine chart where $\alpha\neq 1$, parametrized by $\beta/\alpha,\gamma/\alpha,\delta/\alpha$. The red one is the region carved out by the inequalities (\ref{6d-bounds}). The blue one is the region given by condition that $c_\text{L},c_\text{R}>0$ for almost all minimal surfaces of general type (meaning that we take $\chi$ and $\sigma$ to be large that the inequalities (\ref{min-surface-bounds}) can be replaced by homogeneous ones).}
\end{figure}

Nevertheless one can still ask what the conditions on the 6d anomaly polynomial coefficients $\alpha,\beta,\gamma,\delta$ are so that the unitarity bounds $c_\text{L},c_\text{R}>0$ are satisfied for almost all minimal surfaces of general type (meaning that we take $\chi$ and $\sigma$ to be large that the inequalities (\ref{min-surface-bounds}) can be replaced by homogeneous ones), and whether it has a non-trivial intersection with conditions (\ref{6d-bounds}). The answer turns out to be positive, although neither of two systems of inequalities imply the other. See Figure~\ref{fig:geography-landscape}.

One can check that the (naive) expressions for  supercharges (\ref{2d-unitarity-bounds}) indeed produce positive quantities for $O(-n)$ theories and general rank E-string theories assuming $\chi$ and $\sigma$ satisfy the conditions (\ref{min-surface-bounds}).

Consider $N$ M5-branes probing $\Z_k$ singularities, i.e. $(N,k)$ theories. For the $(N,k)$ theory, we have the following values for the coefficients of the anomaly polynomial \cite{Razamat:2016dpl}
\bea
\alpha &=& \frac{1}{24}(N - 1)(2 + k^2(N^2 + N - 1)), \\
\beta &=& -\frac{1}{48}(N - 1)(k^2 - 2), \\
\gamma &=& \frac{1}{5760}(7(k^2 - 1) + 30(N - 1)), \\
\delta &=& -\frac{1}{1440}((k^2 - 1) + 30(N - 1)).
\eea
The central charges of the effective 2d theory are then given by
\bea
c_\text{R}&=& \frac34 (N-1) \left[\left(4 + k^2 (-2 + 3 N + 3 N^2)\right) \sigma +2 \left(2 + k^2 (N^2 + N -1)\right) \chi\right]\\
c_\text{L}&=& \frac18 \big[\left(12 N-13 +  k^2 (18 N^3 -27 N+10)\right) \sigma \nonumber \\ & & +2 (N-1) \left(10 + k^2 (6 N^2+ 6 N-5)\right) \chi\big]
\eea

One can see that for certain values of the parameters $N,k,\chi$, and $\sigma$, the central charges $c_\text{L}$ and/or $c_\text{R}$ can be negative. In particular, when $M_4=K3$ one has $c_\text{R}=-12(N-1)(k^2)$. However, when both $k$ and $N$ are large and for a generic complex surface one has
\begin{equation}
c_\text{L} \sim c_\text{R} \sim 3/4(2\chi+3\sigma)N^3k^2,
\end{equation}
which is always non-negative for closed K\"ahler manifolds and zero only
for K3, because it is proportional to the square of the canonical
class $c_1^2=2\chi+3\sigma\geq 0$.

\bibliographystyle{JHEP}
\bibliography{4Man}

\providecommand{\href}[2]{#2}\begingroup\raggedright\begin{thebibliography}{10}

\bibitem{Witten:1995zh}
E.~Witten, \emph{{Some comments on string dynamics}},  in \emph{{Future
  perspectives in string theory. Proceedings, Conference, Strings'95, Los
  Angeles, USA, March 13-18, 1995}}, pp.~501--523, 1995,
  \href{https://arxiv.org/abs/hep-th/9507121}{{\ttfamily hep-th/9507121}}.

\bibitem{Strominger:1995ac}
A.~Strominger, \emph{{Open p-branes}},
  \href{https://doi.org/10.1016/0370-2693(96)00712-5}{\emph{Phys. Lett.}
  {\bfseries B383} (1996) 44}
  [\href{https://arxiv.org/abs/hep-th/9512059}{{\ttfamily hep-th/9512059}}].

\bibitem{Heckman:2013pva}
J.~J. Heckman, D.~R. Morrison and C.~Vafa, \emph{{On the Classification of 6D
  SCFTs and Generalized ADE Orbifolds}},
  \href{https://doi.org/10.1007/JHEP06(2015)017,
  10.1007/JHEP05(2014)028}{\emph{JHEP} {\bfseries 05} (2014) 028}
  [\href{https://arxiv.org/abs/1312.5746}{{\ttfamily 1312.5746}}].

\bibitem{Bhardwaj:2015xxa}
L.~Bhardwaj, \emph{{Classification of 6d $ \mathcal{N}=\left(1,0\right) $ gauge
  theories}}, \href{https://doi.org/10.1007/JHEP11(2015)002}{\emph{JHEP}
  {\bfseries 11} (2015) 002}
  [\href{https://arxiv.org/abs/1502.06594}{{\ttfamily 1502.06594}}].

\bibitem{Heckman:2015bfa}
J.~J. Heckman, D.~R. Morrison, T.~Rudelius and C.~Vafa, \emph{{Atomic
  Classification of 6D SCFTs}},
  \href{https://doi.org/10.1002/prop.201500024}{\emph{Fortsch. Phys.}
  {\bfseries 63} (2015) 468}
  [\href{https://arxiv.org/abs/1502.05405}{{\ttfamily 1502.05405}}].

\bibitem{Vafa:1994tf}
C.~Vafa and E.~Witten, \emph{{A Strong coupling test of S-duality}},
  \href{https://doi.org/10.1016/0550-3213(94)90097-3}{\emph{Nucl. Phys.}
  {\bfseries B431} (1994) 3}
  [\href{https://arxiv.org/abs/hep-th/9408074}{{\ttfamily hep-th/9408074}}].

\bibitem{Kapustin:2006pk}
A.~Kapustin and E.~Witten, \emph{{Electric-Magnetic Duality And The Geometric
  Langlands Program}},
  \href{https://doi.org/10.4310/CNTP.2007.v1.n1.a1}{\emph{Commun. Num. Theor.
  Phys.} {\bfseries 1} (2007) 1}
  [\href{https://arxiv.org/abs/hep-th/0604151}{{\ttfamily hep-th/0604151}}].

\bibitem{Witten:2011zz}
E.~Witten, \emph{{Fivebranes and Knots}},
  \href{https://arxiv.org/abs/1101.3216}{{\ttfamily 1101.3216}}.

\bibitem{Gadde:2013sca}
A.~Gadde, S.~Gukov and P.~Putrov, \emph{{Fivebranes and 4-manifolds}},
  \href{https://arxiv.org/abs/1306.4320}{{\ttfamily 1306.4320}}.

\bibitem{Dedushenko:2017tdw}
M.~Dedushenko, S.~Gukov and P.~Putrov, \emph{{Vertex algebras and 4-manifold
  invariants}},  \href{https://arxiv.org/abs/1705.01645}{{\ttfamily
  1705.01645}}.

\bibitem{Dimofte:2011ju}
T.~Dimofte, D.~Gaiotto and S.~Gukov, \emph{{Gauge Theories Labelled by
  Three-Manifolds}},
  \href{https://doi.org/10.1007/s00220-013-1863-2}{\emph{Commun. Math. Phys.}
  {\bfseries 325} (2014) 367}
  [\href{https://arxiv.org/abs/1108.4389}{{\ttfamily 1108.4389}}].

\bibitem{Dimofte:2011py}
T.~Dimofte, D.~Gaiotto and S.~Gukov, \emph{{3-Manifolds and 3d Indices}},
  \href{https://doi.org/10.4310/ATMP.2013.v17.n5.a3}{\emph{Adv. Theor. Math.
  Phys.} {\bfseries 17} (2013) 975}
  [\href{https://arxiv.org/abs/1112.5179}{{\ttfamily 1112.5179}}].

\bibitem{Gukov:2016gkn}
S.~Gukov, P.~Putrov and C.~Vafa, \emph{{Fivebranes and 3-manifold homology}},
  \href{https://arxiv.org/abs/1602.05302}{{\ttfamily 1602.05302}}.

\bibitem{Gukov:2017kmk}
S.~Gukov, D.~Pei, P.~Putrov and C.~Vafa, \emph{{BPS spectra and 3-manifold
  invariants}},  \href{https://arxiv.org/abs/1701.06567}{{\ttfamily
  1701.06567}}.

\bibitem{Witten:1986bf}
E.~Witten, \emph{{Elliptic Genera and Quantum Field Theory}},
  \href{https://doi.org/10.1007/BF01208956}{\emph{Commun. Math. Phys.}
  {\bfseries 109} (1987) 525}.

\bibitem{DelZotto:2015isa}
M.~Del~Zotto, J.~J. Heckman, D.~S. Park and T.~Rudelius, \emph{{On the Defect
  Group of a 6D SCFT}},
  \href{https://doi.org/10.1007/s11005-016-0839-5}{\emph{Lett. Math. Phys.}
  {\bfseries 106} (2016) 765}
  [\href{https://arxiv.org/abs/1503.04806}{{\ttfamily 1503.04806}}].

\bibitem{Bah:2012dg}
I.~Bah, C.~Beem, N.~Bobev and B.~Wecht, \emph{{Four-Dimensional SCFTs from
  M5-Branes}}, \href{https://doi.org/10.1007/JHEP06(2012)005}{\emph{JHEP}
  {\bfseries 06} (2012) 005} [\href{https://arxiv.org/abs/1203.0303}{{\ttfamily
  1203.0303}}].

\bibitem{Razamat:2016dpl}
S.~S. Razamat, C.~Vafa and G.~Zafrir, \emph{{4d $ \mathcal{N}=1 $ from 6d (1,
  0)}}, \href{https://doi.org/10.1007/JHEP04(2017)064}{\emph{JHEP} {\bfseries
  04} (2017) 064} [\href{https://arxiv.org/abs/1610.09178}{{\ttfamily
  1610.09178}}].

\bibitem{Kim:2017toz}
H.-C. Kim, S.~S. Razamat, C.~Vafa and G.~Zafrir, \emph{{E‐String Theory on
  Riemann Surfaces}},
  \href{https://doi.org/10.1002/prop.201700074}{\emph{Fortsch. Phys.}
  {\bfseries 66} (2018) 1700074}
  [\href{https://arxiv.org/abs/1709.02496}{{\ttfamily 1709.02496}}].

\bibitem{Witten:1988ze}
E.~Witten, \emph{{Topological Quantum Field Theory}},
  \href{https://doi.org/10.1007/BF01223371}{\emph{Commun. Math. Phys.}
  {\bfseries 117} (1988) 353}.

\bibitem{ST1}
S.~Stolz and P.~Teichner, \emph{What is an elliptic object?}, p.~247–343.
\newblock London Mathematical Society Lecture Note Series.
\newblock Cambridge University Press, 2004.
\newblock 10.1017/CBO9780511526398.013.

\bibitem{ST2}
S.~{Stolz} and P.~{Teichner}, \emph{{Supersymmetric field theories and
  generalized cohomology}}, {\emph{ArXiv e-prints} (2011) }
  [\href{https://arxiv.org/abs/1108.0189}{{\ttfamily 1108.0189}}].

\bibitem{hopkins1995topological}
M.~J. Hopkins, \emph{{Topological modular forms, the Witten genus, and the
  theorem of the cube}}, {\emph{Proc. ICM, 1994, Birkhauser, Zurich} (1995)
  554}.

\bibitem{bauer2004stable}
S.~Bauer and M.~Furuta, \emph{{A stable cohomotopy refinement of Seiberg-Witten
  invariants: I}}, {\emph{Inventiones mathematicae} {\bfseries 155} (2004) 1}.

\bibitem{bauer2004stable2}
S.~Bauer, \emph{{A stable cohomotopy refinement of Seiberg-Witten invariants:
  II}}, {\emph{Inventiones mathematicae} {\bfseries 155} (2004) 21}.

\bibitem{Kim:2018lfo}
H.-C. Kim, S.~S. Razamat, C.~Vafa and G.~Zafrir, \emph{{Compactifications of
  ADE conformal matter on a torus}},
  \href{https://arxiv.org/abs/1806.07620}{{\ttfamily 1806.07620}}.

\bibitem{Dimofte:2010tz}
T.~Dimofte, S.~Gukov and L.~Hollands, \emph{{Vortex Counting and Lagrangian
  3-manifolds}}, \href{https://doi.org/10.1007/s11005-011-0531-8}{\emph{Lett.
  Math. Phys.} {\bfseries 98} (2011) 225}
  [\href{https://arxiv.org/abs/1006.0977}{{\ttfamily 1006.0977}}].

\bibitem{Ohmori:2014kda}
K.~Ohmori, H.~Shimizu, Y.~Tachikawa and K.~Yonekura, \emph{{Anomaly polynomial
  of general 6d SCFTs}}, \href{https://doi.org/10.1093/ptep/ptu140}{\emph{PTEP}
  {\bfseries 2014} (2014) 103B07}
  [\href{https://arxiv.org/abs/1408.5572}{{\ttfamily 1408.5572}}].

\bibitem{Cordova:2015fha}
C.~Cordova, T.~T. Dumitrescu and K.~Intriligator, \emph{{Anomalies,
  renormalization group flows, and the a-theorem in six-dimensional (1, 0)
  theories}}, \href{https://doi.org/10.1007/JHEP10(2016)080}{\emph{JHEP}
  {\bfseries 10} (2016) 080}
  [\href{https://arxiv.org/abs/1506.03807}{{\ttfamily 1506.03807}}].

\bibitem{Ohmori:2015pua}
K.~Ohmori, H.~Shimizu, Y.~Tachikawa and K.~Yonekura, \emph{{6d
  $\mathcal{N}=(1,0)$ theories on $T^2$ and class S theories: Part I}},
  \href{https://doi.org/10.1007/JHEP07(2015)014}{\emph{JHEP} {\bfseries 07}
  (2015) 014} [\href{https://arxiv.org/abs/1503.06217}{{\ttfamily
  1503.06217}}].

\bibitem{Putrov:2015jpa}
P.~Putrov, J.~Song and W.~Yan, \emph{{(0,4) dualities}},
  \href{https://doi.org/10.1007/JHEP03(2016)185}{\emph{JHEP} {\bfseries 03}
  (2016) 185} [\href{https://arxiv.org/abs/1505.07110}{{\ttfamily
  1505.07110}}].

\bibitem{HenriquesST}
A.~Henriques, \emph{Examples of stolz–teichner cocycles},  in
  \emph{Oberwolfach}, 2015,
  \href{https://www.staff.science.uu.nl/~henri105/PDF/OW-report-STcocycles.pdf}{https://www.staff.science.uu.nl/~henri105/PDF/OW-report-STcocycles.pdf}.

\bibitem{DIVECCHIA1985701}
P.~D. Vecchia, V.~Knizhnik, J.~Petersen and P.~Rossi, \emph{{A supersymmetric
  Wess-Zumino lagrangian in two dimensions}},
  \href{https://doi.org/https://doi.org/10.1016/0550-3213(85)90554-1}{\emph{Nuclear
  Physics B} {\bfseries 253} (1985) 701 }.

\bibitem{Kazama:1988qp}
Y.~Kazama and H.~Suzuki, \emph{{New N=2 Superconformal Field Theories and
  Superstring Compactification}},
  \href{https://doi.org/10.1016/0550-3213(89)90250-2}{\emph{Nucl. Phys.}
  {\bfseries B321} (1989) 232}.

\bibitem{Braun:2004qg}
V.~Braun and S.~Schafer-Nameki, \emph{{Supersymmetric WZW models and twisted
  K-theory of SO(3)}},
  \href{https://doi.org/10.4310/ATMP.2008.v12.n2.a1}{\emph{Adv. Theor. Math.
  Phys.} {\bfseries 12} (2008) 217}
  [\href{https://arxiv.org/abs/hep-th/0403287}{{\ttfamily hep-th/0403287}}].

\bibitem{Freed:2012bs}
D.~S. Freed and C.~Teleman, \emph{{Relative quantum field theory}},
  \href{https://doi.org/10.1007/s00220-013-1880-1}{\emph{Commun. Math. Phys.}
  {\bfseries 326} (2014) 459}
  [\href{https://arxiv.org/abs/1212.1692}{{\ttfamily 1212.1692}}].

\bibitem{Tachikawa:2013hya}
Y.~Tachikawa, \emph{{On the 6d origin of discrete additional data of 4d gauge
  theories}}, \href{https://doi.org/10.1007/JHEP05(2014)020}{\emph{JHEP}
  {\bfseries 05} (2014) 020} [\href{https://arxiv.org/abs/1309.0697}{{\ttfamily
  1309.0697}}].

\bibitem{Witten:2009at}
E.~Witten, \emph{{Geometric Langlands From Six Dimensions}},
  \href{https://arxiv.org/abs/0905.2720}{{\ttfamily 0905.2720}}.

\bibitem{mahowald2009topological}
M.~Mahowald and C.~Rezk, \emph{{Topological Modular Forms of Level 3}},
  {\emph{Pure and Applied Mathematics Quarterly} {\bfseries 5} (2009) 853}.

\bibitem{hill2016topological}
M.~Hill and T.~Lawson, \emph{Topological modular forms with level structure},
  {\emph{Inventiones mathematicae} {\bfseries 203} (2016) 359}.

\bibitem{Witten:1996md}
E.~Witten, \emph{{On flux quantization in M theory and the effective action}},
  \href{https://doi.org/10.1016/S0393-0440(96)00042-3}{\emph{J. Geom. Phys.}
  {\bfseries 22} (1997) 1}
  [\href{https://arxiv.org/abs/hep-th/9609122}{{\ttfamily hep-th/9609122}}].

\bibitem{Sati:2009ic}
H.~Sati, U.~Schreiber and J.~Stasheff, \emph{{Differential twisted String and
  Fivebrane structures}},
  \href{https://doi.org/10.1007/s00220-012-1510-3}{\emph{Commun. Math. Phys.}
  {\bfseries 315} (2012) 169}
  [\href{https://arxiv.org/abs/0910.4001}{{\ttfamily 0910.4001}}].

\bibitem{Monnier:2013kna}
S.~Monnier, \emph{{The global anomaly of the self-dual field in general
  backgrounds}}, \href{https://doi.org/10.1007/s00023-015-0423-z}{\emph{Annales
  Henri Poincare} {\bfseries 17} (2016) 1003}
  [\href{https://arxiv.org/abs/1309.6642}{{\ttfamily 1309.6642}}].

\bibitem{Monnier:2016jlo}
S.~Monnier, \emph{{Topological field theories on manifolds with Wu
  structures}}, \href{https://doi.org/10.1142/S0129055X17500155}{\emph{Rev.
  Math. Phys.} {\bfseries 29} (2017) 1750015}
  [\href{https://arxiv.org/abs/1607.01396}{{\ttfamily 1607.01396}}].

\bibitem{Kapustin:2014tfa}
A.~Kapustin, \emph{{Symmetry Protected Topological Phases, Anomalies, and
  Cobordisms: Beyond Group Cohomology}},
  \href{https://arxiv.org/abs/1403.1467}{{\ttfamily 1403.1467}}.

\bibitem{Kapustin:2014dxa}
A.~Kapustin, R.~Thorngren, A.~Turzillo and Z.~Wang, \emph{{Fermionic Symmetry
  Protected Topological Phases and Cobordisms}},
  \href{https://doi.org/10.1007/JHEP12(2015)052}{\emph{JHEP} {\bfseries 12}
  (2015) 052} [\href{https://arxiv.org/abs/1406.7329}{{\ttfamily 1406.7329}}].

\bibitem{Freed:2016rqq}
D.~S. Freed and M.~J. Hopkins, \emph{{Reflection positivity and invertible
  topological phases}},  \href{https://arxiv.org/abs/1604.06527}{{\ttfamily
  1604.06527}}.

\bibitem{kervaire1963groups}
M.~A. Kervaire and J.~W. Milnor, \emph{{Groups of homotopy spheres: I}},
  {\emph{Annals of Mathematics} (1963) 504}.

\bibitem{Labastida:1995zj}
J.~M.~F. Labastida and M.~Marino, \emph{{Non-Abelian monopoles on four
  manifolds}}, \href{https://doi.org/10.1016/0550-3213(95)00300-H}{\emph{Nucl.
  Phys.} {\bfseries B448} (1995) 373}
  [\href{https://arxiv.org/abs/hep-th/9504010}{{\ttfamily hep-th/9504010}}].

\bibitem{fidkowski2011topological}
L.~Fidkowski and A.~Kitaev, \emph{Topological phases of fermions in one
  dimension}, {\emph{Physical review B} {\bfseries 83} (2011) 075103}.

\bibitem{Feigin:2018bkf}
B.~Feigin and S.~Gukov, \emph{{VOA[$M_4$]}},
  \href{https://arxiv.org/abs/1806.02470}{{\ttfamily 1806.02470}}.

\bibitem{goerss2009topological}
P.~G. Goerss, \emph{Topological modular forms}, {\emph{S{\'e}minaire BOURBAKI}
  (2009) 61}.

\bibitem{lurie2009survey}
J.~Lurie, \emph{A survey of elliptic cohomology},  in \emph{Algebraic
  topology}, pp.~219--277.
\newblock Springer, 2009.

\bibitem{douglas2014topological}
C.~L. Douglas, J.~Francis, A.~G. Henriques and M.~A. Hill, \emph{Topological
  modular forms}, vol.~201. American Mathematical Soc., 2014.

\bibitem{hopkins2002algebraic}
M.~Hopkins, \emph{Algebraic topology and modular forms},  in \emph{Proceedings
  of the International Congress of Mathematicians, Vol. I (Beijing, 2002),
  Higher Ed}, Citeseer, 2002.

\bibitem{mahowald293structure}
M.~Mahowald and M.~Hopkins, \emph{The structure of 24 dimensional manifolds
  having normal bundles which lift to $\mathrm{BO}$[8]. recent progress in
  homotopy theory (baltimore, md, 2000), 89--110}, {\emph{Contemp. Math}
  {\bfseries 293} }.

\bibitem{Gaiotto:2018ypj}
D.~Gaiotto and T.~Johnson-Freyd, \emph{{Holomorphic SCFTs with small index}},
  \href{https://arxiv.org/abs/1811.00589}{{\ttfamily 1811.00589}}.

\bibitem{borcherds1995automorphic}
R.~E. Borcherds, \emph{Automorphic forms on $\mathrm{O}_{s+2,2}(\mathbb{R})$
  and infinite products}, {\emph{Inventiones mathematicae} {\bfseries 120}
  (1995) 161}.

\bibitem{Sati:2008kz}
H.~Sati, U.~Schreiber and J.~Stasheff, \emph{{Fivebrane Structures}},
  \href{https://doi.org/10.1142/S0129055X09003840}{\emph{Rev. Math. Phys.}
  {\bfseries 21} (2009) 1197}
  [\href{https://arxiv.org/abs/0805.0564}{{\ttfamily 0805.0564}}].

\bibitem{ando2010multiplicative}
M.~Ando, M.~J. Hopkins and C.~Rezk, \emph{{Multiplicative orientations of
  KO-theory and of the spectrum of topological modular forms}},
  {\emph{preprint} (2010) }.

\bibitem{bunke2009secondary}
U.~Bunke and N.~Naumann, \emph{Secondary invariants for string bordism and
  tmf}, {\emph{arXiv preprint arXiv:0912.4875} (2009) }.

\bibitem{atiyah1964clifford}
M.~F. Atiyah, R.~Bott and A.~Shapiro, \emph{Clifford modules}, {\emph{Topology}
  {\bfseries 3} (1964) 3}.

\bibitem{AlvarezGaume:1983at}
L.~Alvarez-Gaume, \emph{{Supersymmetry and the Atiyah-Singer Index Theorem}},
  \href{https://doi.org/10.1007/BF01205500}{\emph{Commun. Math. Phys.}
  {\bfseries 90} (1983) 161}.

\bibitem{stolz2004elliptic}
S.~Stolz and P.~Teichner, \emph{What is an elliptic object?}, {\emph{London
  Mathematical Society Lecture Note Series} {\bfseries 308} (2004) 247}.

\bibitem{stolz2011supersymmetric}
S.~Stolz and P.~Teichner, \emph{Supersymmetric field theories and generalized
  cohomology}, {\emph{Mathematical foundations of quantum field theory and
  perturbative string theory} {\bfseries 83} (2011) 279}.

\bibitem{hatcher2000notes}
A.~Hatcher, \emph{Notes on basic 3-manifold topology},
  \href{https://pi.math.cornell.edu/~hatcher/3M/3M.pdf}{https://pi.math.cornell.edu/~hatcher/3M/3M.pdf}.

\bibitem{atiyah1969equivariant}
M.~F. Atiyah, G.~B. Segal et~al., \emph{{Equivariant K-theory and completion}},
  {\emph{J. Differential Geometry} {\bfseries 3} (1969) 9}.

\bibitem{bruner2010connective}
R.~R. Bruner and J.~P.~C. Greenlees, \emph{{Connective real K-theory of finite
  groups}}, no.~169. American Mathematical Soc., 2010.

\bibitem{fok2014real}
C.-K. Fok, \emph{{The Real K-theory of compact Lie groups}}, {\emph{Symmetry,
  Integrability and Geometry: Methods and Applications} {\bfseries 10} (2014)
  22}.

\bibitem{furuta2014equivariant}
M.~Furuta and Y.~Kametani, \emph{{Equivariant version of Rochlin-type
  congruences}}, {\emph{Journal of the Mathematical Society of Japan}
  {\bfseries 66} (2014) 205}.

\bibitem{henriques2007homotopy}
A.~Henriques, \emph{The homotopy groups of tmf and of its localizations},
  {\emph{Topological Mocular Forms, Math. Surveys Monogr} {\bfseries 201}
  (2007) 25}.

\bibitem{eichler1985theory}
M.~Eichler and D.~Zagier, \emph{{The theory of Jacobi forms}}, vol.~55.
  Springer, 1985.

\bibitem{wirthmuller1992root}
K.~Wirthm{\"u}ller, \emph{{Root systems and Jacobi forms}}, {\emph{Compositio
  Math} {\bfseries 82} (1992) 293}.

\bibitem{Bantay:1990yr}
P.~Bantay, \emph{{Orbifolds and Hopf algebras}},
  \href{https://doi.org/10.1016/0370-2693(90)90676-W}{\emph{Phys. Lett.}
  {\bfseries B245} (1990) 477}.

\bibitem{Coste:2000tq}
A.~Coste, T.~Gannon and P.~Ruelle, \emph{{Finite group modular data}},
  \href{https://doi.org/10.1016/S0550-3213(00)00285-6}{\emph{Nucl. Phys.}
  {\bfseries B581} (2000) 679}
  [\href{https://arxiv.org/abs/hep-th/0001158}{{\ttfamily hep-th/0001158}}].

\bibitem{Gaberdiel:2013nya}
M.~R. Gaberdiel, D.~Persson and R.~Volpato, \emph{{Generalised Moonshine and
  Holomorphic Orbifolds}}, {\emph{Proc. Symp. Pure Math.} {\bfseries 90} (2015)
  73} [\href{https://arxiv.org/abs/1302.5425}{{\ttfamily 1302.5425}}].

\bibitem{Persson:2013xpa}
D.~Persson and R.~Volpato, \emph{{Second Quantized Mathieu Moonshine}},
  \href{https://doi.org/10.4310/CNTP.2014.v8.n3.a2}{\emph{Commun. Num. Theor.
  Phys.} {\bfseries 08} (2014) 403}
  [\href{https://arxiv.org/abs/1312.0622}{{\ttfamily 1312.0622}}].

\bibitem{Eckhard:2018raj}
J.~Eckhard, S.~Schäfer-Nameki and J.-M. Wong, \emph{{An $\mathcal{N}=1$ 3d-3d
  Correspondence}}, \href{https://doi.org/10.1007/JHEP07(2018)052}{\emph{JHEP}
  {\bfseries 07} (2018) 052}
  [\href{https://arxiv.org/abs/1804.02368}{{\ttfamily 1804.02368}}].

\bibitem{Benini:2016hjo}
F.~Benini and A.~Zaffaroni, \emph{{Supersymmetric partition functions on
  Riemann surfaces}}, \href{https://doi.org/10.1090/pspum/096}{\emph{Proc.
  Symp. Pure Math.} {\bfseries 96} (2017) 13}
  [\href{https://arxiv.org/abs/1605.06120}{{\ttfamily 1605.06120}}].

\bibitem{joachim2004higher}
M.~Joachim, \emph{Higher coherences for equivariant k-theory},
  {\emph{Structured ring spectra} {\bfseries 315} (2004) 87}.

\bibitem{Seiberg:1996bd}
N.~Seiberg, \emph{{Five-dimensional SUSY field theories, nontrivial fixed
  points and string dynamics}},
  \href{https://doi.org/10.1016/S0370-2693(96)01215-4}{\emph{Phys. Lett.}
  {\bfseries B388} (1996) 753}
  [\href{https://arxiv.org/abs/hep-th/9608111}{{\ttfamily hep-th/9608111}}].

\bibitem{Yankielowicz:2017xkf}
S.~Yankielowicz and Y.~Zhou, \emph{{Supersymmetric Rényi entropy and Anomalies
  in 6d (1,0) SCFTs}},
  \href{https://doi.org/10.1007/JHEP04(2017)128}{\emph{JHEP} {\bfseries 04}
  (2017) 128} [\href{https://arxiv.org/abs/1702.03518}{{\ttfamily
  1702.03518}}].

\end{thebibliography}\endgroup

\end{document}